\documentclass{article}

\usepackage{fullpage}
\usepackage{hyperref}

\usepackage{amsmath,amssymb}

\usepackage{makeidx}         
\usepackage{graphicx}        
\usepackage{wrapfig}
\usepackage{tabularx}

\usepackage{longtable}
\usepackage{multicol}        
\usepackage{multirow}
\usepackage[bottom]{footmisc}
\usepackage{newtxtext}       %
\usepackage[varvw]{newtxmath}       

\usepackage{cite}
\usepackage{caption}
\usepackage{subcaption}
\usepackage{titlesec}

\usepackage{bm}
\usepackage{color}
\usepackage{url}
\usepackage{bigfoot}

\usepackage{setspace}
\usepackage{placeins}
\usepackage{inputenc}
\usepackage{ulem}
\usepackage{upgreek}

\def\##1{{\bf #1}}
\def\=#1{\underline{\underline #1}}

\def\.{\mbox{\tiny{$^\bullet$} }}

\def\le{\left(}
\def\ri{\right)}
\def\les{\left[}
\def\ris{\right]}
\def\lec{\left\{}
\def\ric{\right\}}

\def\ux{\hat{\#u}_{\rm x}}
\def\uy{\hat{\#u}_{\rm y}}
\def\uz{\hat{\#u}_{\rm z}}
\def\um{\hat{\#u}_{\rm m}}
\def\un{\hat{\#u}_{\rm n}}

\def\ualpha{\hat{\#u}_{\alpha}}
\def\ubeta{\hat{\#u}_{\beta}}
\def\ugamma{\hat{\#u}_{\gamma}}

\def\eps{\varepsilon}
\def\teps{\tilde{\varepsilon}}

\def\muo{\mu_{\scriptscriptstyle 0}}
\def\epso{\eps_{\scriptscriptstyle 0}}
\def\etao{\eta_{\scriptscriptstyle 0}}
\def\lambdao{\lambda_{\scriptscriptstyle 0}}
\def\ko{k_{\scriptscriptstyle 0}}

\def\xx{_{\rm xx}}
\def\xy{_{\rm xy}}
\def\xz{_{\rm xz}}
\def\yx{_{\rm yx}}
\def\yy{_{\rm yy}}
\def\yz{_{\rm yz}}
\def\zx{_{\rm zx}}
\def\zy{_{\rm zy}}
\def\zz{_{\rm zz}}

\def\aal{a_{\rm L}}
\def\aar{a_{\rm R}}
\def\bbl{r_{\rm L}}
\def\bbr{r_{\rm R}}
\def\ccl{t_{\rm L}}
\def\ccr{t_{\rm R}}

\def\rLL{r_{\rm LL}}
\def\rLR{r_{\rm LR}}
\def\rRL{r_{\rm RL}}
\def\rRR{r_{\rm RR}}
\def\tLL{t_{\rm LL}}
\def\tLR{t_{\rm LR}}
\def\tRL{t_{\rm RL}}
\def\tRR{t_{\rm RR}}

\def\aas{a_{\rm s}}
\def\aap{a_{\rm p}}
\def\bbs{r_{\rm s}}
\def\bbp{r_{\rm p}}
\def\ccs{t_{\rm s}}
\def\ccp{t_{\rm p}}

\def\rss{r_{\rm ss}}
\def\rsp{r_{\rm sp}}
\def\rps{r_{\rm ps}}
\def\rpp{r_{\rm pp}}
\def\tss{t_{\rm ss}}
\def\tsp{t_{\rm sp}}
\def\tps{t_{\rm ps}}
\def\tpp{t_{\rm pp}}

\def\inc{_{\rm inc}}

\def\rel{_{\rm rel}}

\def\thetainc{\theta\inc}

\def\sp{\#s}
\def\pinc{\#p_+}
\def\pref{\#p_-}

\def\Ei{\#E^{\rm inc}(\#r)}

\def\Er{\#E^{\rm ref}(\#r)}

\def\Et{\#E^{\rm tr}(\#r)}

\def\cpsi{\cos\psi}
\def\spsi{\sin\psi}
\def\stheta{\sin\thetainc}
\def\ctheta{\cos\thetainc}

\def\Re{{\rm Re}}
\def\Im{{\rm Im}}

\def\gn#1{g_{#1}}

\def\Rlin{\les R_{\rm lin}\ris}
\def\Tlin{\les T_{\rm lin}\ris}
\def\Rcirc{\les R_{\rm circ}\ris}
\def\Tcirc{\les T_{\rm circ}\ris}

\def\Sz{{\=S_z}}
\def\matBz{\les\#B_z\ris}

\def\mattK{\les \tilde{\#K}\ris}
\def\matP{\les \#P\ris}
\def\mattP{\les \tilde{\#P}\ris}

\def\matV{\les \#V\ris}
\def\mattV{\les \tilde{\#V}\ris}

\def\mattVone{\les \tilde{\#V}^{(1)}\ris}

\def\mattVtwo{\les \tilde{\#V}^{(2)}\ris}

\def\mattVthree{\les \tilde{\#V}^{(3)}\ris}

\def\matG{\les \#G\ris}

\def\matGone{\les {\#G}^{(1)}\ris}
\def\matGtwo{\les {\#G}^{(2)}\ris}
\def\matGthree{\les {\#G}^{(3)}\ris}

\def\mattM{\les \tilde{\#M}\ris}

\def\mattT{\les \tilde{\#T}\ris}

\def\mattZ{\les \tilde{\#Z}\ris}

\def\matW{\les \tilde{W}(\zeta)\ris}
\def\mattW{\les \tilde{W}(\zeta)\ris}

\def\vecez{\#e(z)}
\def\vechz{\#h(z)}
\def\vectez{\tilde{\#e}(z)}

\def\vecfz{\les\#f(z)\ris}
\def\vectfz{\les\tilde{\#f}(z)\ris}

\def\vecvn{\les{\#v}_n\ris}
\def\vectv{\les{\tilde{\#v}}\ris}
\def\vectvn{\les{\tilde{\#v}}_n\ris}

\def\vectw{\les{\tilde{\#w}}\ris}
\def\vectwn{\les{\tilde{\#w}}_n\ris}

\def\vecvone{\les{\#v}_1\ris}
\def\vecvtwo{\les{\#v}_2\ris}
\def\vecvthree{\les{\#v}_3\ris}
\def\vecvfour{\les{\#v}_4\ris}

\def\vectvone{\les{\tilde{\#v}}_1\ris}
\def\vectvtwo{\les{\tilde{\#v}}_2\ris}
\def\vectvthree{\les{\tilde{\#v}}_3\ris}
\def\vectvfour{\les{\tilde{\#v}}_4\ris}

\def\vectwone{\les{\tilde{\#w}}_1\ris}

\def\vectwthree{\les{\tilde{\#w}}_3\ris}

\def\matA{\les A\ris}

\def\zetau{\zeta_u}
\def\zetav{\zeta_v}
\def\zetaw{\zeta_w}

\def\calP{{\cal P}}
\def\calPone{{\cal P}1}
\def\calPtwo{{\cal P}2}
\def\calPthree{{\cal P}3}
\def\calPthreeS{{\cal P}3S}

\def\calR{{\cal R}}
\def\calRone{{\cal R}1}
\def\calRtwo{{\cal R}2}
\def\calRthree{{\cal R}3}

\def\calT{{\cal T}}
\def\calTone{{\cal T}1}
\def\calTtwo{{\cal T}2}
\def\calTthree{{\cal T}3}

\def\calCone{{\cal C}1}
\def\calCtwo{{\cal C}2}
\def\calCthree{{\cal C}3}
\def\calCfour{{\cal C}4}

\def\betaz{\beta_z}
\def\betay{\beta_y}

\def\Bdc{B_{\rm dc}}

\makeindex

\begin{document}

\begin{center}

\textbf{Theory of Observability of Voigt-Wave Propagation in a Homogeneous Dielectric Material}\\

Akhlesh Lakhtakia$^{1,2,*}$  and Tom G. Mackay$^{3}$ 

$^1$NanoMM~--~Nanoengineered Metamaterials Group,
Department of Engineering Science and Mechanics,
The Pennsylvania State University, University Park, PA 16802, USA\\

$^2$School of Mathematics, University of Edinburgh, Edinburgh EH9 3FD, UK\\

$^3$School of Mathematics and  Maxwell Institute for Mathematical Sciences, University of Edinburgh, Edinburgh EH9 3FD, UK\\

$^*$akhlesh@psu.edu

\end{center}

\begin{abstract}
A Voigt wave can propagate in a dissipative anisotropic dielectric material when a 4$\times$4 propagation matrix is defective, as happens for a uniform plane wave along a singular axis, and the field depends on the propagation distance linearly as well as exponentially. Whether it is observable outside a slab of that material generally depends on whether the defectiveness is inherited by the 2$\times$2 reflection and transmission matrices, which are extracted from the 2$\times$2 blocks of a 4$\times$4 transfer matrix that relates the fields at the two faces of the slab. Explicit tests exist for each type of defectiveness, and a counting argument shows that excitation and observability co-occur at oblique incidence only for isolated combinations of constitutive parameters and slab thickness. Normal incidence is exceptional: Voigt-wave excitation then generally implies a defective transmission matrix for almost all thicknesses, which explains Voigt's 1902 experiment on andalusite. Numerical searches for andalusite, $\beta$-gallium oxide, terbium orthoaluminate
 in a magnetostatic field, and hexagonal boron nitride show that a Voigt wave may be excited without being observable and that a defective reflection or transmission matrix need not signal the excitation of a Voigt wave.\\

\noindent {\bf keywords;} anisotropy, defective matrix, dissipation, exceptional point, propagation matrix, reflection matrix, singular axis, transmission matrix, Voigt wave

\end{abstract}

\section{Introduction}\label{intro}

In any linear homogeneous medium, whether isotropic, anisotropic, or bianisotropic, ordinarily two uniform plane waves can propagate in a specified direction and two more in the opposite direction  \cite{Berry2005,MLProgOpt}. If the two uniform plane waves propagating in a given direction have different complex wavenumbers, the medium is birefringent for that direction, and the two plane waves then necessarily have different polarization states. If  both plane waves have the same complex wavenumber, the medium is unirefringent for that direction. Unirefringence may be true for propagation in all directions, as happens in free space
as well as in any isotropic dielectric material \cite{Chen}. Unirefringence in an anisotropic material may be
confined to specific directions, as for propagation along the optic axis \cite{opticaxis} of a uniaxial dielectric material; or, it may be pathological, holding for all propagation directions \cite{LVV1991}. Even so, the two plane waves in a unirefringence direction possess linearly independent polarization states, so that a plane wave  propagates with its polarization state unchanged. All of the foregoing is ordinarily true.

\subsection{Discovery of Voigt waves}

A remarkable exception, however, was uncovered in 1902 by Voigt \cite{VoigtAP1902} for  Lorentz-reciprocal biaxial \cite{Chen} dielectric materials. Lorentz reciprocity requires the relative permittivity dyadic $\=\eps\rel$ to be symmetric \cite{Krowne}. Voigt
theoretically deduced that dissipation splits each of the two optic axes of such a  material
into two directions along each of which not only is there unirefringence but also only one polarization state can propagate unchanged. He called these directions \textsf{\textit{Singularitäts-} oder \textit{Windungsaxen}}, translated into English as \textsf{singular axes} or \textsf{winding axes}. Voigt concluded that, if a plate cut normal to an optic axis is viewed in convergent circularly polarized light, more light should get through along the singular axis that propagates the incident circular polarization state unchanged than along the other singular axis, which should therefore appear darker. 

For experimentally confirming his theoretical finding, Voigt \cite{VoigtPM1902}  used a plate of andalusite, a pleochroic  crystal that is Lorentz reciprocal, biaxial, and weakly dissipative in the visible spectral regime. The  crystal was cut so that the  normal to both faces of the plate lay very close to one of the two optic axes (which are taken to be the optic axes derivable from $\Re\lec\=\eps\rel\ric$, for a weakly dissipative material such as andalusite \cite{opticaxis}).
 The plate was viewed in convergent blue light through a single polarizer followed by a quarter-wave plate, so that circularly polarized light was incident. Interference rings appeared even without an analyzer, and the field of view was darker on one side of the plane containing the optic axes than on the other. Reversal of the handedness of the incident circularly polarized light interchanged the dark and the bright sides. Voigt interpreted this as light of one circular polarization state being transmitted with little change, and light of the other circular polarization state being partially converted into the first.

 The observation of a dark and a bright spot in the field of view, one on either side of the optic axis, was taken by Voigt as an experimental confirmation of his  theoretical finding \cite{VoigtPM1902}, and he added an appendix to that effect to the theoretical paper \cite{VoigtAP1902}. Half a century later, Pancharatnam \cite{Pancharatnam1955a}   showed that the reverse is true: the singular axis that propagates the incident circular polarization state unchanged should appear darker than the other one. 

Because a single eigenvector cannot span the two-dimensional space of transverse polarization states, the general solution along a singular axis must be completed by a generalized eigenvector \cite{Boyce}, and the fields then depend on the propagation distance not only exponentially but also linearly. Plane waves of this kind are now called Voigt waves \cite{Khapalyuk,FedorovGoncharenko,Ranganath,Gerardin,Borzdov1996}.  Note, however, that Voigt never stated that the
wave propagating along a singular axis has a linear-times-exponential
dependence on the propagation distance, although that dependence can be deduced from
Eqs.~(98), (102), and (109) of his theoretical paper \cite{VoigtAP1902}.

The relative permittivity dyadic $\=\eps\rel$ of a linear homogeneous dielectric material must satisfy two
 conditions for a Voigt wave to propagate in a specified direction \cite{Gerardin}. The first  condition ensures that the two wavenumbers coincide, and the second that only one eigenvector exists for that wavenumber. These conditions show that Voigt waves cannot propagate as uniform plane waves in isotropic or uniaxial dielectric materials, whether dissipative or not, and that they cannot propagate in nondissipative biaxial dielectric materials either \cite{Gerardin,Sturm}. The underlying issue is that of a non-diagonalizable propagation matrix, a matter well known in linear algebra \cite{Bernstein,Golub}.

\subsection{Pancharatnam's correction and insight}
Voigt had worked with the magnetic field phasor $\#H$ because it is always transverse to the propagation direction in a dielectric material. He had then deduced the
dispersion equation, leading to the identification of the singular axes. Pancharatnam approached the issue differently in 1955 \cite{Pancharatnam1955a}. 
Treating the propagation of light as subject to the simultaneous action of linear birefringence and linear dichroism whose principal axes need not coincide, he followed the polarization state on the Poincar\'e sphere. Along a singular axis, the two effects are equal in magnitude and their principal axes are inclined at $\pm45$~deg to each other, so that only one circular-polarization state is propagated unchanged; the oppositely handed circular-polarization state is gradually converted into it, and its transmitted intensity exceeds that of the unchanged state by the factor $1+\varkappa^2d^2$, where $\varkappa$ measures the dichroism and $d$ is the propagation distance. He confirmed his explanation of Voigt waves experimentally on iolite plates in convergent light \cite{Pancharatnam1955b}. 

Pancharatnam \cite{Pancharatnam1955a} also showed that his results agree with those of  electromagnetic theory, in which the 
``vibration" in the wavefront plane is governed by a complex symmetric 2$\times$2 matrix (denoted by $[Q]$ in this paper)
that is formed from the sections of the index 
and absorption ellipsoids. In modern terms, his vibration corresponds to the electric displacement phasor $\#D$, which is always 
transverse to the propagation direction for uniform plane waves in a homogeneous dielectric material \cite{Chen,Borzdov1996,Agranovich}. Furthermore, 
$[Q]$ appears in an eigenvalue equation 
for the two Cartesian components  of $\#D$ in the transverse plane. In a 
nondissipative Lorentz-reciprocal dielectric material, $[Q]$  is real symmetric and therefore always diagonalizable, with two 
orthogonal eigenvectors even along an optic axis. In a dissipative Lorentz-reciprocal
 dielectric material, $[Q]$  is complex symmetric. Therefore, it can be defective, i.e., it can have a repeated eigenvalue with 
 only one eigenvector, so that it is similar to a Jordan block of size $2$
and not to a diagonal matrix. The exponential of a defective matrix leads to a term that is linear in the propagation distance, and the Voigt 
wave then must have a linear-times-exponential dependence on that distance. Since the sole eigenvector $[q_a, q_b]^T$ of a defective 
complex symmetric 2$\times$2 matrix satisfies the relation $q_a^2+q_b^2=0$, the Voigt wave is circularly polarized in a Lorentz-reciprocal  dielectric 
material (such as andalusite or iolite).

Pancharatnam's insight ties up directly with the notion of exceptional points \cite{BerryDennis2003,Berry2004,Heiss2012,MiriAlu2019}. An exceptional point is a point in the parameter space of a non-Hermitian operator at which two (or more) eigenvalues coalesce and so do their eigenvectors, the operator becoming defective there. By contrast, at a diabolical point of a Hermitian operator the eigenvalues coalesce but the eigenvectors remain distinct \cite{Berry2004,Grundmann}. Near an exceptional point, the eigenvalues vary as the square root of the deviation of the parameters, and encircling the exceptional point in parameter space interchanges the two eigenvectors. The optic axes of a nondissipative biaxial dielectric material with symmetric $\=\eps\rel$ are diabolical points in the space of propagation directions. When dissipation is introduced, each diabolical point  generically splits into two exceptional points (but see Sec.~\ref{nonexistence}), which are the singular axes \cite{BerryDennis2003}. Voigt waves are thus the earliest known manifestation of exceptional points in optics \cite{Richter2019}.

Both anisotropy and dissipation are needed for Voigt-wave propagation in a passive dielectric material \cite{Gerardin,MLProgOpt}. Parenthetically, if a dissipative material described by $\=\eps\rel$ supports Voigt-wave propagation, then so does the active material whose relative permittivity dyadic is the  Hermitian conjugate of $\=\eps\rel$, the Voigt wave then exhibiting linear gain with propagation distance \cite{MLactive2016}. Bianisotropic materials offer still greater scope \cite{Berry2005,MLProgOpt}.

\subsection{Current interest}
With new optical materials being investigated, interest in Voigt waves has been renewed. Sturm and Grundmann \cite{Sturm} classified Lorentz-reciprocal biaxial dielectric materials by the degeneracy of the refractive indices of their four singular axes, derived analytical expressions for the orientations of the singular axes of monoclinic and orthorhombic materials, and analyzed the spectral dispersion of those orientations in monoclinic $\beta$-Ga$_2$O$_3$, an ultrawide-bandgap semiconductor of interest for power electronics and ultraviolet optoelectronics. Grundmann \textit{et al.} \cite{Grundmann} reviewed the singular axes of anisotropic crystals and showed that planar microcavities containing anisotropic materials behave as effectively biaxial systems with exceptional points.  Following Voigt \cite{Voigt1902b,VoigtParadox}, they also solved for reflection and transmission of a  plane wave normally incident on a $\beta$-Ga$_2$O$_3$ slab cut perpendicular to a singular axis and determined that the transmission matrix was then defective.

Exceptional points were subsequently demonstrated in ZnO-based microcavities, in which the coupling to the exterior provides the non-Hermiticity even if the constituent materials are transparent \cite{Richter2017,Richter2019}; the interest there lies in circularly polarized eigenmodes that could be harnessed for circularly polarized emission without magnetic fields or spin injection. 
Motivated by lasing near an optic axis of a biaxial gain medium, Brenier \cite{Brenier2015} experimentally investigated the Nd-doped monoclinic  crystal KGd(WO$_4$)$_2$ close to 810-nm free-space wavelength, in which spectral regime it absorbs strongly.  Each optic axis was found to split into two singular axes, each of which supports the propagation of only one circular polarization state unchanged.

Engineered materials open further avenues. A homogenized composite material can support Voigt waves even though none of its constituent materials does \cite{ML2003,ML2004,Mackay2011}, and the directions of the singular axes can be designed through the shapes, orientations, and volume fractions of the constituent particles \cite{Mackay2011}. Infiltration of a porous anisotropic material by a fluid alters the singular axes, which offers a route to tunability and optical sensing \cite{Mackay2014}, as does the electro-optic (Pockels) effect \cite{Mackay2015}. Nanotechnology also enables the fabrication of structurally chiral materials, in which anomalous axial propagation of the Voigt type can arise \cite{Lakhtakia1998}, and of materials with extreme anisotropy \cite{Elser,Sihvola,MackayGiant}. One can therefore envisage Voigt-wave-based devices such as sensors that register small changes in an infiltrant through shifts of the singular axes, compact circular-polarization filters,  circularly polarized light sources that do not rely on structural chirality, electrically switchable polarization converters, and amplifiers with linear as well as exponential gain along designated directions.

\subsection{Observability of Voigt waves}
An important question has, however, been overlooked. Can a Voigt wave be excited but not be observed? Voigt's theoretical treatment in  1902   \cite{VoigtAP1902} of normal incidence neither separated propagation from transmission nor
 accounted for reflection. Voigt considered reflection separately and only for normal incidence on
  a weakly dissipative dielectric crystal cut perpendicular (or nearly so) to a singular axis \cite{Voigt1902b,VoigtParadox}. 
Grundmann \textit{et al.} \cite{Grundmann} solved the boundary-value problem exactly for the same configuration
but for a strongly dissipative dielectric crystal.

A Voigt wave exists  inside a material. Its presence can be inferred only from measurements of the reflected and transmitted fields when a plane wave is incident on a slab of finite thickness $L$ of that material. A 4$\times$4  matrix $\mattP$
can be formulated for the propagation of electromagnetic fields from one face of the slab to the other, and the
matrix  $\mattM=\exp\lec{i\mattP L}\ric$ then relates the fields at the two faces of the slab, with $i=\sqrt{-1}$. Then, 2$\times$2
 reflection and transmission matrices can be extracted from 2$\times$2 blocks of a 4$\times$4 transfer matrix $\mattT$ derived from $\mattM$.
 
Voigt's own observations were premised on the transmission matrix in the circular-polarization basis being (very nearly) defective \cite{VoigtPM1902}. As block extraction does not necessarily preserve Jordan structure, the defectiveness of $\mattP$ (and therefore of $\mattM$)  need not be inherited by the reflection and transmission matrices, and a Voigt wave could be excited inside the slab without leaving an unambiguous signature outside. Conversely, although defective reflection and transmission matrices can be engineered \cite{Gallant}, they do not imply that a Voigt wave propagates inside the slab. The mere excitation of a Voigt wave therefore does not guarantee its observability, and an apparent observation can be misleading. 

We have therefore initiated here a theory of observability of Voigt-wave propagation. We deem a Voigt wave observable in transmission (reflection) only when the transmission (reflection) matrix is also defective, i.e., when the exceptional point of 
$\mattP$ coincides with an exceptional point of an externally measurable  matrix. The rationale for adopting this criterion
and the sense in which it is not essential are discussed  in Sec.~\ref{essential}.
We have confined ourselves here to a slab of a homogeneous dissipative dielectric material. Our theory of observability of Voigt waves  is straightforwardly
extensible to active dielectric materials \cite{MLactive2016} and even to bianisotropic materials \cite{Berry2005,MLProgOpt}.

\subsection{Plan and notation}
The plan of this paper is as follows. The boundary-value problem (BVP) of reflection and transmission of an obliquely incident plane wave by a homogeneous anisotropic dielectric slab is formulated in Sec.~\ref{bvp}, culminating in expressions for reflection and transmission matrices. Section~\ref{vwe} is devoted to Voigt-wave excitation. The characteristic polynomial of $\mattP$ is presented and the three types of defectiveness of that matrix (and a specialization of one of them) are identified together with explicit tests for each; the special nature of normal incidence is established;   the defectiveness of $\mattP$ is shown to be inherited by   $\mattT$ relating the amplitudes of the incident, reflected, and transmitted plane waves; and the differences between uniform and nonuniform Voigt waves are  described. In Sec.~\ref{observability}, Voigt's experimental observations are revisited theoretically, conditions for the defectiveness of the transmission and reflection matrices are formulated, the exceptional role of normal incidence is proved,  a counting argument is used to establish when Voigt-wave excitation and defectiveness of the transmission and/or reflection matrices can co-occur, and the role of that defectiveness as the criterion of observability is examined. Illustrative numerical results for andalusite (an orthorhombic material), $\beta$-Ga$_2$O$_3$ (a monoclinic material), TbAlO$_3$ (an orthorhombic material) in a magnetostatic field, and hexagonal boron nitride (a uniaxial material) are presented in Sec.~\ref{ie}. Concluding remarks are provided in Sec.~\ref{cr}.

An $ \exp(-i \omega t)$  dependence on time $t$ is implicit, where $ \omega$ is the angular frequency.
With $ \epso$ and $\muo$, respectively,
denoting the permittivity and permeability  of free space,
the free-space wavenumber is denoted by $\ko = \omega \sqrt{\epso \muo}$,    $\lambdao=2\pi/\ko$ is the free-space
wavelength, and $\etao=\sqrt{\muo/\epso}$ is the intrinsic impedance of free space.

Vectors are in boldface and unit vectors are additionally decorated by a caret on top. 
The Cartesian coordinate system $(x,y,z)$ is adopted
with unit vectors $\ux$, $\uy$, and $\uz$. Dyadics are double underlined,
with $\=0$ as the null dyadic and $\=I$ as the identity dyadic. 
Column vectors and square matrices are denoted by a  symbol enclosed
within square brackets. That symbol is boldface for  column 4-vectors and 4$\times$4 matrices.
 $[I]$ and $[\#I]$ are identity matrices. $[0]$ and $[\#0]$ denote either null column vectors or
 null square matrices.
The superscript $^T$ denotes the transpose.

\section{Boundary-value problem}\label{bvp}
The half-space $z < 0$ is the region of incidence and reflection, while
the half-space $z > L$ is the region of transmission. Both half-spaces are vacuous.
The region $0<z<L$ is occupied by a homogeneous anisotropic dielectric material 
with relative
permittivity dyadic
\begin{equation}
\=\eps\rel=\displaystyle{\sum_{m\in\lec{x,y,z}\ric}\sum_{n\in\lec{x,y,z}\ric}  \eps_{\rm mn}\,\um\un}
 \in\mathbb{C}^{3\times3}\,.
\label{eq:eps}
\end{equation}
Addition of the condition $\eps \zz\ne0$ allows elimination of $\uz\.\#E$ to formulate a 4$\times$4 matrix treatment of the 
BVP for the reflection and transmission of a plane wave incident on a slab \cite{MLProgOpt}.

\subsection{Incident, reflected, and transmitted plane waves}
A plane
wave, propagating in the half-space
$z < 0$ at an angle $\thetainc\in[0,\pi/2)$ with respect to the $z$ axis and at an angle $\psi\in[0,2\pi)$ with respect
to the $x$ axis in the $xy$ plane, is incident on the dielectric slab.
The electric  field phasor associated
with the incident plane wave is represented as  
\begin{subequations}
\begin{eqnarray}
\label{eqEi-lin}
\Ei&=& \le \aas \sp + \aap\pinc\ri  \exp{\les i\kappa \le x\cos\psi + y\sin\psi \ri
\ris} \exp{\left(i\ko z{\cos\thetainc}\right)}\,, \qquad   z < 0\,,
\\
\nonumber
&=&
\les \frac{\le i\sp - \pinc \ri}{\sqrt{2}} \, \aal -\, \frac{\le i\sp +\pinc
\ri}{\sqrt{2}} \, \aar \ris
\exp{\les i\kappa \le x\cos\psi + y\sin\psi \ri
\ris}\\
 && \times \exp{\left(i\ko z{\cos\thetainc}\right)}\,, \qquad   z < 0\,,
 \label{eqEi-circ}
\end{eqnarray}
\end{subequations}
where
 \begin{equation}
 \left.\begin{array}{l}
\kappa =
\ko\sin{\thetainc}\\[5pt]
\sp=-\ux\sin\psi + \uy \cos\psi \\[5pt]
\#p_\pm=\mp\le \ux \cos\psi + \uy \sin\psi \ri \cos{\thetainc} 
 + \uz \sin{\thetainc}
\end{array}
\right\}
\, .
\end{equation}
The amplitudes  of the perpendicular- and parallel-polarized
components, respectively, are denoted by $\aas$ and $\aap$ in Eq.~(\ref{eqEi-lin}).
The amplitudes  of the left-circularly polarized
(LCP)  and  the right-circularly polarized (RCP)
components of the incident plane wave are denoted by 
$\aal$ and
$\aar$, respectively, in Eq.~(\ref{eqEi-circ}).  

The   electric   field phasor of the reflected plane wave is expressed as
\begin{subequations}
\begin{eqnarray}
\label{eqEr-lin}
\Er&=& \le \bbs\sp+\bbp\pref\ri \exp{\les i\kappa \le x\cos{\psi} + y\sin{\psi} \ri
\ris} \exp\left({-i\ko   z\cos{\thetainc}}\right) \,,\qquad z < 0\,,
\\
\nonumber
&=&-\, \les \frac{\le i\sp - \pref \ri}{\sqrt{2}} \, {\bbl} -\, \frac{\le i\sp +
\pref \ri}{\sqrt{2}} \, {\bbr} \ris  \exp{\les i\kappa \le x\cos{\psi} + y\sin{\psi} \ri
\ris} \\
&&\times\,\exp\left({-i\ko   z\cos{\thetainc}}\right)\,,\qquad z < 0 \, ,
\label{eqEr-circ}
\end{eqnarray}
\end{subequations}
and the electric   field phasor of the transmitted plane wave is represented as
\begin{subequations}
\begin{eqnarray}
\label{eqEt-lin}
\Et&=&\le \ccs\sp +\ccp\pinc\ri    \exp{\les i\kappa \le x\cos{\psi} + y\sin{\psi} \ri
\ris}  \exp\les{i\ko   (z-L)\cos{\thetainc}}\ris \,,\qquad z >L\,,
\\
\nonumber
&=& \les \frac{\le i\sp - \pinc \ri}{\sqrt{2}} \, {\ccl} -\, \frac{\le i\sp +\pinc
\ri}{\sqrt{2}} \, {\ccr} \ris  \exp{\les i\kappa \le x\cos{\psi} + y\sin{\psi} \ri
\ris}\\
\label{eqEt-circ}
& &  \times \,  \exp\les{i\ko   (z-L)\cos{\thetainc}}\ris \, , \qquad z > L \, .
\end{eqnarray}
\end{subequations}
Linear reflection amplitudes are denoted by $\bbs$ and $\bbp$  in Eq.~(\ref{eqEr-lin}), whereas the
circular reflection amplitudes are denoted by $\bbl$ and $\bbr$ in Eq.~(\ref{eqEr-circ}).
Similarly,  $\ccs$ and $\ccp$ are the linear transmission amplitudes in Eq.~(\ref{eqEt-lin}),
whereas $\ccl$ and $\ccr$ are the circular transmission amplitudes in Eq.~(\ref{eqEt-circ}).

\subsection{Field phasors in the region $0<z<L$}
In the region $0<z<L$   occupied by  the anisotropic dielectric material with relative
permittivity dyadic $\=\eps\rel$,   we define the auxiliary field phasors
\begin{equation}
\left.\begin{array}{l}
\vecez =\#E(\#r) \exp\les -i\kappa (x\cos\psi + 
y\sin\psi)\ris
\\[5pt]
\vechz = \#H(\#r)  \,\exp\les- i\kappa (x\cos\psi + 
y\sin\psi)\ris
\end{array}\right\}\,, \qquad z\in(0,L)\,.
\end{equation}
Then the electromagnetic field phasors in this  region  satisfy the  matrix ordinary differential equation 
\begin{equation}
\label{eq7.12}
\frac{d}{dz}\, \vecfz = i \matP\. \vecfz\,,\quad z\in(0,L)\,,
\end{equation}
where
\begin{equation}
\label{def-f}
\vecfz =
\les {e_x(z),\ e_y(z),\, h_x(z),\,h_y(z)}\ris^T
\end{equation}
is a column 4-vector and the 4$\times$4 propagation matrix $\matP$ is independent of $z$.

In order to express $\matP$ compactly, we define the rotated relative permittivity dyadic
\begin{equation}
\label{def-teps}
\=\teps\rel =  \Sz^T   \. \=\eps\rel \. \Sz 
\end{equation}
with components $\teps\xx$, etc.,
where the rotation dyadic
\begin{equation}
\label{def-Sz}
\Sz=\uz\uz+\left(\ux\ux+\uy\uy\right)\cpsi+\left(\uy\ux-\ux\uy\right)\spsi
\end{equation}
describes a rotation by angle $\psi$ about the $z$ axis. Then,
\begin{equation}
\matP=\matBz\.\mattP\.\matBz^T\,,
\end{equation}
where the 4$\times$4 rotation matrix
\begin{equation}
\label{def-Bz}
\matBz=
\les
\begin{array}{cccc}
 \cpsi & -\spsi & 0  & 0\\
 \spsi & \cpsi & 0 & 0\\
0 & 0 &  \cpsi & -\spsi \\
0 & 0 &  \spsi & \cpsi 
\end{array}
\ris\,
\end{equation}
and the 4$\times$4 matrix
\begin{equation}
\label{def-tP}
\mattP=
\les
\begin{array}{cccc}
 -\frac{\kappa  \teps \zx}{\teps \zz} & -\frac{\kappa   \teps \zy}{\teps\zz} & 0 & \omega \muo-\frac{\kappa^2}{\omega\epso\teps\zz}  \\
0 & 0 & -\omega \muo & 0 \\
\epso \left(\frac{\omega  \teps \yz \teps \zx}{\teps \zz}-\omega  \teps
   \yx\right) & \frac{\kappa^2  }{\omega \muo}+\epso
   \left(\frac{\omega  \teps \yz \teps \zy}{\teps \zz}-\omega  \teps \yy\right) & 0 & \frac{\kappa   \teps \yz}{\teps \zz} \\
 \omega \epso \left(\teps \xx-\frac{\teps \xz \teps \zx}{\teps \zz}\right) &  \omega \epso \left(\teps
   \xy-\frac{\teps \xz \teps \zy}{\teps \zz}\right) & 0 & -\frac{\kappa  \teps \xz}{\teps \zz} \\
\end{array}
\ris\,,
\end{equation}
with $\teps \zz = \eps \zz \neq 0$.

The solution of Eq.~\eqref{eq7.12} can be written straightforwardly as
\begin{equation}
\label{soln}
\vecfz =\matBz\.\exp\lec i \mattP (z-z_0)\ric\.\matBz^T\. \les\#f(z_0)\ris \,,
\end{equation}
for all $z\in(0,L)$ and all $z_0\in(0,L)$, since  $\mattP$ is independent of $z$.

\subsection{BVP solution}
Equation~\eqref{soln} straightforwardly delivers
\begin{subequations}
\begin{equation}
[\#f(L^-)] =\matBz\.\mattM\.\matBz^T\. [\#f(0^+)] \,.
\end{equation}
Since the planes $z=0$ and $z=L$ are  current-free, the
continuity conditions $[\#f(L^+)]=[\#f(L^-)]$ and $[\#f(0^-)]=[\#f(0^+)]$ hold, leading to
\begin{equation}
[\#f(L^+)] =\matBz\.\mattM\.\matBz^T\. [\#f(0^-)] \,.
\end{equation}
\end{subequations}
But
\begin{subequations}
\begin{equation}
\label{f0}
\les\#f(0^-)\ris =
\matBz\. \mattK\.
\les \aas,\, \aap,\,\bbs,\,\bbp\ris^T
\,,
\end{equation}
follows from Eqs.~\eqref{eqEi-lin} and \eqref{eqEr-lin}, whereas
\begin{equation}
\label{fL}
\les\#f(L^+)\ris =
\matBz\. \mattK\.
\les \ccs,\, \ccp,\,0,\,0 \ris^T
\,,
\end{equation}
\end{subequations}
follows from
Eq.~\eqref{eqEt-lin}, with
 the 4$\times$4 matrix
\begin{equation}
\mattK
=
\les\begin{array}{cccc}
0& - \ctheta & 0 &  \ctheta\\[2pt]
1 & 0 & 1 & 0\\[2pt]
-\etao^{-1}\, \ctheta &0&\etao^{-1}\, \ctheta &0\\[2pt]
0 &-\etao^{-1}  & 0 & -\etao^{-1} 
\end{array}\ris\,.
\label{eq7.39}
\end{equation}

The next result is the matrix equation
\begin{subequations}
\begin{equation}
\label{BVPalgebra}
\les \ccs,\, \ccp,\,0,\,0 \ris^T
 = 
\mattT\.
\les \aas,\, \aap,\,\bbs,\,\bbp\ris^T
\,,
\end{equation}
where the 4$\times$4 transfer matrix
\begin{equation}
\mattT= 
\mattK^{-1}\.\mattM\. \mattK \,.
\end{equation}
\end{subequations}
Equation~\eqref{BVPalgebra} can be solved to deliver the reflection amplitudes $\lec\bbs,\bbp\ric$ and transmission
amplitudes $\lec\ccs,\ccp\ric$ in terms of the incidence amplitudes $\lec\aas,\aap\ric$.

\subsection{Reflection and transmission matrices}
Our current interest generally lies in determining the reflection and transmission coefficients
entering the 2$\times$2 matrices  in each of the following four relations:
\begin{subequations}
\begin{equation}
\label{eq13a}
\les \begin{array}{c} \bbs \\ \bbp  \end{array}\ris  =
\Rlin \.
\les \begin{array}{c} \aas \\ \aap \end{array}\ris  =
\les \begin{array}{cc} \rss &\quad \rsp \\ \rps &\quad \rpp \end{array}\ris \.
\les \begin{array}{c} \aas \\ \aap  \end{array}\ris
\, , \end{equation}
\begin{equation}
\label{eq13b}
\les \begin{array}{c} \ccs \\ \ccp  \end{array}\ris =
\Tlin \.
\les \begin{array}{c} \aas \\ \aap \end{array}\ris =
\les \begin{array}{cc} \tss &\quad \tsp \\ \tps &\quad \tpp \end{array}\ris \.
\les \begin{array}{c} \aas \\ \aap \end{array}\ris 
\, ,
\end{equation}
\begin{equation}
\label{eq13c}
\les \begin{array}{c} \bbl \\ \bbr  \end{array}\ris = 
\Rcirc \.
\les \begin{array}{c} \aal \\ \aar  \end{array}\ris =
\les \begin{array}{cc} \rLL&\quad \rLR \\ \rRL &\quad \rRR \end{array}\ris \.
\les \begin{array}{c} \aal \\ \aar  \end{array}\ris 
\, , 
\end{equation}
and
\begin{equation}
\label{eq13d}
\les \begin{array}{c} \ccl \\ \ccr  \end{array}\ris  = 
\Tcirc \.
\les \begin{array}{c} \aal \\ \aar  \end{array}\ris=
\les \begin{array}{cc} \tLL &\quad \tLR \\ \tRL &\quad \tRR \end{array}\ris \.
\les \begin{array}{c} \aal \\ \aar  \end{array}\ris 
\, .
\end{equation}
\end{subequations}
From Eqs.~(\ref{eqEi-lin})--(\ref{eqEt-circ}),
\begin{subequations}
\begin{equation}
\label{Rlin-to-Rcirc}
\left.\begin{array}{l}
\Rcirc=-\matA^{-1}\.\Rlin\.\matA
\\
 \Tcirc=\matA^{-1}\.\Tlin\.\matA
 \end{array}\right\}\,,
\end{equation}
where
\begin{equation}
\label{def-matA}
\matA=\tfrac1{\sqrt2}\les\begin{array}{cc}i&-i\\-1&-1\end{array}\ris\,.
\end{equation}
\end{subequations}

If   the transfer matrix $\mattT$ is partitioned into four 2$\times$2 blocks as
\begin{equation}
\label{T-partition}
\mattT=\les\begin{array}{c|c}
\les T^{(11)}\ris \quad & \les T^{(12)}\ris
\\[5pt]
----- &----- 
\\[5pt]
\les T^{(21)}\ris \quad & \les T^{(22)}\ris
\end{array}\ris\,,
\end{equation}
then,
\begin{equation}
\label{RlinTlin}
\left.\begin{array}{l}
\Rlin=-\les T^{(22)}\ris^{-1}\.\les T^{(21)}\ris
\\[5pt]
\Tlin=\les T^{(11)}\ris- \les T^{(12)}\ris\.\les T^{(22)}\ris^{-1}\.\les T^{(21)}\ris
\end{array}\ric\,
\end{equation}
follow from Eq.~\eqref{BVPalgebra}. Thus, nonsingularity of the 2$\times$2 matrix $\les T^{(22)}\ris$
 is the solvability condition of Eq.~\eqref{BVPalgebra}, 
 and $\Tlin$ is the Schur complement \cite{Bernstein,Golub} of $\les T^{(22)}\ris$ in $\mattT$.

\section{Voigt-wave excitation}\label{vwe}
\subsection{Characteristic polynomial}
Eigenvalue analysis of the matrix $\matP$ is required for analytical progress, in view of
the differential equation \eqref{eq7.12}.  The matrices $\matP$ and $\mattP$
have the same eigenvalues, and
the eigenvectors  of $\matP$  are simply related to the eigenvectors   of $\mattP$.
Therefore, we can work with the eigenvalue equation $\mattP\.\vectv =g\vectv$ for $\mattP$, with
$g$ standing for the eigenvalues and $\vectv$ for the eigenvectors of $\mattP$.

The characteristic polynomial of  
 $\mattP$, normalized as $\tilde{\Gamma}(\zeta)=\left(\teps\zz/\ko^4\right)\det\left(\mattP-\ko\zeta\les\#I\ris\right)$, is 
\begin{eqnarray}
\nonumber
\tilde{\Gamma}(\zeta)&=&
\zeta^4\teps\zz+ \zeta^3 (\teps\xz+\teps\zx)\stheta
\\
\nonumber
&&
\quad+\zeta^2 \les(\teps\xx+\teps\zz)\sin ^2\thetainc -(\teps\xx\teps\zz-\teps\xz\teps\zx)-(\teps\yy\teps\zz-\teps\yz\teps\zy)\ris
\\
\nonumber
&&
\quad+
 \zeta \les\teps\xy\teps\yz+\teps\yx\teps\zy-(\teps\xz+\teps\zx)(\teps\yy-\sin^2\thetainc)\ris\stheta
\\
\label{charpoly}
&&
\quad+
\det\les \=\teps\rel\ris
-\big[\teps\xx(\teps\yy+\teps\zz)-\teps\xy\teps\yx-\teps\xz\teps\zx\big]\sin^2\thetainc+\teps\xx\,\sin^4\thetainc\,,
\end{eqnarray}
where $\zeta=g/\ko$. Equivalently, $\tilde{\Gamma}(\zeta)=\det\mattW$, where
\begin{equation}
\label{3x3 matrix}
\mattW=
 \begin{bmatrix}
 \teps\xx-\zeta^2&\teps\xy&\teps\xz+\zeta\sin\thetainc\\
 \teps\yx&\teps\yy-\sin^2\thetainc-\zeta^2&\teps\yz\\
 \teps\zx+\zeta\sin\thetainc&\teps\zy&\teps\zz-\sin^2\thetainc
 \end{bmatrix} 
\end{equation}
is a 3$\times$3 matrix emerging from the Maxwell equations.

The four eigenvalues $\lec \gn1,\gn2,\gn3,\gn4\ric$ of $\mattP$  
emerge from the four solutions of the equation
\begin{equation}
\tilde{\Gamma}(\zeta)=0\,.
\label{charpolyeqn}
\end{equation}
Restricting ourselves to  dissipative (and therefore passive) materials, we label the eigenvalues so that $\Im\lec g_n\ric>0$ for $n\in\lec1,2\ric$ and $\Im\lec g_n\ric<0$ for $n\in\lec3,4\ric$.
Thus, $g_1$ and $g_2$ lie in the upper half of the complex plane, and $g_3$ and $g_4$ in the lower half of that plane.

Ordinarily, each eigenvalue $g_n$, $n\in\lec1,2,3,4\ric$, has geometric multiplicity $1$ and algebraic multiplicity
$1$. The   eigenvector $\vecvn$ of $\matP$  is related to the eigenvector $\vectvn$ of $\mattP$ simply as
$\vecvn=  \matBz\.\vectvn$,
where
\begin{equation}
\label{eig}
\left.\begin{array}{l}
\left(\matP-g_n \les\#I\ris\right)\.\vecvn= [\#0]\\[5pt]
\left(\mattP-g_n \les\#I\ris\right)\.\vectvn= [\#0]
\end{array}\ric\,, \quad n\in\lec1,2,3,4\ric\,.
\end{equation}
With the matrices
\begin{equation}
\label{deg-G}
\matG = {\rm Diag}\les\gn1,\,\gn2,\,\gn3,\,\gn4\ris\,
\end{equation}
and
\begin{equation}
\label{def-V}
\left.\begin{array}{l}
\matV =  \matBz\.\mattV= \big[\,\vecvone,\,\vecvtwo,\,\vecvthree,\,\vecvfour\,\big]
\\[5pt]
\mattV =  \big[\,\vectvone,\,\vectvtwo,\,\vectvthree,\,\vectvfour\,\big]\,
\end{array}\right\}\,,
\end{equation}
we can use the factorizations
\begin{equation}
\label{diag}
\left.\begin{array}{l}
\matP =  \matV\.\matG\.\matV^{-1}
\\[5pt]
\mattP =  \mattV\.\matG\.\mattV^{-1}
\end{array}\right\}\,
\end{equation}
for computational ease.

\subsection{Defectiveness of $\matP$ and $\mattP$}

A Voigt wave can propagate inside the dielectric slab if and only if $\matP$  is defective in the sense
that at least one of its eigenvalues has algebraic multiplicity greater than its geometric multiplicity. A defective
matrix cannot be diagonalized \cite{Bernstein,Golub}. We use $\calP$ to denote the defectiveness of $\matP$.
Since two eigenvalues of $\matP$ lie in the upper half of the complex plane and the other two in the lower half for a dissipative material, only eigenvalues in the same half-plane can coalesce. As a result, $\matP$ cannot contain 
Jordan blocks of size greater
than $2$ and
$\calP$ encompasses three types of defectiveness denoted by $\calPone$, $\calPtwo$, and $\calPthree$.

Defectiveness $\calPone$ arises when  $g_1 = g_2$ (algebraic multiplicity 2, geometric multiplicity 1) but $g_3\ne g_4$. Then,
\begin{equation}
\label{CaseP1}
\left.\begin{array}{l}
\mattP =  \mattVone\.\matGone\.\mattVone^{-1}
\\[5pt]
\mattVone =  \big[\,\vectvone,\,\vectwone,\,\vectvthree,\,\vectvfour\,\big]\,
\end{array}\right\}\,
\end{equation}
and
\begin{equation}
\label{def-matGone}
\matGone =\les\begin{array}{cccc}
\gn1 & 1 & 0 & 0\\
0 & \gn1 & 0 & 0\\
0 & 0 & \gn3 & 0\\
0  & 0 & 0 & \gn4
\end{array}\ris\,,
\end{equation}
where
\begin{equation}
\left.\begin{array}{l}
\left(\mattP-\gn1 \les\#I\ris\right)\.\vectvone= \les\#0\ris\\[5pt]
\left(\mattP-\gn1 \les\#I\ris\right)\.\vectwone= \vectvone
\end{array}\ric\,,
\end{equation}
and Eqs.~\eqref{eig} can be used to find $\vectvthree$ and $\vectvfour$.  
If $\matGone$ is partitioned into four 2$\times$2 blocks in the same way as $\mattT$ in Eq.~\eqref{T-partition}, then its upper-left block is a Jordan block, i.e., the defectiveness resides in the pair of modes with $\Im\lec g\ric>0$. The upper-left 2$\times$2 block of $\mattP$ itself is not defective in general.

Defectiveness $\calPtwo$ arises when $g_1 \ne g_2$ but $g_3= g_4$ (algebraic multiplicity 2, geometric multiplicity 1).
Then,
\begin{equation}
\label{CaseP2}
\left.\begin{array}{l}
\mattP =  \mattVtwo\.\matGtwo\.\mattVtwo^{-1}
\\[5pt]
\mattVtwo =  \big[\,\vectvone,\,\vectvtwo,\,\vectvthree,\,\vectwthree\,\big]\,
\end{array}\right\}\,
\end{equation}
and
\begin{equation}
\label{def-matGtwo}
\matGtwo =\les\begin{array}{cccc}
\gn1 & 0 & 0 & 0\\
0 & \gn2 & 0 & 0\\
0 & 0 & \gn3 & 1\\
0  & 0 & 0 & \gn3
\end{array}\ris\,,
\end{equation}
where
\begin{equation}
\left.\begin{array}{l}
\left(\mattP-\gn3 \les\#I\ris\right)\.\vectvthree= \les\#0\ris\\[5pt]
\left(\mattP-\gn3 \les\#I\ris\right)\.\vectwthree= \vectvthree
\end{array}\ric\,,
\end{equation}
and Eqs.~\eqref{eig} can be used to find $\vectvone$ and $\vectvtwo$.
If $\matGtwo$ is partitioned likewise, then its lower-right 2$\times$2 block is a Jordan block, i.e., the defectiveness resides in the pair of modes with $\Im\lec g\ric<0$.

Finally, defectiveness $\calPthree$ arises when  $g_1 = g_2$ (algebraic multiplicity 2, geometric multiplicity 1) and $g_3= g_4$ (algebraic multiplicity 2, geometric multiplicity 1). Then,
\begin{equation}
\label{CaseP3}
\left.\begin{array}{l}
\mattP =  \mattVthree\.\matGthree\.\mattVthree^{-1}
\\[5pt]
\mattVthree =  \big[\,\vectvone,\,\vectwone,\,\vectvthree,\,\vectwthree\,\big]\,
\end{array}\right\}\,
\end{equation}
and
\begin{equation}
\label{def-matGthree}
\matGthree =\les\begin{array}{cccc}
\gn1 & 1 & 0 & 0\\
0 & \gn1 & 0 & 0\\
0 & 0 & \gn3 & 1\\
0  & 0 & 0 & \gn3
\end{array}\ris\,,
\end{equation}
where
\begin{equation}
\left.\begin{array}{l}
\left(\mattP-g_{n} \les\#I\ris\right)\.\vectvn= \les\#0\ris\\[5pt]
\left(\mattP-g_{n} \les\#I\ris\right)\.\vectwn= \vectvn
\end{array}\ric\,,\quad n\in\lec 1,3\ric\,.
\end{equation}
Both   diagonal 2$\times$2 blocks of $\matGthree$ are then Jordan blocks.
Defectiveness $\calPthree$ has a specialization $\calPthreeS$:  $g_1 = g_2=-g_3= -g_4$.

We first have to  determine whether $\matP$ satisfies the requirements
for $\calPone$, $\calPtwo$, or $\calPthree$,  for a particular choice of the incidence direction specified by $\lec\thetainc,\psi\ric$. 
Since   $\matP$ and $\mattP$ are similar,  they have
the same  Jordan structure.  
Equation~\eqref{eq7.12} can be written as
\begin{equation}
\label{eq7.12-1}
\frac{d}{dz}\, \vectfz= i \mattP\.\vectfz\,,\quad z\in(0,L)\,,
\end{equation}
where
\begin{equation}
\label{def-tf}
\vectfz =\matBz^T\.\vecfz\,.
\end{equation}
If we also define $\vectez=\Sz^T\.\vecez$ in the same spirit, then
\begin{equation}
\tilde{\=W}(\zeta)\.\vectez=\#0\,
\end{equation}
follows from the Maxwell equations for a solution $\vectez\propto\exp(i\ko\zeta z)$ [equivalently, from Eq.~\eqref{eq7.12-1}],
with $\tilde{\=W}(\zeta)$ as the dyadic equivalent to the matrix $\mattW$.

\subsection{Tests for defectiveness of $\matP$ and $\mattP$}
In order for propagation to occur,  $\det\mattW$
must equal zero; i.e., $\zeta$ must be a root of $\tilde{\Gamma}(\zeta)$. Since $\mattW$ is a  3$\times$3 matrix, its 
adjugate comprises 
the following $3\times3=9$ cofactors:
\begin{subequations}
\label{eq:direct-minors}
\begin{eqnarray}
 & (\teps\yy-\sin^2\thetainc-\zeta^2)
   (\teps\zz-\sin^2\thetainc)-\teps\yz\teps\zy\,,
\\
 & (\teps\xz+\zeta\sin\thetainc)\teps\zy
    -\teps\xy(\teps\zz-\sin^2\thetainc)\,,
\\
 & \teps\xy\teps\yz
    -(\teps\xz+\zeta\sin\thetainc)
      (\teps\yy-\sin^2\thetainc-\zeta^2)\,,
 \\
 & \teps\yz(\teps\zx+\zeta\sin\thetainc)
    -\teps\yx(\teps\zz-\sin^2\thetainc)\,,
\\
 & (\teps\xx-\zeta^2)(\teps\zz-\sin^2\thetainc)
    -(\teps\xz+\zeta\sin\thetainc)
      (\teps\zx+\zeta\sin\thetainc)\,,
 \\
 & (\teps\xz+\zeta\sin\thetainc)\teps\yx
    -(\teps\xx-\zeta^2)\teps\yz\,,
\\
 & \teps\yx\teps\zy
    -(\teps\yy-\sin^2\thetainc-\zeta^2)
      (\teps\zx+\zeta\sin\thetainc)\,,
\\
 & \teps\xy(\teps\zx+\zeta\sin\thetainc)
    -(\teps\xx-\zeta^2)\teps\zy\,,
\\
 & (\teps\xx-\zeta^2)
      (\teps\yy-\sin^2\thetainc-\zeta^2)
    -\teps\xy\teps\yx\,.
\end{eqnarray}
\end{subequations}
For any candidate repeated root $\zeta$, geometric multiplicity one
requires that at least one of those nine cofactors be nonzero. For a root of algebraic multiplicity exactly two, that condition is necessary
and sufficient for defectiveness.  If all nine expressions vanish at a double root, that root is
semisimple and must not be counted as a Voigt-wave degeneracy.

A nonzero cofactor
makes the rank of $\mattW$  equal to two, 
and so its nullspace is one-dimensional. The Maxwell
equations identify this nullspace with the eigenspace of $\mattP$, and
the hypothesis $\teps\zz\ne0$ makes that correspondence one-to-one.

{$\calPone$}
holds if and only if there exist pairwise distinct roots $\zetau$, $\zetav$, and $\zetaw$  of $\tilde{\Gamma}(\zeta)=0$
such that  $\Im\lec \zetau\ric>0$, $\Im \lec\zetav\ric<0$, and $\Im\lec \zetaw\ric<0$;
at least one of the nine expressions in
Eq.~\eqref{eq:direct-minors} is nonzero at $\zeta=\zetau$; and all four of the
following equations are satisfied:
\begin{subequations}
 \label{eq:direct-case12}
\begin{eqnarray}
 &{\teps\zz}\left(2\zetau+\zetav+\zetaw\right)=- {\sin\thetainc(\teps\xz+\teps\zx)}\,,
 \\
& \teps\zz\big[\zetau^2+2\zetau(\zetav+\zetaw)+\zetav\zetaw\big]=
 (\teps\xx+\teps\zz)\sin^2\thetainc
 -(\teps\xx\teps\zz-\teps\xz\teps\zx)
 -(\teps\yy\teps\zz-\teps\yz\teps\zy)\,,
 \\
 &-\teps\zz\big[\zetau^2(\zetav+\zetaw)+2\zetau\zetav\zetaw\big]=
 \sin\thetainc\big[\teps\xy\teps\yz+\teps\yx\teps\zy 
 -(\teps\xz+\teps\zx)(\teps\yy-\sin^2\thetainc)\big]\,,
 \\
 &\teps\zz \zetau^2\zetav\zetaw=
 \teps\xx\teps\yy\teps\zz
 +\teps\xy\teps\yz\teps\zx
 +\teps\xz\teps\yx\teps\zy 
 -\teps\xz\teps\yy\teps\zx
 -\teps\xy\teps\yx\teps\zz
 -\teps\xx\teps\yz\teps\zy\nonumber\\
 &-\big[\teps\xx(\teps\yy+\teps\zz)
 -\teps\xy\teps\yx-\teps\xz\teps\zx\big]\sin^2\thetainc 
 +\teps\xx\sin^4\thetainc\,.
\end{eqnarray}
\end{subequations}
Equations~\eqref{eq:direct-case12} arise from equating the coefficients
associated with the factorization of the
characteristic polynomial $\tilde{\Gamma}(\zeta)$ as
$\teps\zz(\zeta-\zetau)^2(\zeta-\zetav)(\zeta-\zetaw)$.
Satisfaction of the foregoing conditions yields $g_1=g_2=\ko \zetau$, $g_3=\ko \zetav$, and
$g_4=\ko \zetaw$, with the repeated eigenvalue defective and the other two
eigenvalues simple.

{$\calPtwo$}
is possible if and only if pairwise distinct roots $\zetau$, $\zetav$, and $\zetaw$ satisfy
Eqs.~\eqref{eq:direct-case12}; at least one of the nine expressions in
Eq.~\eqref{eq:direct-minors} is nonzero at $\zeta=\zetau$; and 
$\Im \lec\zetau\ric<0$, $\Im \lec\zetav\ric>0$, and $\Im\lec \zetaw\ric>0$.
Then,  $g_1=\ko \zetav$,  
$g_2=\ko \zetaw$, and $g_3=g_4=\ko \zetau$.  

{$\calPthree$}
requires the existence of distinct roots $\zetau\ne\zetav$ of $\tilde{\Gamma}(\zeta)=0$
such that  $\Im \lec\zetau\ric>0$ and $\Im \lec\zetav\ric<0$;
at least one of the nine expressions in
Eq.~\eqref{eq:direct-minors} is nonzero at $\zeta=\zetau$; 
at least one of the nine expressions in
Eq.~\eqref{eq:direct-minors} is nonzero at $\zeta=\zetav$; and all four of the
following equations are satisfied:
\begin{subequations}
 \label{eq:direct-case3}
\begin{eqnarray}
\label{eq:direct-case3a}
 &2{\teps\zz}\left(\zetau+\zetav\right)=- {\sin\thetainc(\teps\xz+\teps\zx)}\,,
 \\
 \label{eq:direct-case3b}
& \teps\zz(\zetau^2+4\zetau\zetav+\zetav^2)=
 (\teps\xx+\teps\zz)\sin^2\thetainc
 -(\teps\xx\teps\zz-\teps\xz\teps\zx)
 -(\teps\yy\teps\zz-\teps\yz\teps\zy)\,,
 \\
 \label{eq:direct-case3c}
 & -2\teps\zz \zetau\zetav(\zetau+\zetav)=
 \sin\thetainc\big[\teps\xy\teps\yz+\teps\yx\teps\zy 
 -(\teps\xz+\teps\zx)(\teps\yy-\sin^2\thetainc)\big]\,,
 \\
 \label{eq:direct-case3d}
 &\teps\zz \zetau^2\zetav^2=
 \teps\xx\teps\yy\teps\zz
 +\teps\xy\teps\yz\teps\zx
 +\teps\xz\teps\yx\teps\zy 
 -\teps\xz\teps\yy\teps\zx
 -\teps\xy\teps\yx\teps\zz
 -\teps\xx\teps\yz\teps\zy\nonumber\\
 &-\big[\teps\xx(\teps\yy+\teps\zz)
 -\teps\xy\teps\yx-\teps\xz\teps\zx\big]\sin^2\thetainc 
 +\teps\xx\sin^4\thetainc\,.
\end{eqnarray}
\end{subequations}
Satisfaction of the foregoing conditions yields
 $g_1=g_2=\ko \zetau$ and $g_3=g_4=\ko \zetav$, and both double
eigenvalues have geometric multiplicity one. 

{$\calPthreeS$} requires the satisfaction of all conditions for
$\calPthree$ along with the additional condition $\zetav=-\zetau$. For oblique incidence, then
Eqs.~\eqref{eq:direct-case3a} and  \eqref{eq:direct-case3c} 
reduce to
\begin{subequations}
 \label{eq:direct-case3S}
\begin{equation}
\label{eq:direct-case3Sa}
 \teps\xz+\teps\zx=0
\end{equation}
and
\begin{equation}
 \label{eq:direct-case3Sb}
\teps\xy\teps\yz+\teps\yx\teps\zy=0\,,
\end{equation}
\end{subequations}
respectively.
For Lorentz-reciprocal materials \cite{Krowne}, they further reduce to
$\teps\xz=\teps\zx=0$ and $\teps\xy\teps\yz=0$, respectively.

\subsection{Normal incidence} \label{normalinc}
The case of    $\calPthreeS$ for normal incidence ($\thetainc=0$~deg) deserves special consideration, because of direct relevance to the experimental observations of Voigt \cite{VoigtPM1902} and Pancharatnam \cite{Pancharatnam1955b}.
Equations~\eqref{eq:direct-case3a} and \eqref{eq:direct-case3c} are satisfied identically with $\zetav=-\zetau$, so that Eqs.~\eqref{eq:direct-case3Sa} and \eqref{eq:direct-case3Sb} are not required, and Eqs.~\eqref{eq:direct-case3b} and \eqref{eq:direct-case3d} reduce to $2\teps\zz\zetau^2=(\teps\xx\teps\zz-\teps\xz\teps\zx)$ $+(\teps\yy\teps\zz-\teps\yz\teps\zy)$ and $\teps\zz\zetau^4=\det\les\=\teps\rel\ris$.
The remaining conditions can be written more compactly as
\begin{equation}
\label{P3Scondition1-theta0}
 \left[\teps\xx-\teps\yy
 -\frac{\teps\xz\teps\zx-\teps\yz\teps\zy}{\teps\zz}\right]^2
 +4\left(\teps\xy-\frac{\teps\xz\teps\zy}{\teps\zz}\right)
 \left(\teps\yx-\frac{\teps\yz\teps\zx}{\teps\zz}\right)=0\,,
\end{equation}
with at least one of
\begin{equation}
\label{P3Scondition2-theta0}
 \teps\xx-\teps\yy
 -\frac{\teps\xz\teps\zx-\teps\yz\teps\zy}{\teps\zz},
 \qquad
 \teps\xy-\frac{\teps\xz\teps\zy}{\teps\zz},
 \qquad
 \teps\yx-\frac{\teps\yz\teps\zx}{\teps\zz}
\end{equation}
nonzero, and
\begin{equation}
\label{P3Scondition3-theta0}
\teps\xx+\teps\yy
 - \frac{\teps\xz\teps\zx+\teps\yz\teps\zy}{\teps\zz}
 \notin[0,\infty)\,.
\end{equation}
These normal-incidence conditions replace the general cofactor and
dissipation tests. 

Normal incidence  presents a nuance that facilitated Voigt's experimental observations \cite{VoigtPM1902}.
If $[\teps\rel]$  
denotes the 3$\times$3 matrix equivalent to $\=\teps\rel$, the 2$\times$2 matrix
\begin{equation}
\label{def-C}
[C]=\begin{bmatrix}
\teps\xx-\frac{\teps\xz\teps\zx}{\teps\zz} &
\teps\xy-\frac{\teps\xz\teps\zy}{\teps\zz}\\[5pt]
\teps\yx-\frac{\teps\yz\teps\zx}{\teps\zz}&
\teps\yy-\frac{\teps\yz\teps\zy}{\teps\zz}
\end{bmatrix}\,
\end{equation}
is the Schur complement of $\teps\zz$ in $[\teps\rel]$.  For $\kappa=0$, elimination of $\tilde{e}_z$ and $\tilde{h}_z$ from the Maxwell equations yields 
\begin{subequations}
\label{normal-ODE}
\begin{equation}
\left([I]\frac{d^2}{dz^2} +\ko^2[C]\right)\.\begin{bmatrix}\tilde e_x(z)\\ \tilde e_y(z)\end{bmatrix}=[0] 
\end{equation}
with
\begin{equation}
\begin{bmatrix}\tilde h_x(z)\\ \tilde h_y(z)\end{bmatrix}=\frac{1}{i\omega\muo}[J]\.\frac{d}{dz}\begin{bmatrix}\tilde e_x(z)\\ \tilde e_y(z)\end{bmatrix}\,,
\end{equation}
\end{subequations}
where $[J]=\begin{bmatrix}0&-1\\1&0\end{bmatrix}$. Indeed, Eq.~\eqref{def-tP} yields $\mattP=\mattZ$ for $\kappa=0$, where
\begin{subequations}
\begin{equation}
\label{def-mattZ}
\mattZ=\begin{bmatrix}{[0]}&\omega\muo[J]^{-1}\\ \omega\epso[J]\.[C]&{[0]}\end{bmatrix}\,
\end{equation}
is a 4$\times$4 matrix partitioned into 2$\times$2 blocks.  Since $\matP$ and $\mattP$ are similar to each other,
$\matP$ is similar to $\mattZ$ for $\kappa=0$.
Furthermore,
\begin{equation}
\label{mattZsq}
\mattZ^2=\ko^2\begin{bmatrix}{[C]}&[0]\\ [0] & [J]\.[C]\.[J]^{-1} \end{bmatrix}
\end{equation}
\end{subequations}
is block diagonal. 
As one diagonal block of $\mattZ^2$ is equal to $\ko^2[C]$ and the other diagonal block of $\mattZ^2$
is similar to  $\ko^2[C]$,  the Jordan structure of $\mattZ^2$ is that of $[C]$ but repeated twice. 
A null eigenvalue of $[C]$ would lead to  $\mattZ$ having a null eigenvalue.  
As that would be contrary to $\Im\lec g_n\ric\ne0$, $n\in\lec1,2,3,4\ric$, for a dissipative material,
$\det[C]\neq0$ and  $\mattZ$ is nonsingular.

The matrix $[C]$ can be either
defective or diagonalizable.
\begin{itemize}
\item[(i)] If $[C]$ is defective, then so is $\mattZ^2$ and hence $\mattZ$, because the square of a diagonalizable matrix is diagonalizable. 
\item[(ii)] If $[C]$ is diagonalizable with eigenvalues $c_1\ne0$ and $c_2\ne0$, then
$\mattZ^2$ is diagonalizable with eigenvalues $\ko^2c_1$ and $\ko^2c_2$.  Hence $\mathfrak{q}(\mattZ^2)=[\#0]$, where the polynomial $\mathfrak{q}(\eta)=(\eta-\ko^2c_1)(\eta-\ko^2c_2)$ when $c_1\ne c_2$ and $\mathfrak{q}(\eta)=\eta-\ko^2c_1$ when $c_1=c_2$. Therefore, $\mattZ$ is annihilated by the polynomial $\mathfrak{p}(\eta)=\mathfrak{q}(\eta^2)$, whose roots $\pm\ko\sqrt{c_1}$ and $\pm\ko\sqrt{c_2}$ are simple because $c_1\ne0$ and $c_2\ne0$. The minimal polynomial of $\mattZ$ divides $\mathfrak{p}(\eta)$, and so it too has only simple roots, which is necessary and sufficient for $\mattZ$ to be diagonalizable \cite{Bernstein}.
\end{itemize}
Hence, $\mattZ$ and $\mattP$ are diagonalizable.

Thus, $\matP$ is defective if and only if $[C]$
is defective, that is, if and only if $[C]=c_a[I]+[N]$, where the scalar $c_a\in\mathbb{C}$ and the 2$\times$2
matrix $[N]\neq[0]$ is such that $[N]^2=[0]$. 
This form of $[C]$ is correct  for the following  three reasons. 
\begin{itemize}
\item[(i)]
A 2$\times$2 matrix is defective if and only if it has a double eigenvalue but is not a multiple of $[I]$. If $[C]$ is defective with the double eigenvalue $c_a$, then $[N]=[C]-c_a[I]\ne[0]$ has the double eigenvalue $0$, so that ${\rm trace}[N]=\det[N]=0$ and the Cayley--Hamilton theorem gives $[N]^2=[0]$. Conversely, if $[C]=c_a[I]+[N]$ with $[N]\ne[0]$ and $[N]^2=[0]$, then every eigenvalue $\nu$ of $[N]$ satisfies $\nu^2=0$, so that $c_a$ is a double eigenvalue of $[C]$. 
\item[(ii)]
$[C]$ is not diagonalizable, because a diagonalizable 2$\times$2 matrix with a double eigenvalue $c_a$ equals $c_a[I]$, whereas $[N]\ne[0]$. 
\item[(iii)]
$c_a\ne0$, because $\det[C]=c_a^2$ and $\det[C]\ne0$.
\end{itemize}

Since the eigenvalues of $\mattZ$ occur in pairs $\pm\ko\sqrt{c_j}$ and $[C]$ is nonsingular, $\calPone$,
  $\calPtwo$, and $\calPthree$ with $\zetav\ne-\zetau$ cannot occur at
normal incidence. However $\calPthreeS$ can occur, but only if conditions
\eqref{P3Scondition1-theta0}--\eqref{P3Scondition3-theta0}  hold.

Not every dissipative biaxial dielectric material admits $\calPthreeS$ at normal incidence. If $[C]^{-1}$ is the sum of a complex multiple of a real symmetric matrix and a multiple of $[I]$, then neither $[C]$ nor $\matP$ can be defective when $\thetainc=0$~deg, as discussed in Sec.~\ref{nonexistence}.

\subsection{Defectiveness of $\mattT$}

For $L>0$, $\mattT$ is defective if and only if $\mattP$ (equivalently, $\matP$) is defective.
Since $\mattT$ is similar to  $\mattM=\exp\lec i\mattP L\ric$, it suffices to examine the latter. A diagonal $\matG$ gives a diagonal exponential $\exp\lec i\matG L\ric$. A $2\times2$ Jordan block, as in Eqs.~\eqref{def-matGone},
\eqref{def-matGtwo}, or \eqref{def-matGthree}
gives
$\exp(i g L)\begin{bmatrix}1&iL\\0&1\end{bmatrix}$, which is not diagonalizable for $L\neq0$.

\subsection{Uniform and nonuniform Voigt waves}\label{nonuni}
Suppose that $g=\ko\zetau$ is a defective double eigenvalue of $\mattP$, with eigenvector $\vectv$ and generalized eigenvector $\vectw$. Then Eq.~\eqref{eq7.12-1} has the two independent solutions $\exp(igz)\,\vectv$ and $\exp(igz)\,(\vectw+iz\vectv)$, and the Voigt-wave contribution to the electric field phasor in the slab is of the form
\begin{equation}
\label{VW-field}
\#E(\#r)=\les \gamma_1\#e_v+\gamma_2\left(\#e_w+iz\,\#e_v\right)\ris\exp\left(i\ko {\#a}\.\#r\right)\,,\qquad z\in(0,L)\,,
\end{equation}
where the vector
\begin{equation}
\label{def-kVW}
{\#a}=\left(\ux\cos\psi+\uy\sin\psi\right)\sin\thetainc+\uz\,\zetau\in\mathbb{C}^3\,;
\end{equation}
the vectors $\#e_v$ and $\#e_w$ 
correspond to $\vectv$ and $\vectw$, respectively; and
$\gamma_1$ and $\gamma_2$ are constants fixed by the boundary conditions.
The Voigt wave is uniform if and only if $\Re\lec {\#a}\ric$ and $\Im\lec {\#a}\ric$ are parallel. Since $\Im\lec {\#a}\ric=\uz\,\Im\lec\zetau\ric\neq\#0$ for a dissipative material, a Voigt wave excited in the slab is uniform if and only if $\thetainc=0$. In that case,  $\calPthreeS$ occurs for normal incidence and the singular axis is parallel to the
thickness direction  \cite{VoigtAP1902,Pancharatnam1955b,Grundmann}. The Voigt wave has to be nonuniform when either $\calPone$, $\calPtwo$, or $\calPthree$ holds for oblique incidence.

\subsection{Possible non-existence of singular axes}\label{nonexistence}
It is widely held that singular axes exist in every dissipative biaxial dielectric material. Voigt wrote in 1908 that ``in every pleochroic crystal there are four directions, the winding axes, in which two circularly oscillating waves, rotating in the same sense, propagate with equal speed and equal absorption,'' even though he had largely left triclinic and partly left monoclinic crystals aside  \cite{Voigt1908}. Born and Wolf stated that ``in every absorbing crystal there exist four directions, lying in pairs near the optic axes, for which the polarization is circular'' \cite{BornWolf}.   Sturm and Grundmann recently concluded that ``the singular optical axis can be found in all absorbing biaxial materials'' \cite{Sturm}.  These statements hold for almost all dissipative biaxial materials, but not for all of them.

Before proving that assertion, we revisit Pancharatnam's criterion \cite{Pancharatnam1955a} for a singular axis. Suppose that a uniform plane wave is propagating parallel to the unit vector $\hat{\#e}_3\in\mathbb{R}^3$ inside a material with relative impermittivity dyadic $\=\kappa\rel$.  Then, the Maxwell equations require that $\#D= \left(D_1\hat{\#e}_1+D_2\hat{\#e}_2\right)\exp(i\ko \xi\hat{\#e}_3\.\#r)$ is wholly transverse, with $\xi\in\mathbb{C}$ and  the unit vectors $\lec\hat{\#e}_1,\hat{\#e}_2,\hat{\#e}_3\ric$ forming a basis \cite{Kong1974}.
The  components $D_1$ and $D_2$ satisfy
\begin{equation}
\label{Q-eig}
[Q]\begin{bmatrix}D_1\\ D_2\end{bmatrix}=\frac{1}{\xi^2}\begin{bmatrix}D_1\\ D_2\end{bmatrix}\,,
\end{equation}
where
\begin{equation}
\label{def-Q}
[Q]=\begin{bmatrix}
\hat{\#e}_1\.\=\kappa\rel\.\hat{\#e}_1 & \hat{\#e}_1\.\=\kappa\rel\.\hat{\#e}_2\\[5pt]
\hat{\#e}_2\.\=\kappa\rel\.\hat{\#e}_1 & \hat{\#e}_2\.\=\kappa\rel\.\hat{\#e}_2
\end{bmatrix}\,.
\end{equation}
The unit vector $\hat{\#e}_3$ is parallel to a singular axis if and only if $[Q]$ is defective.

The 2$\times$2 matrix $[Q]$ is related to the 3$\times$3  matrix $\mattW$ of Eq.~\eqref{3x3 matrix} as follows. With the  vector $\xi\hat{\#e}_3$ in lieu of $\ux\sin\thetainc+\uz\zeta$ in Eq.~\eqref{3x3 matrix}, let $\les\tilde{U}\ris$ denote the 3$\times$3 matrix with elements $\tilde{U}_{mn}=\hat{\#e}_m\.\tilde{\=W}\.\hat{\#e}_n$, $m\in\lec1,2,3\ric$ and $n\in\lec1,2,3\ric$. Then
\begin{equation}
\label{QandU}
[Q]=\left(\les\tilde{V}\ris+\xi^2[I]\right)^{-1}\,,
\end{equation}
where  $\les\tilde{V}\ris$ is the Schur complement of $\tilde{U}_{33}$ in $\les\tilde{U}\ris$.
When $\hat{\#e}_3=\uz$, $[Q]=[C]^{-1}$, with $[C]$ given by Eq.~\eqref{def-C}.

As a counter-example to the universality implicit in the statements quoted at the beginning of Sec.~\ref{nonexistence}, consider a material for which
$\=\kappa\rel= b_1(\varphi_x \ux\ux+ \varphi_y\uy\uy+\varphi_z\uz\uz) +b_2\=I$, where $b_1\in\mathbb{C}$ is such that $\vert{b_1}\vert>0$, $b_2\in\mathbb{C}$, and $\varphi_{x,y,z}\in\mathbb{R}$ are pairwise distinct. Equivalently, the reciprocals of its three principal relative permittivities are collinear in the complex plane. This material is  biaxial dielectric, and it is also dissipative if $\Im\lec(b_1\varphi_j+b_2)^{-1}\ric>0$ for $j\in\lec x,y,z\ric$. Equation~\eqref{def-Q} then gives $[Q]=b_1[\Phi]+b_2[I]$, where $[\Phi]$ is the real symmetric 2$\times$2 matrix with elements $\hat{\#e}_m\.(\varphi_x\ux\ux+\varphi_y\uy\uy+\varphi_z\uz\uz)\.\hat{\#e}_n$, $m\in\lec1,2\ric$ and $n\in\lec1,2\ric$. Hence, $[Q]$ is diagonalizable for every  $\hat{\#e}_3
\in\mathbb{R}^3$, and singular axes do not exist in this counter-example.

Collinearity of the   reciprocals of the three principal relative permittivities in the complex plane is a single real condition \cite{real-condition} on the constitutive parameters. The counter-example is therefore exceptional rather than generic: almost any small perturbation of the constitutive parameters destroys the collinearity and restores the singular axes. Andalusite illustrates the generic case. The reciprocals of its three principal relative permittivities are nearly, but not exactly, collinear, deviating from a straight line by an angle of only about $2\times10^{-4}$~rad. That deviation causes andalusite to have singular axes---which lie about $10^{-4}$~rad from its optic axes (Secs.~\ref{Voigt-expt1} and \ref{Voigt-expt2}). A uniaxial dielectric material is a degenerate counter-example, in which two of $\varphi_x$, $\varphi_y$, and $\varphi_z$ are equal (Sec.~\ref{hBN-detailed}).
  
Even in the absence of singular axes for this counter-example, $\matP$ can be defective for oblique incidence.  The Voigt wave is then a nonuniform plane wave (Sec.~\ref{nonuni}), just as for the uniaxial material considered  in Sec.~\ref{hBN-detailed}.

\section{Observability}\label{observability}
\subsection{Voigt's experimental observations}\label{Voigt-expt1}
A defective double eigenvalue supplies a 2$\times$2 Jordan block   and
permits a field contribution proportional to $z\exp(igz)$ inside the anisotropic dielectric slab.
$\calPone$ and $\calPtwo$ have one such block and two simple eigenvalues.
$\calPthree$ and $\calPthreeS$ have two such blocks. However, the defectiveness of
$\mattP$ for a specific incidence direction $(\thetainc,\psi)$  does not
ensure the excitation of a Voigt wave for every  polarization state of the incident plane wave.
It is appropriate therefore to briefly revisit Voigt's 1902 experiment \cite{VoigtPM1902}.

He used blue light incident normally on an andalusite crystal with entry face $z=0$ and exit face $z=L$. The relative permittivity dyadic of the
crystal can be estimated at $\lambdao=450$~nm as
\begin{equation}
\=\eps\rel=\eps_\alpha\,\ualpha\ualpha+\eps_\beta\,\ubeta\ubeta+\eps_\gamma\,\ugamma\ugamma\,,
\label{eq:andalusite}
\end{equation}
with $\eps_\alpha=2.666689+i3.27\times10^{-6}$, 
$\eps_\beta=2.686321+i4.92\times10^{-6}$, and
$\eps_\gamma=2.702736+i1.315\times10^{-5}$ as the principal relative permittivities.
These quantities are the squares of the principal complex refractive indices $n_\alpha=1.633+i1.0\times10^{-6}$, $n_\beta=1.639+i1.5\times10^{-6}$, and $n_\gamma=1.644+i4.0\times10^{-6}$. The real parts of $n_{\alpha,\beta,\gamma}$ lie within the ranges of the principal refractive indices reported for andalusite \cite{DHZ}, whereas the imaginary parts are estimates chosen to represent the weak dissipation of andalusite in the visible spectral regime.

Furthermore, Voigt oriented the andalusite plate  such that
\begin{equation}
\left.\begin{array}{l}
\ualpha=-\uy\cos \vartheta+\uz\sin\vartheta
\\[5pt]
\ubeta=\ux
\\[5pt]
\ugamma=\uy\sin \vartheta+\uz\cos\vartheta
\end{array}\right\}\,;
\label{eq:axesVoigt}
\end{equation}
hence, the only non-zero components of $\=\eps\rel$ in Eq.~\eqref{eq:eps} are
$\eps_{\rm xx}=\eps_\beta$, 
$\eps_{\rm yy}=\eps_\alpha\cos^2\vartheta+\eps_\gamma\sin^2\vartheta$,
$\eps_{\rm zz}=\eps_\alpha\sin^2\vartheta+\eps_\gamma\cos^2\vartheta$, and
$\eps_{\rm yz}=\eps_{\rm zy}=(\eps_\gamma-\eps_\alpha)\sin\vartheta\cos\vartheta$.
As stated in Sec.~\ref{intro}, the optic axes of andalusite are taken to be those of the nondissipative biaxial material whose relative permittivity dyadic is $\Re\lec\=\eps\rel\ric$ \cite{opticaxis}. Since $\Re\lec\eps_\alpha\ric<\Re\lec\eps_\beta\ric<\Re\lec\eps_\gamma\ric$, the two optic axes lie in the plane spanned by $\ualpha$ and $\ugamma$ (i.e., the $yz$ plane), each inclined to $\ugamma$ at the angle $\theta_v=47.7516$~deg given by
$\cos^2\theta_v=\left(1/\Re\lec\eps_\beta\ric-1/\Re\lec\eps_\gamma\ric\right)/\left(1/\Re\lec\eps_\alpha\ric-1/\Re\lec\eps_\gamma\ric\right)$.
According to Eqs.~\eqref{eq:axesVoigt}, the $z$ axis is inclined to $\ugamma$ at the angle $\vert\vartheta\vert$ in the $yz$ plane. Hence, the choice $\vartheta=-47.7516$~deg aligns one optic axis with the $z$ axis, and the choice $\vartheta=47.7516$~deg aligns the other optic axis with the $z$ axis. Two of the four singular axes lie about $10^{-4}$~rad away from each optic axis (Sec.~\ref{Voigt-expt2}).
The incident light in the experiment was 
a normally incident convergent beam instead of a normally incident plane wave. 

With $\vartheta=-47.7516$~deg so that one optic axis of andalusite  is aligned with the $z$ axis,
Voigt's observations imply that $\vert\tLL\vert=\vert\tRR\vert$ is very high, $\vert\tLR\vert \propto L$, and $\vert\tRL\vert$ 
is  very low.  This  suggests that $\Tcirc$ could be nearly defective (non-diagonalizable).
In order to replicate  this observation theoretically, we set $\lambdao=450$~nm, 
$\thetainc= 0.0096674$~deg, and $\psi=-0.2466$~deg. We calculated   $\zeta_1 = \zeta_2 = 1.63899999115 + i 2.06503\times10^{-6}$ and $\zeta_3\approx\zeta_4\approx-\zeta_1$, with $\vert\zeta_3-\zeta_4\vert\approx1.5\times10^{-7}$,
where $\zeta_n=g_n/\ko$, $n\in \lec1,2,3,4\ric$.
 Therefore,
 $\matP$  satisfies the conditions for $\calPone$ exactly and those for $\calPthreeS$ very nearly. 
 Our calculations for $L\approx5$~mm yielded
$\vert\tLL\vert=\vert\tRR\vert\approx0.82$, $\vert\tLR\vert \approx \vert\tLL\vert L/\Lambda$, and $\vert\tRL\vert\ll\vert\tLR\vert$, with the characteristic length $\Lambda\approx63$~mm.
Parenthetically,  for $\psi=0$~deg and $L=5$~mm, 
$\ko L\vert\zeta_1-\zeta_2\vert\approx0.007$ indicates almost a double root.

With $\vartheta=47.7516$~deg so that the other  optic axis of andalusite is aligned with the $z$ axis,
Voigt's observations imply that $\vert\tLL\vert=\vert\tRR\vert$ is very high, $\vert\tRL\vert \propto L$, and $\vert\tLR\vert$ 
is very low.  Again,  $\Tcirc$ could be nearly defective.
With $\lambdao=450$~nm, $\vartheta= 47.7516$~deg,
$\thetainc= 0.0096674$~deg, and $\psi=0.2466$~deg, our calculations  for $L\approx 5$~mm yielded
 $\vert\tLL\vert=\vert\tRR\vert\approx0.82$, $\vert\tRL\vert \approx\vert\tLL\vert L/\Lambda$, and $\vert\tLR\vert\ll\vert\tRL\vert$.
In the first configuration, LCP light propagates almost unchanged along the singular axis, and yet the transmittance $\vert\tLL\vert^2+\vert\tRL\vert^2\approx0.673$ for LCP incidence is smaller than the transmittance $\vert\tLR\vert^2+\vert\tRR\vert^2\approx0.677$ for RCP incidence, because RCP incidence also delivers the LCP contribution that grows linearly with $L$. The singular axis that propagates the incident circular polarization state unchanged is thus the darker one, in agreement with Pancharatnam \cite{Pancharatnam1955a} but contrary to Voigt's expectation \cite{VoigtPM1902,VoigtAP1902}. The same conclusion holds for the second configuration with LCP and RCP interchanged.

\begin{figure}[htbp]
\centering
\includegraphics[width=\textwidth]{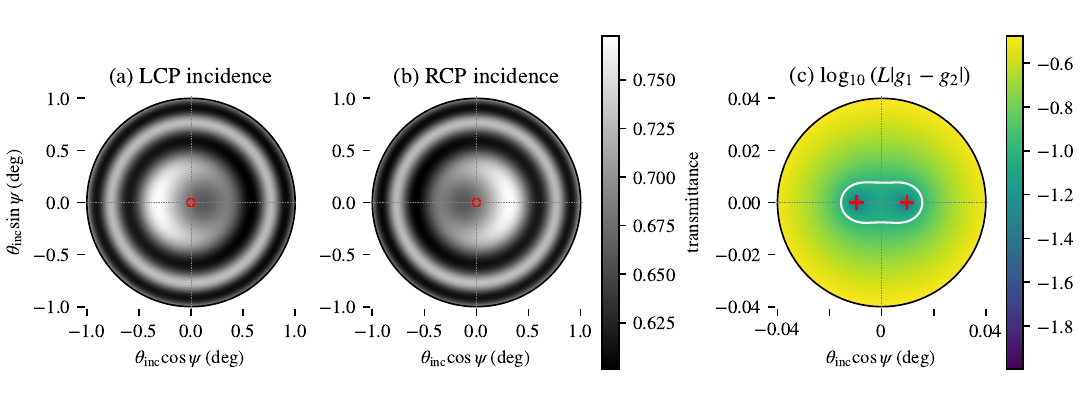}
\caption{Computed transmittances (a)  $\vert\tLL\vert^2+\vert\tRL\vert^2$ for LCP incidence and (b)
 $\vert\tLR\vert^2+\vert\tRR\vert^2$ for RCP incidence
 of a $5$-mm-thick andalusite plate ($\vartheta=-47.7516$~deg, $\lambdao=450$~nm)  as  functions of $\thetainc$ and $\psi$. The small red circle marks the region $\thetainc\leq0.04$~deg. (c) $\log_{10}(L\vert\gn1-\gn2\vert)$ in the region
$\thetainc\leq0.04$~deg. The white contour delineates $L\vert\gn1-\gn2\vert=0.1$ and the red crosses mark the singular axes  $\hat{\#s}_{\rm A}$ ($\psi=-0.2466$~deg) and $\hat{\#s}_{\rm B}$ ($\psi=180.2466$~deg) of Eqs.~\eqref{sAsB}.}
\label{fig:cono}
\end{figure}

Voigt viewed the andalusite plate in convergent light with a single polarizer, followed by a quarter-wave plate when circularly polarized light was required \cite{VoigtPM1902,VoigtAP1902}. Figure~\ref{fig:cono} shows the computed transmittances $\vert\tLL\vert^2+\vert\tRL\vert^2$ for LCP incidence and $\vert\tLR\vert^2+\vert\tRR\vert^2$ for RCP incidence for all incidence directions within 1~deg of the $z$ axis, when $\vartheta=-47.7516$~deg, $\lambdao=450$~nm, and $L=5$~mm. Rings are seen even though there is no analyzer, and the field of view is darker on one side of the plane of the optic axes (the $yz$ plane) than on the other; reversing the handedness of the incident light interchanges the dark and bright sides. Both features are exactly as Voigt reported \cite{VoigtPM1902}. The dark and bright spots are a few tenths of a degree across. Their size is set by the rings, i.e., by where $L\vert\gn1-\gn2\vert\sim\pi$, and not by the region in which $\matP$ is nearly defective. This becomes
clear from Fig.~\ref{fig:cono}(c), which shows that $L\vert\gn1-\gn2\vert\leq0.1$ only within about $0.01$~deg of the two singular axes, located at $\thetainc=0.0096674$~deg with $\psi=-0.2466$~deg and $\psi=180.2466$~deg, at both of which $\calPone$ holds. ($\calPtwo$ holds at $\psi=0.2466$~deg and $\psi=179.7534$~deg for the same $\thetainc$, i.e., for directions of incidence within $8.3\times10^{-5}$~deg of those for $\calPone$, and these are indistinguishable at the scale of the figure.) The singular axes thus decide which side of the field of view is dark for a given circular polarization state, while the spots that Voigt saw are broadened over the ring structure and, in addition, by the angular spread of his convergent beam.

\subsection{A tale of three defective matrices: $[Q]$, $\mattP$, and $\Tcirc$}\label{Voigt-expt2}
Three different matrices are involved in the interpretation of Voigt's observations, and all three are defective for slightly different directions.

First, Pancharatnam's matrix $[Q]$ of Eq.~\eqref{def-Q} refers to a uniform plane wave propagating along a real direction $\hat{\#e}_3$. For the andalusite plate with $\vartheta=-47.7516$~deg, $[Q]$ is defective for $\hat{\#e}_3\in\lec\pm\hat{\#s}_{\rm A},\pm\hat{\#s}_{\rm B}\ric$, where
\begin{equation}
\label{sAsB}
\left.\begin{array}{l}
\hat{\#s}_{\rm A}\simeq  1.029446\times10^{-4}\,\ux-4.432\times10^{-7}\,\uy+\uz
\\[5pt]
\hat{\#s}_{\rm B}\simeq -1.029446\times10^{-4}\,\ux-4.432\times10^{-7}\,\uy+\uz
\end{array}\right\}
\end{equation}
are the two singular axes next to the optic axis that is almost aligned with the $z$ axis. They are mirror images of each other in the plane of the optic axes (the $yz$ plane), and $\xi=1.6390000+i2.0650\times10^{-6}$  for either of them. Since $[Q]$ is the same for $\hat{\#e}_3$ and $-\hat{\#e}_3$, uniform plane waves along $\hat{\#s}_{\rm A}$ and $-\hat{\#s}_{\rm A}$ are both Voigt waves.

Second, $\mattP$ refers to the plane waves refracted into the slab, for which the normalized wave vector $\#a$ in Eq.~\eqref{def-kVW} is complex with $\Im\lec\#a\ric\parallel\uz$. $\calPone$ holds for $\thetainc=0.0096674$~deg and $\psi\in\lec-0.2466, 180.2466\ric$~deg, the vector $\Re\lec\#a\ric$ of the forward-propagating Voigt wave then being parallel to $\hat{\#s}_{\rm A}$ and $\hat{\#s}_{\rm B}$, respectively, for the two values of $\psi$. 
$\calPtwo$ holds for $\thetainc=0.0096674$~deg and $\psi\in\lec179.7534,0.2466\ric$~deg, the backward-propagating Voigt wave then traveling along $-\hat{\#s}_{\rm A}$ and $-\hat{\#s}_{\rm B}$, respectively, for the two values of $\psi$. 
 The backward-propagating plane waves that accompany a forward-propagating plane wave along $\hat{\#s}_{\rm A}$ travel along the mirror image of $\hat{\#s}_{\rm A}$ in the plane $z=0$, which is $8.9\times10^{-7}$~rad away from $-\hat{\#s}_{\rm B}$. That is why $\calPone$ holds exactly but $\calPthreeS$ only very nearly.
 
$\calPthreeS$ would require a singular axis to be exactly parallel to the $z$ axis, which is impossible in Voigt's orientation because $\hat{\#s}_{\rm A}$ and $\hat{\#s}_{\rm B}$ lie on either side of the $yz$ plane. It is achievable by rotating the plate through $0.0059$~deg about the $y$ axis so that either $\hat{\#s}_{\rm A}$ or $\hat{\#s}_{\rm B}$ becomes parallel to $\uz$
 (i.e., $\betaz=90\pm0.00797$~deg with $\betay=47.75157$~deg in the notation of Sec.~\ref{sec:andalusite}). Table~\ref{tab1} lists the equivalent orientations with $\betay=132.24843$~deg, which correspond to Voigt's second configuration. 
 
Third, $\Tcirc$ is defective for $L=5$~mm when $\thetainc=0.0098165$~deg and $\psi=4.7091$~deg, i.e., for a direction of incidence $8.6\times10^{-4}$~deg away from the one for $\calPone$. This direction  is given by $\thetainc=0.0087120$~deg and $\psi=1.9302$~deg for $L=1$~mm, and by $\thetainc=0.0090873$~deg and $\psi=1.0507$~deg for $L=10$~mm. 

All three of the directions for defectiveness of the three matrices lie within $10^{-3}$~deg of each other, whereas the dark and bright regions seen by Voigt are a few tenths of a degree across. $\Tcirc$ is nearly defective in the vicinities of $\hat{\#s}_{\rm A}$ and $\hat{\#s}_{\rm B}$. The singular axes $\hat{\#s}_{\rm A}$ and $\hat{\#s}_{\rm B}$, which have opposite handedness, determine on which side of the $yz$ plane the field of view is darker for a given circular polarization state, and hence why the dark and bright sides are interchanged when the handedness of the incident light is reversed. However, convergent-light observations at Voigt's angular resolution can distinguish neither a defective $[Q]$ from a defective $\mattP$, nor either of them from a defective $\Tcirc$.

The two key takeaways from  Secs.~\ref{Voigt-expt1} and \ref{Voigt-expt2} are that 
\begin{itemize}
\item  $\Tcirc$ is very nearly defective within about $0.01$~deg of each singular axis, and 
\item the singular axes determine the dark and bright sides that Voigt observed.
\end{itemize}
Voigt took his experimental observations to confirm his theoretical predictions \cite{VoigtAP1902}, even though his expectation about which singular axis appears darker was mistaken \cite{Pancharatnam1955a}.

\subsection{Defectiveness of $\Tcirc$ and $\Tlin$}
We must now explore the defectiveness of $\Tcirc$ and $\Tlin$, to be denoted by $\calT$. 

Neither $\Tcirc$ nor $\Tlin$
can be nilpotent for our BVP. To prove that assertion, we begin
with the identities $\det\Tcirc=\det\Tlin=\det\mattT/\det\les T^{(22)}\ris$ that follow from Eqs.~\eqref{Rlin-to-Rcirc} and \eqref{RlinTlin}.
But $\mattT$ is nonsingular, because $\det \mattT =\exp\left(iL \,{\rm trace} \left\{\mattP\right\}\right)\ne0$. Therefore,
neither $\Tcirc$ nor $\Tlin$ can be singular. A nilpotent matrix has to be singular.

As for any 2$\times$2 matrix,
$\Tcirc$
 can be defective  if and only if
\begin{itemize}
\item $\calT$: 
$\Delta_T\equiv(\tLL-\tRR)^2+4\tLR\tRL=0$ and $\Tcirc\not\propto [I]$\,.
\end{itemize}
That is possible  in exactly 
one of the following three mutually exclusive ways:
\begin{itemize} 
\item $\calTone$: $\tLR=0$, $\tRL\ne0$, and $\tLL=\tRR \ne0$. The only eigenvector is $(0,1)^T$. The sole transmission eigenpolarization is RCP, and LCP incidence leaks into RCP  transmission.
\item $\calTtwo$: $\tRL=0$, $\tLR\ne0$, and $\tLL=\tRR \ne0$. The only eigenvector is $(1,0)^T$. The sole transmission eigenpolarization is LCP, and RCP incidence leaks into LCP  transmission.
\item $\calTthree$:  $\tLR\tRL\ne0$ and $(\tLL-\tRR)^2=-4\tLR\tRL\implies\tLL\ne \tRR$. The only eigenvector is $\big(2\tLR,\,\tRR-\tLL\big)^T$, which is a polarization state that is neither LCP nor RCP and is elliptical in general.
\end{itemize}
If $\Tcirc$ is defective, its sole eigenvalue $\gamma_T=(\tLL+\tRR)/2$ is  nonzero.

$\calP$ does not guarantee $\calT$ in general. The defectiveness of $\matP$ is inherited by  $\mattT$.
Although $\Tcirc=\matA^{-1}\Tlin\matA$ inherits the defectiveness of $\Tlin$, $\Tlin=\les T^{(11)}\ris- \les T^{(12)}\ris\.\les T^{(22)}\ris^{-1}\.\les T^{(21)}\ris$ 
does not necessarily inherit the defectiveness of $\mattT$, because
block extraction does not necessarily preserve Jordan structure. The converse fails as well; i.e., $\calT$ does not guarantee
$\calP$ in general.

\subsection{Defectiveness of $\Rcirc$ and $\Rlin$}

Voigt's experiment was conducted for almost-normal incidence. It would be nearly impossible to separate the reflected beam
from the incident beam for almost-normal
incidence. But reflection can be observed for sufficiently oblique incidence, which raises the question whether the defectiveness of $\matP$
can be inherited by $\Rcirc=-\matA^{-1}\Rlin\matA$ and $\Rlin$.

Let $\calR$   denote the defectiveness of $\Rcirc$ and $\Rlin$.
$\Rcirc$ can be defective   if and only if
\begin{itemize}
\item $\calR$: 
$\Delta_R\equiv(\rLL-\rRR)^2+4\rLR\rRL=0$ and $\Rcirc\not\propto [I]$\,,
\end{itemize}
which is possible  in exactly 
one of the following three mutually exclusive ways:
\begin{itemize} 
\item $\calRone$: $\rLR=0$, $\rRL\ne0$, and $\rLL=\rRR$. The only eigenvector is $(0,1)^T$. The sole reflection eigenpolarization is RCP, and LCP incidence leaks into RCP reflection.
\item $\calRtwo$: $\rRL=0$, $\rLR\ne0$, and $\rLL=\rRR$. The only eigenvector is $(1,0)^T$. The sole reflection eigenpolarization is LCP, and RCP incidence leaks into LCP reflection.
\item $\calRthree$: $\rLR\rRL\ne0$ and $(\rLL-\rRR)^2=-4\rLR\rRL\implies\rLL\ne \rRR$. The only eigenvector is $\big(2\rLR,\,\rRR-\rLL\big)^T$, which is  a polarization state that is neither LCP nor RCP and is elliptical in general.
\end{itemize}

Unlike $\Tcirc$, $\Rcirc$ can be nilpotent, i.e.,  simultaneously defective and singular.  The sole eigenvalue $\gamma_R$ of $\Rcirc$ equals $(\rLL+\rRR)/2$.
 From Eqs.~\eqref{Rlin-to-Rcirc} and \eqref{RlinTlin},
$\det\Rcirc=\det\Rlin=\left(\det \les T^{(21)}\ris\right)/\left(\det \les T^{(22)}\ris\right)$  vanishes whenever $\det \les  T^{(21)}\ris=0$. 
Since $\Delta_R=({\rm trace}\Rcirc)^2-4\det\Rcirc$, a defective $\Rcirc$ with $\det\Rcirc=0$ also has ${\rm trace}\Rcirc=0$ and hence $\gamma_R=0$. By virtue of the Cayley--Hamilton theorem,  $\Rcirc^2=[0]$, i.e., $\Rcirc$ is then nilpotent 
(necessarily with index of nilpotency equal to 2, as a nonzero nilpotent 2$\times$2 matrix
cannot have any other index).
  
$\calP$ does not guarantee $\calR$ in general. 
Although $\Rcirc$ inherits the defectiveness of $\Rlin = - \les T^{(22)}\ris^{-1}\.\les T^{(21)}\ris$,  $\Rcirc$ and $\Rlin$ 
do not necessarily inherit the defectiveness of $\mattT$, because
block extraction does not necessarily preserve Jordan structure. Additionally,  $\calR$ does not guarantee
$\calP$.

\subsection{Normal incidence}\label{normalincidence}
The defectiveness of $\matP$ can be inherited by $\Tcirc$ for almost all $L>0$, and by $\Rcirc$ at isolated values of $L$,
for normal incidence ($\thetainc=0$~deg). As stated in Sec.~\ref{normalinc}, only $\calPthreeS$ can then occur, and it occurs if and only if the 2$\times$2 matrix
$[C]$ of Eq.~\eqref{def-C} satisfies $[C]=c_a[I]+[N]$, where $c_a\in\mathbb{C}$, $c_a\ne0$, $[N]\ne[0]$, and $[N]^2=[0]$.

Normal incidence is special because the anisotropic slab then behaves as an isotropic dielectric slab whose scalar relative permittivity has been replaced by the 2$\times$2 matrix $[C]$.  The general solution of Eq.~(\ref{normal-ODE}a) is
\begin{equation}
\label{normal-sol}
\begin{bmatrix}\tilde e_x(z)\\ \tilde e_y(z)\end{bmatrix}=\exp\left(i\ko z[C]^{1/2}\right)\.[\tilde a^+]+\exp\left(-i\ko z[C]^{1/2}\right)\.[\tilde a^-]\,,
\end{equation}
where the column vectors $[\tilde a^\pm]$ are fixed by the boundary conditions at $z=0$ and $z=L$. 

In Eq.~\eqref{normal-sol}, $[C]^{1/2}$ denotes a primary (or principal) square root of $[C]$ and can be written as
a linear superposition of $[I]$ and $[C]$ \cite{Higham}.
 A primary square root exists because $\det[C]\ne0$, as discussed inSec.~\ref{normalinc}, but it is not unique.
If $[C]$ has two distinct eigenvalues $c_1$ and $c_2$, then $[C]$ has exactly four primary square roots,
namely
$[C]^{1/2}=\lec s_1\left([C]-c_2[I]\right)-s_2\left([C]-c_1[I]\right)\ric/(c_1-c_2)$ with $s_j\in\lec\sqrt{c_j},-\sqrt{c_j}\ric$ for $j\in\lec1,2\ric$.
If $[C]$ is diagonalizable but $c_1=c_2$, then $[C]=c_1[I]$; although $c_1[I]$ has infinitely many square roots, only two of them, namely $\pm\sqrt{c_1}[I]$, are primary.
If $[C]$ is defective, then Sec.~\ref{normalinc} shows that $[C]=c_a[I]+[N]$ with $c_a\ne0$. In that
case, $[C]$ has exactly two distinct square roots, both primary, namely $[C]^{1/2}=\pm\lec\sqrt{c_a}[I]+[N]/(2\sqrt{c_a})\ric$.
 The choice of branch is immaterial, because reversing the sign of $s_j$  merely interchanges the corresponding parts of $[\tilde a^+]$ and $[\tilde a^-]$ in Eq.~\eqref{normal-sol}. For definiteness, the branches with $\Im\{\sqrt{c_j}\}>0$ may be chosen, so that $\exp\left(i\ko z[C]^{1/2}\right)$ represents waves that attenuate as they propagate in the $+z$ direction.

Inside an isotropic dielectric slab of   relative permittivity $\eps_{\rm iso}\in\mathbb{C}$ and thickness $L$, 
Eqs.~\eqref{normal-ODE} and \eqref{normal-sol} hold with $[C]$ replaced by $\eps_{\rm iso}[I]$. The normal-incidence transmission and reflection coefficients of the isotropic dielectric slab are well-known as \cite{BornWolf}
\begin{equation}
\label{tau-rho-iso}
\left.\begin{array}{l}
\tau(\eps_{\rm iso})= 2\sqrt{\eps_{\rm iso}}\,\Upsilon^{-1}(\eps_{\rm iso},L)
\\[5pt]
\rho(\eps_{\rm iso})= {i(\eps_{\rm iso}-1)\sin(\sqrt{\eps_{\rm iso}}\ko L)}
\,\Upsilon^{-1}(\eps_{\rm iso},L)
\end{array}\right\}\,,
\end{equation}  
where $\Upsilon(\eps_{\rm iso},L)={2\sqrt{\eps_{\rm iso}}\cos(\sqrt{\eps_{\rm iso}}\ko L)-i(1+\eps_{\rm iso})\sin(\sqrt{\eps_{\rm iso}}\ko L)}$.

The  matrices $\exp\left(i\ko z[C]^{1/2}\right)$ and $\exp\left(-i\ko z[C]^{1/2}\right)$ in Eq.~\eqref{normal-sol}
commute with  one another. Hence, the scalar $\eps_{\rm iso}$ can be replaced by the matrix $[C]$ in Eqs.~\eqref{tau-rho-iso}
to create the 2$\times$2 matrices
\begin{equation}
\label{tau-rho-C}
\left.\begin{array}{l}
[\tau_C]= 2[C]^{1/2}\.[\Upsilon_C]^{-1} 
\\[5pt]
[\rho_C]= -{i\left([I]-[C]\right)\.\sin\left([C]^{1/2}\ko L\right)}\.[\Upsilon_C]^{-1} 
\\[5pt]
[\Upsilon_C]=2[C]^{1/2}\.\cos\left([C]^{1/2}\ko L\right)-i\left([I]+[C]\right) \.\sin\left([C]^{1/2}\ko L\right)
\end{array}\right\}\,.
\end{equation} 

The existence of $[\Upsilon_C]^{-1}$ is guaranteed for the following reason.
The eigenvalues of  $[\Upsilon_C]$  are $\Upsilon(c_j,L)$. Therefore, $[\Upsilon_C]^{-1}$ can exist if and only if $\Upsilon(c_j,L)\ne0$ for every eigenvalue $c_j$ of $[C]$. If $[C]$ is defective, that condition reduces to $\Upsilon(c_a,L)\ne0$. If $[\Upsilon_C]$ were singular, a nonzero transmitted plane wave could exist without an incident plane wave, so that the slab would support a source-free field radiating power into the vacuum on both sides at the real angular frequency $\omega$. Poynting's theorem forbids that for a dissipative material, since the power absorbed inside the slab would then have to be negative. 

The upshot is that
\begin{subequations}
\label{TR-normal}
\begin{equation}
\Tlin=[J]^{-1}\.[\tau_C]\.[J]
\end{equation}
is the transmission matrix and
\begin{equation}
 \Rlin=[\Sigma]\.[J]^{-1}\.[\rho_C]   \.[J]
\end{equation}
\end{subequations}
is the   reflection matrix for    normal incidence on the anisotropic dielectric slab. Note that
$[\Sigma]={\rm Diag}[1,-1]$ in Eqs.~\eqref{TR-normal} accounts for the reversal of the direction of propagation on reflection.

In the defective case, $[C]=c_a[I]+[N]$ with $c_a\ne0$, $[N]\ne[0]$, and $[N]^2=[0]$ (Sec.~\ref{normalinc}), so that $f([C])=f(c_a)[I]+f^\prime(c_a)[N]$ for every function $f$ that is analytic at $c_a$. 
Application to Eq.~\eqref{normal-sol} gives
\begin{equation}
\label{normal-sol-def}
\exp\left(\pm i\ko z[C]^{1/2}\right)=\exp\left(\pm i\ko z\sqrt{c_a}\right)\left([I]\pm\frac{i\ko z}{2\sqrt{c_a}}[N]\right)\,.
\end{equation}
Thus, the Voigt wave has the propagation constant $\ko\sqrt{c_a}$ of an isotropic dielectric medium of relative permittivity $c_a$, but its amplitude has a component that grows linearly with $z$ along the sole eigenvector of $[N]$. Since $\tau(\eps_{\rm iso})$ and $\rho(\eps_{\rm iso})$ are even functions of $\sqrt{\eps_{\rm iso}}$, they are analytic functions of $\eps_{\rm iso}$ in the neighborhood of $c_a$. Hence, Eqs.~\eqref{TR-normal} yield
\begin{equation}
\label{TR-normal-2}
\left.\begin{array}{l}
\Tlin=\tau(c_a)[I]+\alpha_\tau(c_a){[\tilde N]}
\\[5pt]
\Rlin=[\Sigma]\.\lec\rho(c_a)[I]+\alpha_\rho(c_a){[\tilde N]}\ric
 \end{array}\right\}\,,
\end{equation}
where $\alpha_\tau(\eps_{\rm iso})=d\tau(\eps_{\rm iso})/d\eps_{\rm iso}$, $\alpha_\rho(\eps_{\rm iso})=d\rho(\eps_{\rm iso})/d\eps_{\rm iso}$, and ${[\tilde N]}=[J]^{-1}\.[N]\.[J]$ satisfies ${[\tilde N]}\ne[0]$ and ${[\tilde N]}^2=[0]$.  Thus, for normal incidence, the reflection (transmission) matrix of a slab supporting a Voigt wave is isomorphic to that of an isotropic slab of relative permittivity $c_a$, augmented by a nilpotent term whose strength is the sensitivity of the isotropic reflection (transmission) coefficient to its relative permittivity. The derivation of 
Eqs.~\eqref{TR-normal-2} is provided in Sec.~S1 of the Supplementary Material.

Equations~\eqref{TR-normal-2} lead to the following conditions for co-occurrence, which are  proved in Sec.~S1 of the Supplementary Material. First,
$\calT$ can co-occur  with $\calPthreeS$ if and only if 
\begin{equation}
\alpha_\tau(c_a)\ne0\,.
\end{equation}
Second, $\calR$ can co-occur  with $\calPthreeS$ if and only if 
\begin{equation}
\label{R-normal-cond}
\left.\begin{array}{l}
\alpha_\rho(c_a)\ne0
\\[5pt]
\alpha_\rho^2(c_a)\left({\tilde N_{11}}-  {\tilde N_{22}}\right)^2+4\rho^2(c_a)=0
\end{array}\right\}\,,
\end{equation}
where ${\tilde N_{11}}$ and ${\tilde N_{22}}$ are the diagonal elements of
 ${[\tilde N]}$.

The following four consequences emerge for normal incidence:
\begin{itemize}
\item[$\calCone$.] $\alpha_\tau(c_a)$ is an analytic function of $L$ which does not vanish identically for $L>0$,
 since $\alpha_\tau(c_a)\approx i\ko L/2$ for $\ko L\vert\sqrt{c_a}\vert\ll1$. 
Therefore, the zeros of $\alpha_\tau(c_a)$ as a function of $L$ are isolated. Hence, $\calPthreeS\implies\calT$ for all $L>0$ except possibly a discrete set. 

\item[$\calCtwo$.] Conversely, $[\tau_C]$ is diagonalizable whenever $[C]$ is. Hence, $\calT\implies\calPthreeS$.

\item[$\calCthree$.] The sole eigenvector of $\Tcirc$ is $\matA^{-1}\.[\tilde v]$, where ${[\tilde N]}\.[\tilde v]=[0]$; i.e., it is the polarization state of the Voigt wave itself.
 Thus $\calTone$  occurs if and only if the Voigt wave is RCP,   $\calTtwo$ occurs if and only if the Voigt wave is  LCP, 
 and $\calTthree$ occurs otherwise. 
 
 If $\=\eps\rel$ is symmetric (i.e., the anisotropic dielectric
 material is Lorentz reciprocal \cite{Krowne}), then $[C]$ and $[N]$ are symmetric, so that $[N]$ is proportional to $\begin{bmatrix}1&\pm i\\ \pm i&-1\end{bmatrix}$. Therefore, the Voigt wave has to be circularly polarized and $\calTthree$ cannot occur.
 
 If $\=\eps\rel$ is not symmetric, however, $\calTthree$ can co-occur with $\calPthreeS$. For instance, a magnetostatic field of flux density $\Bdc$ applied along the $z$ axis adds the gyrotropic term $i\eps_g\=I\times\uz$ to the relative
 permittivity dyadic, where the gyrotropy factor  $\eps_g\approx2\bar{n}V\Bdc/\ko$, the scalar
 $\bar{n}\in\mathbb{C}$ is a mean refractive index in the absence of gyrotropy, and $V$ is the Verdet constant \cite{Carothers}. 
 Switching on $\Bdc$ will destroy the defectiveness of $[C]$, which can be restored by a slight reorientation of the newly gyrotropic material provided that the in-plane linear
dichroism of the material for $\Bdc=0$ is not weaker than  $\vert \eps_g\vert$; the resulting Voigt wave will be elliptically polarized in general.

\item[$\calCfour$.] $\Rcirc$ is nilpotent  only if ${\tilde N_{11}}=0$ and $\rho(c_a)=0$.
 The latter requires $\sin(\ko L\sqrt{c_a})=0$, which is impossible for real $L >0$ if $\Im\lec\sqrt{c_a}\ric\ne0$. A dissipative slab therefore cannot have a simultaneously defective and singular $\Rcirc$ at normal incidence.
\end{itemize}

\subsection{Simultaneous $\calP$ and $\calT$}\label{PTtogether}
$\calP$ can arise only from $\mattP$. For a specific $\=\eps\rel$ at a specific $\lambdao$, the discriminant of $\tilde\Gamma(\zeta)$
must vanish for $\calPone$ or $\calPtwo$ or $\calPthreeS$. Thus, $\calP$ imposes two real conditions on $\lec\thetainc,\psi\ric$, independently of $L$, for $\calPone$ and $\calPtwo$.
 Two double roots ($\calPthree$) impose four real conditions in general, but only two for $\calPthreeS$ when symmetry renders $\tilde\Gamma(\zeta)$ even [Eqs.~\eqref{eq:direct-case3Sa} and \eqref{eq:direct-case3Sb}], as for  andalusite in Voigt's experiment with $\psi\in\lec0,180\ric$~deg (Sec.~\ref{Voigt-expt1}). 

$\calT$ imposes the  conditions
\begin{equation}
\label{DeltaT}
\left.\begin{array}{l}
\Delta_T =0
\\
\Tcirc\ne\gamma_T[I]
\end{array}\right\}\,,
\end{equation}
which involve $L$ as well. At oblique incidence, the co-occurrence of $\calP$ and $\calT$ therefore requires four real conditions to be satisfied by three real parameters $\lec\thetainc,\psi,L\ric$. That can happen only if a fourth parameter (such as $\lambdao$ or the orientation of the crystal) is also tuned, 
and then only for isolated values of $L$.

Normal incidence is exceptional, because the slab then responds as a function of the single matrix $[C]$. $\calPthreeS$ then delivers $\calT$ for almost all $L>0$
and $\calT$ requires $\calPthreeS$ (Sec.~\ref{normalincidence}). By continuity, $\vert\Delta_T\vert$ remains small near normal incidence over a continuum of values of $L$, and this is what enabled Voigt's experimental observations \cite{VoigtPM1902} in the immediate vicinity of a singular axis. For the andalusite parameters of Sec.~\ref{Voigt-expt1} with $\vartheta=-47.7516$~deg, $\psi=-0.2466$~deg, and $\thetainc=0.0096674$~deg, $\calPone$ holds exactly and $\calPthreeS$ very nearly, but $\Delta_T\ne0$. Nevertheless, $\vert\Delta_T\vert/\vert\tLL\vert^2$ lies between $5\times10^{-5}$ and $3\times10^{-3}$ and $\vert\tRL\vert/\vert\tLR\vert$ 
lies between $0.03$ and $0.06$ for $L\in[1,10]$~mm, so that $\Tcirc$ is very nearly of type $\calTtwo$. Consistently with $\calCthree$, the Voigt wave 
is then almost exactly LCP, the ratio of its RCP and LCP amplitudes being $6\times10^{-5}$ in magnitude.

\subsection{Simultaneous $\calP$ and $\calR$}
$\calR$ imposes the  conditions
\begin{equation}
\label{DeltaR}
\left.\begin{array}{l}
\Delta_R =0
\\
\Rcirc\ne\gamma_R[I]
\end{array}\right\}\,,
\end{equation}
so that the counting argument of Sec.~\ref{PTtogether} applies unchanged at oblique incidence. In other words,
the co-occurrence of $\calP$ and $\calR$ requires the tuning of a fourth parameter, and then holds only for isolated values of $L$.

Unlike $\calT$, $\calR$ is not favored by normal incidence. The factor $[\Sigma]$ in Eq.~\eqref{TR-normal}, which arises from the reversal of the propagation direction upon reflection, prevents $\Rlin$ from being similar to a function of $[C]$. Hence, $\calPthreeS$ does not deliver $\calR$ for a continuum of values of $L$. Rotation of the slab about the $z$ axis changes ${\tilde N_{11}}$ but not $c_a$, and the condition $\rho(c_a)=\pm i\alpha_\rho(c_a){\tilde N_{11}}$ is one complex equation in the two real variables $\psi$ and $L$. Accordingly, $\calPthreeS$ and $\calR$ can co-occur only at isolated points in the $(\psi,L)$ plane. For a passive, Lorentz-reciprocal dielectric
material, $\vert {\tilde N_{11}}\vert$ is independent of $\psi$ and cannot exceed $\Im\lec c_a\ric$. Our numerical exploration indicates that $\vert\rho(c_a)/\alpha_\rho(c_a)\vert>\Im\lec c_a\ric$ for all $L$ unless $\Re\lec c_a\ric$ is close to unity, so that $\calPthreeS$ and $\calR$ co-occur at normal incidence only for weakly contrasting (i.e., $\Re\lec c_a\ric\approx1$) and strongly dissipative materials.

\subsection{Simultaneous $\calP$, $\calT$, and $\calR$}
At oblique incidence, $\calP$, $\calT$, and $\calR$ together impose six real conditions on $\lec\thetainc,\psi,L\ric$. Their co-occurrence thus requires three additional parameters to be tuned, and cannot be expected to be commonplace in practice.

At normal incidence, $\calPthreeS$ is a condition on the cut of the crystal alone, independent of $\psi$ and $L$, and  it implies $\calT$ for almost all $L$. All three properties therefore co-occur at the isolated points in the $(\psi,L)$ plane at which $\rho(c_a)=\pm i\alpha_\rho(c_a) {\tilde N_{11}}$, $\alpha_\rho(c_a)\ne0$, and $\alpha_\tau(c_a)\ne0$. As an example, the passive and Lorentz-reciprocal material with $\eps\xx=1.65+i0.5$, $\eps\yy=0.75+i0.5$, $\eps\zz=1.5+i0.5$, $\eps\xy=\eps\yx=i0.45$, and all other components null has $c_a=1.2+i0.5$ and a defective $[C]$. For $\psi=6.0915$~deg and $L=0.41022\lambdao$, $\Tcirc$ is of type $\calTtwo$ and $\Rcirc$ is of type $\calRthree$ with $\rLL=0$ and $\rLR=\rRL$, as confirmed by solving Eq.~\eqref{BVPalgebra} directly. In conformity with $\calCfour$, $\Rcirc$ is not nilpotent.

\section{Illustrative Examples}\label{ie}
In order to search for the occurrences of  $\calPone$, $\calPtwo$, $\calPthree$, $\calPthreeS$, $\calTone$, $\calTtwo$, $\calTthree$,
$\calRone$, $\calRtwo$, $\calRthree$, and their combinations, we used a Mathematica\texttrademark~v15.0 program to calculate
$g_n$ for $n\in\lec1,2,3,4\ric$, $\Tcirc$, and $\Rcirc$ as functions of $\thetainc$, $\psi$, and $\eps\rel$ for specific values of $\lambdao$
and $L$. The output of the program was a table of occurrences for every type and combination searched. 
If no occurrence was found, then that was also recorded. A combination (e.g., $\calPthree\,\&\,\calRone$) can occur only if each of its constituents occurs by itself for the same $\thetainc$ and $L$, and a threefold combination only if each of its three constituent pairs occurs.
The search for a combination was therefore skipped whenever a constituent had not been found, and the reason was recorded.  Rows that agree in every column and differ only in one member of a family have been merged; e.g., $\calPthree\,\&\,\calR$ stands for $\calPthree\,\&\,\calRone$, $\calPthree\,\&\,\calRtwo$, and $\calPthree\,\&\,\calRthree$.

For each type of defectiveness, complex normalized \textit{residuals} (called residuals for brevity) were defined that vanish if and only if the defining equalities hold. Each residual is of order $0.1$ to $1$ for   generic values of the Euler angles $\betaz$ and $\betay$ defined in Sec.~\ref{sec:andalusite}.
 These are $(\gn1-\gn2)^2/(\ko^2\sigma_P)$ for $\calPone$, $(\gn3-\gn4)^2/(\ko^2\sigma_P)$ for $\calPtwo$, both of them for $\calPthree$, and both together with $(\gn1+\gn2+\gn3+\gn4)/(\ko\sigma_S)$ for $\calPthreeS$. Here $\sigma_P=\delta^2/(4\bar\eps)$ and $\sigma_S=\delta/\bar\eps$, where $\delta$ is the largest magnitudinal difference between the three principal relative permittivities and $\bar\eps$ is the magnitude of their mean, so that the normalization reflects the degree of the anisotropy. The residuals are $\tLR/s_T$ and $(\tLL-\tRR)/s_T$ for $\calTone$, $\tRL/s_T$ and $(\tLL-\tRR)/s_T$ for $\calTtwo$, and $\Delta_T/s_T^2$ for $\calTthree$, with $s_T=\tLL+\tRR$.  With $s_R=\vert\rLL\vert+\vert\rLR\vert+\vert\rRL\vert+\vert\rRR\vert$, the residuals are $\rLR/s_R$ and $(\rLL-\rRR)/s_R$ for $\calRone$, $\rRL/s_R$ and $(\rLL-\rRR)/s_R$ for $\calRtwo$, and $\Delta_R/s_R^2$ for $\calRthree$.
The set of residuals of a combination is the union of the sets of residuals of its constituents, and all of the residuals of all of the constituents   are required to vanish simultaneously.

The searches were implemented in Mathematica\texttrademark\ with the built-in function \texttt{FindMinimum}. For fixed $\thetainc$, the sum of the squared magnitudes of the normalized residuals was minimized with respect to $\psi$, $\betaz$, and $\betay$ by the default quasi-Newton (Broyden--Fletcher--Goldfarb--Shanno) method \cite{Ge1983,Fletcher}, with gradients computed by finite differences and a line search along each step. The starting points were the best points of a uniform grid in $(\psi,\betaz,\betay)$, together with the orientations already found for other types of defectiveness. When $\thetainc$ was an additional unknown,   the Levenberg--Marquardt method \cite{Fletcher} was used, the residual vector comprising the real and imaginary parts of the normalized residuals. Starting points were chosen randomly, and the minimization was carried out first for the residuals of $\calP$ alone and then for all residuals of the combination.  Every minimum found was then subjected to the tests (a)--(d).

 A minimum was accepted as an occurrence only if 
\begin{itemize}
\item[(a)] every residual of $\calP$ [i.e., $(\gn1-\gn2)^2/(\ko^2\sigma_P)$, $(\gn3-\gn4)^2/(\ko^2\sigma_P)$, and $(\gn1+\gn2+\gn3+\gn4)/(\ko\sigma_S)$, as applicable] and every 
$\vert\Delta_{T,R}\vert/\vert{s_{T,R}}\vert^2$ was less than $10^{-10}$, and a vanishing off-diagonal element of $\Tcirc$ or $\Rcirc$ was below $10^{-5}$ relative to $\vert{s_{T,R}}\vert$; 

\item[(b)] the double root of $\tilde\Gamma(\zeta)$ was defective, i.e., some cofactor of $\mattW$ in Eqs.~\eqref{eq:direct-minors} was nonzero  at that root; 

\item[(c)] the other two roots of $\tilde\Gamma(\zeta)$ were distinct for $\calPone$ and $\calPtwo$; and 

\item[(d)] $\Tcirc$ or $\Rcirc$ was not a multiple of $[I]$, the nonvanishing off-diagonal elements exceeding $10^{-6}$ relative to  $\vert s_{T,R}\vert$. 
\end{itemize}
Otherwise, the   closest residual (i.e., the smallest Euclidean norm of the residuals reached by the search) was recorded. If the closest residual
was driven below $10^{-8}$ but the relevant test  (a)--(d) failed, the candidate solution was rejected. A candidate solution
that was not accepted but came close to satisfying the defining equalities is recorded as a \textsf{near miss} in the Supplementary Material, where the near-miss criteria are stated.

\subsection{Andalusite}\label{sec:andalusite}
Following Voigt, we chose andalusite as an orthorhombic dielectric material, its relative permittivity dyadic being
\begin{eqnarray}
\nonumber
\=\eps\rel&=& \=R_y\. \=R_z\. \left[\left(2.686321+i4.92\times10^{-6}\right)\ux\ux+\left(2.666689+i3.27\times10^{-6}\right)\uy\uy
\right.
\\[5pt]
&&\left.
+\left(2.702736+i1.315\times10^{-5}\right)\uz\uz
\right]\.\=R_z^T\.\=R_y^T\,
\label{andalusite}
\end{eqnarray}
at $\lambdao=450$~nm.
The rotation dyadics
\begin{subequations}
\begin{equation}
 \=R_z 
 = (\ux\ux+\uy\uy)\cos\betaz +(\ux\uy-\uy\ux)\sin\betaz +\uz\uz
\end{equation}
and
\begin{equation}
 \=R_y 
 =(\ux\ux+\uz\uz)\cos\betay+(\uz\ux-\ux\uz)\sin\betay+\uy\uy,
\end{equation}
\end{subequations}
involve the Euler  angles $\betaz$ and $\betay$.
With $\lambdao=450$~nm and $\thetainc=0.0096674$~deg, Voigt’s first configuration ($\vartheta=-47.7516$~deg) corresponds to $\psi=269.7534$~deg, $\betaz=90$~deg, 
and $\betay=47.7516$~deg; and his second ($\vartheta=+47.7516$~deg) to $\psi=270.2466$~deg, $\betaz=90$~deg, and $\betay=132.2484$~deg.

Let us begin with values of $\psi$, $\betaz$, and $\betay$ for $\calP$, $\calT$, and $\calR$ to occur for  $\theta\inc\in\lec0,45\ric$~deg. These are recorded in Tables~\ref{tab1} and \ref{tab2}, respectively, for $L=5$~mm. Tables~S1 and S2 of the Supplementary Material list the results for every type and combination searched, including those not found.

{\small\setlength{\tabcolsep}{4pt}
\begin{longtable}{@{}lccc>{\raggedright\arraybackslash}p{5.6cm}@{}}
\caption{Occurrences of   $\calPthree$, $\calPthreeS$, $\calTone$,  $\calTtwo$, 
 and their combinations for andalusite described by Eq.~\eqref{andalusite} with stated
values of $\psi$, $\betaz$, and $\betay$ for normal incidence ($\theta\inc=0$~deg) when $\lambdao=450$~nm and $L=5$~mm.
No occurrences of $\calPone$, $\calPtwo$, $\calTthree$, and $\calR$ were found.
The complete results of the search, including the unsuccessful searches, the near misses, and the reasons why a combination was not searched, are given in Table~S1 of the Supplementary Material.}
\label{tab1}\\
\hline
Defectiveness & $\psi$  & $\betaz$   & $\betay$  & Remark \\
  &   (deg) &   (deg) &   (deg) &   \\
\hline
\endfirsthead
\multicolumn{5}{l}{Table~\ref{tab1} (continued)}\\
\hline
Defectiveness & $\psi$  & $\betaz$   & $\betay$  & Remark \\
  &   (deg) &   (deg) &   (deg) &   \\
\hline
\endhead
$\calPthree$ & $[0,360)$ & $89.99203$ & $132.24843$ & Voigt wave excited\\
$\calPthree$ & {$[0,360)$} & $90.00797$ & $132.24843$ & Voigt wave excited \\
$\calPthreeS$ & $[0,360)$ & $89.99203$ & $132.24843$ & Voigt wave excited \\
$\calPthreeS$ & {$[0,360)$} & $90.00797$ & $132.24843$ & Voigt wave excited \\
$\calTone$ & {$[0,360)$} & $89.99203$ & $132.24843$ & co-occurs with $\calPthreeS$   \\
$\calTtwo$ & {$[0,360)$} & $90.00797$ & $132.24843$ & co-occurs with $\calPthreeS$   \\
$\calPthree\,\&\,\calTone$ & {$[0,360)$} & $89.99203$ & $132.24843$ & Voigt wave seen in transmission \\
$\calPthree\,\&\,\calTtwo$ & {$[0,360)$} & $90.00797$ & $132.24843$ & Voigt wave seen in transmission \\
$\calPthreeS\,\&\,\calTone$ & {$[0,360)$} & $89.99203$ & $132.24843$ & Voigt wave seen in transmission \\
$\calPthreeS\,\&\,\calTtwo$ & {$[0,360)$} & $90.00797$ & $132.24843$ & Voigt wave seen in transmission \\
\hline
\end{longtable}}

A Voigt wave is excited only if $\calP$ occurs. It can be observed in transmission if $\calP$ and $\calT$ co-occur, and in reflection if $\calP$ and $\calR$ co-occur. In Table~\ref{tab1} for normal incidence, $\calPthreeS$ holds for $\betaz=90\pm0.00797$~deg and $\betay=132.24843$~deg for every $\psi$, and $\calT$ co-occurs for almost all $L$, in accord with Sec.~\ref{normalincidence}.  $\calPthree$ and $\calPthreeS$ are the same (Sec.~\ref{normalinc}), so that every orientation is listed for both in the table. $\calTone$ is found only when $\betaz=89.99203$~deg and $\calTtwo$ only when $\betaz=90.00797$~deg, where $\calPthreeS$ also holds, as required by $\calCtwo$. The Voigt wave is RCP at the former and LCP at the latter, in accord with $\calCthree$. Thus, at normal incidence, every orientation that excites a Voigt wave also makes it observable in transmission. Voigt's experiment is a slightly perturbed version of this case. 

$\calR$ cannot co-occur with $\calPthreeS$ at normal incidence because $c_a=2.686321+i6.77\times10^{-6}$ and $\vert {\tilde N_{11}}\vert=1.85\times10^{-6}$, whereas $\vert\rho(c_a)/\alpha_\rho(c_a)\vert\ge6.77\times10^{-6}$ for all $L$, so that the second of Eqs.~\eqref{R-normal-cond} cannot be satisfied.

{\small\setlength{\tabcolsep}{4pt}
\begin{longtable}{@{}lccc>{\raggedright\arraybackslash}p{5.6cm}@{}}
\caption{Occurrences of   $\calPone$, $\calPtwo$, $\calTone$,  $\calTtwo$, 
 and $\calRthree$ for andalusite described by Eq.~\eqref{andalusite} with stated
values of $\psi$, $\betaz$, and $\betay$ for  oblique incidence ($\theta\inc=45$~deg) when $\lambdao=450$~nm and $L=5$~mm.
No occurrences of $\calPthree$, $\calPthreeS$, $\calTthree$,  $\calRone$, and $\calRtwo$ were found.
See Table~S2 of the Supplementary Material for the complete results of the search.}
\label{tab2}\\
\hline
Defectiveness & $\psi$  & $\betaz$   & $\betay$  & Remark \\
  &   (deg) &   (deg) &   (deg) &   \\
\hline
\endfirsthead
\multicolumn{5}{l}{Table~\ref{tab2} (continued)}\\
\hline
Defectiveness & $\psi$  & $\betaz$   & $\betay$  & Remark \\
  &   (deg) &   (deg) &   (deg) &   \\
\hline
\endhead
$\calPone$ & $339.12697$ & $101.99317$ & $108.80073$ & Voigt wave excited \\
$\calPtwo$ & $350.11693$ & $84.26687$ & $72.83438$ & Voigt wave excited \\
$\calTone$ & $339.14016$ & $101.97211$ & $108.80773$ & no Voigt wave excited \\
$\calTtwo$ & $339.09364$ & $101.99770$ & $108.80519$ & no Voigt wave excited \\
$\calRthree$ & $24.68554$ & $144.85993$ & $151.11252$ & no Voigt wave excited \\
\hline
\end{longtable}}

In Table~\ref{tab2} for $\thetainc=45$~deg, occurrences of  $\calP$, $\calT$, and $\calR$ are recorded, but no co-occurrence of
$\calP$ and either $\calT$ or $\calR$ was found.  Thus, even though a Voigt wave can be excited in andalusite for $\thetainc=45$~deg,
it cannot   be observed in reflection. It can be observed in transmission, but only approximately because $\calPone\,\&\,\calTtwo$ is a \textsf{near miss} (Table~S2).
An exact co-occurrence of $\calPone$ and $\calTone$ does exist at the nearby angle $\thetainc=44.59719$~deg (Table~\ref{tab3}). Conversely,  any recorded defectiveness of $\calT$ or $\calR$ does
not \textit{per se} indicate the excitation of a Voigt wave.

Representative values of $\thetainc$, $\psi$, $\betaz$, and $\betay$ for which $\calP$ co-occurs with $\calT$ or $\calR$ are given in Table~\ref{tab3} for $L=5$~mm.
For oblique incidence, $\thetainc$ was treated as a fourth unknown because, e.g.,
 $\calPone\,\&\,\calTone$ imposes four real conditions (Sec.~\ref{PTtogether}).  The co-occurrences then found are isolated points for fixed $L$.  
 
According to Table~\ref{tab3}, $\calP$ and $\calT$ co-occur over a wide range of $\thetainc$, including normal incidence and
near-normal incidence ($\thetainc=2.69$~deg). $\calP$ and $\calR$ co-occur only for  $\thetainc>74$~deg, where the reflectances are large enough to be measured; e.g., for the $\calPone\,\&\,\calRthree$ entry  with $\thetainc=84.64196$~deg, $\vert\rLL\vert=0.91$, $\vert\rRR\vert=0.92$, $\vert\rRL\vert=0.092$, and $\vert\rLR\vert=0.0008$. 

Every $\calPone\,\&\,\calRthree$ solution has a $\calPtwo\,\&\,\calRthree$ counterpart with the same $\thetainc$, $\betaz$, and $\betay$ but $\psi+180$~deg, because $\psi\to\psi+180$~deg reverses the component of the direction of incidence parallel to the faces of the slab, interchanges $\calPone$ and $\calPtwo$, and leaves $\calRthree$ unchanged. For instance, the first $\calPone\,\&\,\calRthree$ entry yields $\calPtwo\,\&\,\calRthree$ at $\psi=156.86124$~deg. The $\calPtwo\,\&\,\calRthree$ entries in Table~\ref{tab3} are other solutions found by the search.

Because the andalusite slab is $\sim10^4\lambdao$ thick, the occurrence conditions are very sensitive to the incidence angles $\thetainc$ and $\psi$. Rounded to five decimal places as in Table~\ref{tab3}, the angles give $L\vert\gn{1}-\gn{2}\vert$ or $L\vert\gn3-\gn4\vert\lesssim0.03$ and relative off-diagonal amplitudes $\lesssim
1.3\times10^{-4}$ for $\calT$, i.e., they are     classified as \textsf{near misses}. For $\calRthree$, however, $\sqrt{\vert\Delta_R\vert}/s_R$ then lies between $1.6\times10^{-3}$ and $10^{-2}$, which exceeds the near-miss threshold of $10^{-3}$; seven decimal places are needed to bring the $\calP\,\&\,\calR$ entries within it. The co-occurrence of $\calP$, $\calT$, and $\calR$ imposes six real conditions on four unknowns for fixed $L$, and was therefore not sought.


{\small\setlength{\tabcolsep}{4pt}
\begin{longtable}{lccccl}
\caption{Representative values of $\thetainc$, $\psi$, $\betaz$, and $\betay$ for which a Voigt wave can be seen in transmission ($\calP\,\&\,\calT$) or reflection ($\calP\,\&\,\calR$) for andalusite, when $\lambdao=450$~nm and $L=5$~mm. See Table~S3 of the Supplementary Material for the complete results of the search.}
\label{tab3}\\
\hline
Combination & $\thetainc$ & $\psi$ & $\betaz$ & $\betay$ & Voigt wave \\
 & (deg) & (deg) & (deg) & (deg) & seen in \\
\hline
\endfirsthead
\multicolumn{6}{l}{Table~\ref{tab3} (continued)}\\
\hline
Combination & $\thetainc$ & $\psi$ & $\betaz$ & $\betay$ & Voigt wave \\
 & (deg) & (deg) & (deg) & (deg) & seen in \\
\hline
\endhead
$\calPthree\,\&\,\calTone$ &  $0$ &  $[0,360)$ & $89.99203$ &  $132.24843$ &  transmission \\
$\calPthree\,\&\,\calTtwo$ &  $0$ &  $[0,360)$ &  $90.00797$ &  $132.24843$ &  transmission \\
$\calPthreeS\,\&\,\calTone$ & $0$ & {$[0,360)$} & $89.99203$ & $132.24843$ & transmission \\
$\calPthreeS\,\&\,\calTtwo$ & $0$ & {$[0,360)$} & $90.00797$ & $132.24843$ & transmission \\
$\calPthree\,\&\,\calR$ &  $0$ & -- & --& -- & impossible \\
$\calPthreeS\,\&\,\calR$ & $0$ & -- & -- & -- & impossible \\
$\calPone\,\&\,\calTone$ & $2.69040$ & $113.54460$ & $87.95953$ & $132.92210$ & transmission \\
$\calPone\,\&\,\calTone$ & $5.40308$ & $79.93596$ & $85.60914$ & $131.75601$ & transmission \\
$\calPone\,\&\,\calTone$ & $44.59719$ & $169.16521$ & $83.74697$ & $72.54991$ & transmission \\
$\calPtwo\,\&\,\calTone$ & $55.66771$ & $176.86352$ & $92.12620$ & $102.05168$ & transmission \\
$\calPtwo\,\&\,\calTone$ & $64.93410$ & $22.97800$ & $106.93727$ & $77.88882$ & transmission \\
$\calPone\,\&\,\calRthree$ & $76.61650$ & $336.86124$ & $108.35846$ & $99.59717$ & reflection \\
$\calPone\,\&\,\calRthree$ & $84.64196$ & $104.71313$ & $37.47307$ & $157.18123$ & reflection \\
$\calPtwo\,\&\,\calRthree$ & $74.72141$ & $124.33726$ & $131.04467$ & $117.96669$ & reflection \\
$\calPtwo\,\&\,\calRthree$ & $78.24458$ & $343.59750$ & $76.83644$ & $82.53609$ & reflection \\
\hline
\end{longtable}}


\subsection{Gallium oxide ($\beta$-Ga$_2$O$_3$)}

As a monoclinic dielectric material, we choose
$\beta$-gallium oxide ($\beta$-Ga$_2$O$_3$). At  $\lambdao= 200$~nm, 
this material is characterized by the  
relative permittivity dyadic \cite{Sturm,Grundmann,Anatomy}
\begin{eqnarray}
 \=\eps\rel&=& 
 \=R_y\. \=R_z\. \left[(4.366 + i1.592  )\ux\ux+ (5.016 + i2.15 )\uy\uy+
 (4.686 + i1.46 )\uz\uz\right.
 \nonumber 
 \\ [5pt]
 &&
 \left. + (0.094 + i0.017 )
 (\ux\uz+\uz\ux)
 \right]\.\=R_z^T\.\=R_y^T\,.
 \label{galliumoxide}
\end{eqnarray} 
 
Tables~\ref{tab4} and \ref{tab5}, respectively, provide values of $\psi$, $\betaz$, and $\betay$ so that $\calP$, $\calT$, and $\calR$ occur for  $\theta\inc=0$~deg and
$\theta\inc = 45$~deg, when $L=110$~nm. The normalized thickness $L/\lambdao=0.55$ is several orders of magnitude smaller than for andalusite in Tables~\ref{tab1}--\ref{tab3}
because $\beta$-gallium oxide is highly dissipative whereas andalusite is very weakly dissipative.

{\small\setlength{\tabcolsep}{4pt}
\begin{longtable}{@{}lccc>{\raggedright\arraybackslash}p{5.6cm}@{}}
\caption{Occurrences of  $\calPthree$, $\calPthreeS$, $\calTone$, $\calTtwo$,  
and their combinations for gallium oxide ($\beta$-Ga$_2$O$_3$) described by Eq.~\eqref{galliumoxide} with stated
values of $\psi$, $\betaz$, and $\betay$ for normal incidence ($\theta\inc=0$~deg) when $\lambdao=200$~nm and $L=110$~nm.
No occurrences of $\calPone$, $\calPtwo$, $\calTthree$, and $\calR$ were found.
See Table~S4 of the Supplementary Material for the complete results of the search.}
\label{tab4}\\
\hline
Defectiveness & $\psi$  & $\betaz$   & $\betay$  & Remark \\
  &   (deg) &   (deg) &   (deg) &   \\
\hline
\endfirsthead
\multicolumn{5}{l}{Table~\ref{tab4} (continued)}\\
\hline
Defectiveness & $\psi$  & $\betaz$   & $\betay$  & Remark \\
  &   (deg) &   (deg) &   (deg) &   \\
\hline
\endhead
$\calPthree$ &  $[0,360)$ &  $54.80706$ &  $119.83239$ &  Voigt wave excited \\
$\calPthree$ & {$[0,360)$} & $55.64033$ & $79.45063$ & Voigt wave excited \\
$\calPthreeS$ & $[0,360)$ & $54.80706$ & $119.83239$ & Voigt wave excited \\
$\calPthreeS$ & {$[0,360)$} & $55.64033$ & $79.45063$ & Voigt wave excited \\
$\calTone$ & {$[0,360)$} & $54.80706$ & $119.83239$ &  co-occurs with $\calPthreeS$   \\
$\calTtwo$ & {$[0,360)$} & $55.64033$ & $79.45063$ & co-occurs with $\calPthreeS$  \\
$\calPthree\,\&\,\calTone$ & {$[0,360)$} & $54.80706$ & $119.83239$ & Voigt wave seen in transmission \\
$\calPthree\,\&\,\calTtwo$ & {$[0,360)$} & $55.64033$ & $79.45063$ & Voigt wave seen in transmission \\
$\calPthreeS\,\&\,\calTone$ & {$[0,360)$} & $54.80706$ & $119.83239$ & Voigt wave seen in transmission \\
$\calPthreeS\,\&\,\calTtwo$ & {$[0,360)$} & $55.64033$ & $79.45063$ & Voigt wave seen in transmission \\
\hline
\end{longtable}}

For normal incidence, $\calPthreeS$ (and therefore $\calPthree$) occurs for two orientations of $\beta$-gallium oxide
in Table~\ref{tab4}, each for $\psi\in[0,360)$~deg. $\calPthreeS$ co-occurs with $\calTone$ at the first orientation and with $\calTtwo$ at the second. Furthermore, in accord with $\calCtwo$, $\calTone$ and $\calTtwo$ are found only for those orientations $\lec\betaz,\betay\ric$. Thus, every orientation for which a Voigt wave is excited at normal incidence also makes it observable in transmission. The orientations $\lec\betaz+180~{\rm deg},180~{\rm deg}-\betay\ric$
give the same $\calPthreeS$ but interchange $\calTone$ and $\calTtwo$, as exemplified by  $\lec\betaz=235.64033~{\rm deg}, \betay=100.54937~{\rm deg}\ric$.
These findings agree with $\tLR$, etc., computed by Grundmann \textit{et al.} \cite{Grundmann} for normal incidence on a $\beta$-Ga$_2$O$_3$ slab cut perpendicular to a singular axis at the same wavelength. Their
work provides an instance of the co-occurrence of $\calPthreeS$ and $\calT$ at normal incidence.

$\calR$ cannot co-occur with $\calPthreeS$ at normal incidence because 
$\vert {\tilde N_{11}}\vert\approx 0.13$ and $ 0.15$ for the two orientations
for $\calPthreeS$, whereas $\vert\rho(c_a)/\alpha_\rho(c_a)\vert\ge2.6$ for all $L$. 

{\small\setlength{\tabcolsep}{4pt}
\begin{longtable}{@{}lccc>{\raggedright\arraybackslash}p{5.6cm}@{}}
\caption{Occurrences of  $\calPone$, $\calPtwo$, $\calTone$, and $\calTtwo$  
 for gallium oxide ($\beta$-Ga$_2$O$_3$) described by Eq.~\eqref{galliumoxide} with stated
values of $\psi$, $\betaz$, and $\betay$ for  oblique incidence ($\theta\inc=45$~deg) when $\lambdao=200$~nm and $L=110$~nm.
No occurrences of $\calPthree$, $\calPthreeS$, $\calTthree$, and $\calR$ were found.
See Table~S5 of the Supplementary Material for the complete results of the search.}
\label{tab5}\\
\hline
Defectiveness & $\psi$  & $\betaz$   & $\betay$  & Remark \\
  &   (deg) &   (deg) &   (deg) &   \\
\hline
\endfirsthead
\multicolumn{5}{l}{Table~\ref{tab5} (continued)}\\
\hline
Defectiveness & $\psi$  & $\betaz$   & $\betay$  & Remark \\
  &   (deg) &   (deg) &   (deg) &   \\
\hline
\endhead
$\calPone$ & $134.25433$ & $112.94510$ & $74.79045$ & Voigt wave excited \\
$\calPtwo$ & $314.25609$ & $112.94564$ & $74.79082$ & Voigt wave excited \\
$\calTone$ & $45.08679$ & $90.93329$ & $36.80185$ & no Voigt wave excited \\
$\calTtwo$ & $136.83906$ & $65.11573$ & $92.15653$ & no Voigt wave excited \\
\hline
\end{longtable}}

For $\thetainc=45$~deg, $\calPone$ occurs once and so does $\calPtwo$, but not in combination with either $\calT$ or $\calR$. Therefore, the excited
Voigt wave remains unobservable. The single occurrences of  $\calTone$ and $\calTtwo$ in Table~\ref{tab5} could lead to the conclusion that
Voigt-wave excitation is observable in transmission, but that conclusion would be wrong.


{\small\setlength{\tabcolsep}{4pt}
\begin{longtable}{lccccl}
\caption{Same as Table~\ref{tab3} but for $\beta$-Ga$_2$O$_3$, when $\lambdao=200$~nm and $L=110$~nm. See Table~S6 of the Supplementary Material for the complete results of the search.}\label{tab6}\\
\hline
Combination & $\thetainc$ & $\psi$ & $\betaz$ & $\betay$ & Voigt wave \\
 & (deg) & (deg) & (deg) & (deg) & seen in \\
\hline
\endfirsthead
\multicolumn{6}{l}{Table~\ref{tab6} (continued)}\\
\hline
Combination & $\thetainc$ & $\psi$ & $\betaz$ & $\betay$ & Voigt wave \\
 & (deg) & (deg) & (deg) & (deg) & seen in \\
\hline
\endhead
$\calPthree\,\&\,\calTone$ & $0$ & $[0,360)$ & $54.80706$ & $119.83239$ & transmission \\
$\calPthree\,\&\,\calTtwo$ & $0$ & $[0,360)$ & $55.64033$ & $79.45063$ & transmission \\
$\calPthreeS\,\&\,\calTone$ & $0$ & $[0,360)$ & $54.80706$ & $119.83239$ & transmission \\
$\calPthreeS\,\&\,\calTtwo$ & $0$ & $[0,360)$ & $55.64033$ & $79.45063$ & transmission \\
$\calPthree\,\&\,\calR$ & $0$ & -- & --  & --  & impossible \\
$\calPthreeS\,\&\,\calR$ & $0$ &--  & -- & --  &  impossible \\
$\calPone\,\&\,\calTone$ & $3.88077$ & $87.21476$ & $52.81390$ & $120.06205$ & transmission \\
$\calPone\,\&\,\calTone$ & $55.02777$ & $68.85288$ & $284.03730$ & $109.93957$ & transmission \\
$\calPone\,\&\,\calTtwo$ & $55.02777$ & $111.14712$ & $104.03730$ & $70.06043$ & transmission \\
$\calPone\,\&\,\calTtwo$ & $58.25981$ & $93.64501$ & $282.10842$ & $84.29836$ & transmission \\
$\calPtwo\,\&\,\calTone$ & $47.52530$ & $231.09885$ & $290.77002$ & $105.80787$ & transmission \\
$\calPtwo\,\&\,\calTone$ & $65.87349$ & $109.49068$ & $148.80419$ & $97.63551$ & transmission \\
$\calPtwo\,\&\,\calTtwo$ & $63.75773$ & $342.33450$ & $113.16642$ & $122.13639$ & transmission \\
$\calPtwo\,\&\,\calTtwo$ & $69.31938$ & $152.87416$ & $252.13568$ & $38.90741$ & transmission \\
\hline
\end{longtable}}

Table~\ref{tab6} is the counterpart of Table~\ref{tab3} for $\beta$-gallium oxide with $L=110$~nm.
$\calP$ co-occurs with $\calTone$ or $\calTtwo$ for many values of $\thetainc$. No co-occurrence of $\calP$ and $\calR$ was found for $\thetainc\in(0,89)$~deg, even when $L\in[10,400]$~nm was also varied; the smallest residual ($6\times10^{-4}$) was always approached as $\thetainc\to90$~deg, where $\Rcirc$ tends to a multiple of $[I]$.

\subsection{Terbium orthoaluminate (TbAlO$_3$) in a magnetostatic field} 

Both andalusite and $\beta$-gallium oxide are Lorentz-reciprocal because $\=\eps\rel^T=\=\eps\rel$. Nonreciprocity can be created by adding a gyrotropic term $i\eps_g\=I\times\hat{\#u}$ to the relative permittivity dyadic, $\hat{\#u}$ being a unit vector.
Rare-earth orthoaluminates are promising candidate materials for investigating Voigt waves. These dielectric
materials are biaxial, they absorb strongly in the bands of the rare-earth ions \cite{Sardar2006}, and their Faraday effect has been measured despite the large natural birefringence \cite{Valiev2008,Valiev2023}.

In order to illustrate the effects of non-reciprocity, we chose terbium orthoaluminate. At $\lambdao=485$~nm, which lies in the $^7$F$_6\to{}^5$D$_4$ absorption band, we took the relative permittivity dyadic as
\begin{equation}
\=\eps\rel= \=R_y\. \=R_z\. \left[\eps_a\,\ux\ux+\eps_b\,\uy\uy+\eps_c\,\uz\uz+i\eps_g\=I\times\uz\right]\.\=R_z^T\.\=R_y^T\,,
\label{tbalo3}
\end{equation}
where the  magnetostatic field is directed along $\=R_y\.\=R_z\.\uz=\uz\cos\betay-\ux\sin\betay$, which makes an angle $\betay$ with the normal to both faces of the slab. The mean refractive index $\bar n=1.76840$ follows from the Sellmeier equation for TbAlO$_3$ \cite{Sardar2006}, and the principal refractive indices 
(in the absence of gyrotropy) were taken as $\bar n-0.009$, $\bar n$, and $\bar n+0.009$, in accord with a natural birefringence of about $0.018$ near $500$~nm \cite{Valiev2008}. The mean extinction coefficient $\bar\kappa=8.89\times10^{-6}$ was estimated from the measured line strength of the absorption band \cite{Sardar2006} for an assumed effective bandwidth of $5$~nm, and the three principal extinction coefficients were assumed to be in the ratio $1.5:0.5:1$, since polarized absorption data are not available. Hence, $\eps_a=3.0954884+i4.6923\times10^{-5}$, $\eps_b=3.1272386+i1.5721\times10^{-5}$, and $\eps_c=3.1591508+i3.1602\times10^{-5}$, the real parts of $\eps_{a,b,c}$ being given to seven decimal places because the orientations in Table~\ref{tab7} are sensitive to the sixth. The Verdet constant $V=0.4$~min~cm$^{-1}$~Oe$^{-1}$ measured along an optic axis at $\lambdao=532$~nm \cite{Valiev2023} was scaled to $V=0.50$~min~cm$^{-1}$~Oe$^{-1}$ at $\lambdao=485$~nm on the assumption that $V\propto(\lambdao^2-\lambda_t^2)^{-1}$ with $\lambda_t=220$~nm, and the magneto-optic response was taken to be isotropic, so that $\eps_g=2\bar nV\Bdc/\ko$. We chose $\Bdc=0.2$~T, for which $\eps_g=7.98\times10^{-6}$ and the Faraday rotation is $1.7$~deg~mm$^{-1}$ for propagation along the magnetostatic field. Thus the non-reciprocity is appreciable but readily achievable with a laboratory electromagnet. The thickness $L=1$~mm was fixed. Since $\=\eps\rel$ is not symmetric, $\calPthreeS$ at normal incidence and every co-occurrence with $\calT$ or $\calR$ were searched for afresh, as for andalusite and $\beta$-Ga$_2$O$_3$.

{\small\setlength{\tabcolsep}{4pt}
\begin{longtable}{lccccl}
\caption{Same as Table~\ref{tab3} but for TbAlO$_3$ described by Eq.~\eqref{tbalo3} with $\Bdc=0.2$~T, when $\lambdao=485$~nm and $L=1$~mm. See Table~S7 of the Supplementary Material for the complete results of the search.}\label{tab7}\\
\hline
Combination & $\thetainc$ & $\psi$ & $\betaz$ & $\betay$ & Voigt wave \\
 & (deg) & (deg) & (deg) & (deg) & seen in \\
\hline
\endfirsthead
\multicolumn{6}{l}{Table~\ref{tab7} (continued)}\\
\hline
Combination & $\thetainc$ & $\psi$ & $\betaz$ & $\betay$ & Voigt wave \\
 & (deg) & (deg) & (deg) & (deg) & seen in \\
\hline
\endhead
$\calPthree\,\&\,\calTthree$ & $0$ & $[0,360)$ & $0.02630$ & $45.21871$ & transmission \\
$\calPthree\,\&\,\calTthree$ & $0$ & $[0,360)$ & $0.02630$ & $134.78129$ & transmission \\
$\calPthree\,\&\,\calTthree$ & $0$ & $[0,360)$ & $179.97370$ & $45.21871$ & transmission \\
$\calPthree\,\&\,\calTthree$ & $0$ & $[0,360)$ & $179.97370$ & $134.78129$ & transmission \\
$\calPthreeS\,\&\,\calTthree$ & $0$ & $[0,360)$ & $0.02630$ & $45.21871$ & transmission \\
$\calPthreeS\,\&\,\calTthree$ & $0$ & $[0,360)$ & $0.02630$ & $134.78129$ & transmission \\
$\calPthreeS\,\&\,\calTthree$ & $0$ & $[0,360)$ & $179.97370$ & $45.21871$ & transmission \\
$\calPthreeS\,\&\,\calTthree$ & $0$ & $[0,360)$ & $179.97370$ & $134.78129$ & transmission \\
$\calPthree\,\&\,\calR$ & $0$ & -- & -- & -- & impossible \\
$\calPthreeS\,\&\,\calR$ & $0$ & -- & -- & -- & impossible \\
$\calPone\,\&\,\calTthree$ & $20.83493$ & $153.35062$ & $172.72458$ & $55.38430$ & transmission \\
$\calPone\,\&\,\calTthree$ & $41.78026$ & $106.08769$ & $149.36107$ & $47.34809$ & transmission \\
$\calPone\,\&\,\calTthree$ & $82.83557$ & $9.62627$ & $172.37777$ & $101.28038$ & transmission \\
$\calPone\,\&\,\calTthree$ & $88.97145$ & $277.83047$ & $52.07758$ & $26.41978$ & transmission \\
$\calPtwo\,\&\,\calTthree$ & $31.53697$ & $140.00930$ & $15.55956$ & $30.80769$ & transmission \\
$\calPtwo\,\&\,\calTthree$ & $44.27969$ & $59.05027$ & $28.46441$ & $53.99074$ & transmission \\
$\calPtwo\,\&\,\calTthree$ & $76.21707$ & $98.67111$ & $49.92397$ & $27.32874$ & transmission \\
$\calPone\,\&\,\calRthree$ & $63.19634$ & $330.75283$ & $20.35563$ & $16.34958$ & reflection \\
$\calPone\,\&\,\calRthree$ & $82.85379$ & $9.68658$ & $172.33005$ & $101.28677$ & reflection \\
$\calPtwo\,\&\,\calRthree$ & $64.33726$ & $176.45250$ & $2.52060$ & $14.59444$ & reflection \\
$\calPtwo\,\&\,\calRthree$ & $84.68558$ & $85.02128$ & $52.23637$ & $35.07752$ & reflection \\
\hline
\end{longtable}}


Table~\ref{tab7} shows that a modest magnetostatic field changes the observability of Voigt waves qualitatively. At normal incidence, $\calPthreeS$ (and therefore $\calPthree$) holds for $\betaz\in\lec0.02630,179.97370\ric$~deg and $\betay\in\lec45.21871,134.78129\ric$~deg, for every $\psi$, and $\calT$ co-occurs, in accord with $\calCone$ and $\calCtwo$. The magnetostatic field is then oriented at angle $\betay$ with respect to the $z$ axis. Since the gyrotropic term changes sign when $\betay$ is replaced by $\betay+180$~deg, the orientations $\lec\psi,\betaz,\betay\ric$ and $\lec\psi,-\betaz,\betay+180~{\rm deg}\ric$ are not equivalent for TbAlO$_3$ in a magnetostatic field. The orientations $\lec\psi,\betaz,\betay\ric$ and $\lec\psi+180~{\rm deg},\betaz,-\betay\ric$ are equivalent, however, and $\betay\in[0,180]$~deg has therefore been reported in Table~\ref{tab7}. Because $\=\eps\rel$ is not symmetric, the Voigt wave is elliptically polarized, the ratio of the magnitudes of its RCP and LCP amplitudes being $3.9$ for $\lec\betaz,\betay\ric=\lec0.02630,45.21871\ric$~deg and $\lec179.97370,134.78129\ric$~deg, and $0.26$ for the other two orientations. Consequently, $\Tcirc$ is of type $\calTthree$, in accord with $\calCthree$. Indeed, due to non-reciprocity, neither $\calTone$ nor $\calTtwo$ can co-occur with $\calPthreeS$ at normal incidence.

At oblique incidence, $\calPone$ and $\calPtwo$ co-occur with $\calTthree$ over a wide range of $\thetainc$ ($\thetainc>20$~deg in our searches), whereas no co-occurrence with $\calTone$ or $\calTtwo$ was found. The roles of $\calTthree$ and of $\calTone$ and $\calTtwo$ are thus interchanged in comparison with andalusite and $\beta$-Ga$_2$O$_3$. For a Lorentz-reciprocal material, $\tLL=\tRR$, so that $\calTthree$ is excluded whereas $\calTone$ and $\calTtwo$ each require only the vanishing of one off-diagonal element of $\Tcirc$. For TbAlO$_3$ in a magnetostatic field, $\tLL\ne\tRR$ in general, so that $\calTthree$ imposes only two real conditions whereas $\calTone$ and $\calTtwo$ impose four. 

$\calPone$ and $\calPtwo$ co-occur with $\calRthree$ for $\thetainc>63$~deg; e.g., for the $\calPtwo\,\&\,\calRthree$ entry with $\thetainc=84.68558$~deg, $\vert\rLL\vert=0.93$, $\vert\rRR\vert=0.92$, $\vert\rRL\vert=0.098$, and $\vert\rLR\vert=0.0008$. No co-occurrence of $\calP$ with $\calRone$ or $\calRtwo$ was found. 

$\calPthreeS$ and $\calR$ cannot co-occur at normal incidence, because $\vert {\tilde N_{11}}\vert\leq1.2\times10^{-5}$ for all $\psi$, whereas $\vert\rho(c_a)/\alpha_\rho(c_a)\vert\geq2.76\times10^{-5}$ for all $L$. Parenthetically, unlike for a Lorentz-reciprocal material, $\vert {\tilde N_{11}}\vert$ now depends on $\psi$. Furthermore, $\calPthree$ was not found at oblique incidence, not even when $L\in[0.1,3]$~mm and $\Bdc\in[-1,1]$~T were also treated as unknowns, with every search converging to $\thetainc\to0$. The reason is that $\calPthree$ at oblique incidence requires the directions of propagation of both the forward and the backward  waves inside the slab, which are mirror images of each other in the plane of the slab, to be singular axes. Hence, $\calPthree$ co-occurs with $\calTthree$ only at normal incidence, and $\calPthree$ does not co-occur with $\calRthree$ at all.

\subsection{Hexagonal boron nitride (hBN)}\label{hBN-detailed}

As our final example, we chose hexagonal boron nitride, a uniaxial dielectric material whose principal relative permittivities have real parts of opposite signs in two mid-infrared bands, so that it is a natural hyperbolic material \cite{Caldwell2014}. Its relative permittivity dyadic is
\begin{equation}
\=\eps\rel= \=R_y\. \=R_z\. \left[\eps_t\left(\ux\ux+\uy\uy\right)+\eps_z\,\uz\uz\right]\.\=R_z^T\.\=R_y^T\,,
\label{hBN}
\end{equation}
with
\begin{equation}
\eps_j=\eps_{\infty,j}\left(1+\frac{\tilde\nu_{{\rm LO},j}^2-\tilde\nu_{{\rm TO},j}^2}{\tilde\nu_{{\rm TO},j}^2-\tilde\nu^2-i\Gamma_j\tilde\nu}\right)\,,\qquad j\in\lec t,z\ric\,,
\label{hBN-Lorentz}
\end{equation}
where $\tilde\nu=1/\lambdao$. We used the parameters $\eps_{\infty,t}=4.98$, $\tilde\nu_{{\rm TO},t}=1362.7$~cm$^{-1}$, $\tilde\nu_{{\rm LO},t}=1616.9$~cm$^{-1}$, $\Gamma_t=7.3$~cm$^{-1}$, $\eps_{\infty,z}=2.95$, $\tilde\nu_{{\rm TO},z}=760$~cm$^{-1}$, $\tilde\nu_{{\rm LO},z}=825$~cm$^{-1}$, and $\Gamma_z=3$~cm$^{-1}$, obtained by Caldwell \textit{et al.} \cite{Caldwell2014} by fitting measured infrared reflectance spectra and tabulated by Giles \textit{et al.} \cite{Giles2018}. Thus, $\eps_t=8.0794+i0.01487$ and $\eps_z=-1.9134+i0.18705$ at $\lambdao=12.5$~$\mu$m (type-I hyperbolic), whereas $\eps_t=-4.6091+i0.26715$ and $\eps_z=2.7683+i0.000489$ at $\lambdao=6.6667$~$\mu$m (type-II hyperbolic).

Since hBN is uniaxial, $\=\teps\rel=\eps_t\=I+(\eps_z-\eps_t)\,\hat{\#c}\hat{\#c}$, where the unit vector
\begin{equation}
\hat{\#c}=-\ux\cos\psi\sin\betay+\uy\sin\psi\sin\betay+\uz\cos\betay
\end{equation}
is parallel to the optic axis and does not depend on $\betaz$. Only $\psi$, $\betay$, and $\thetainc$ are therefore available, whereas, e.g., $\calPone\,\&\,\calTone$ imposes four real conditions. Hence, the thickness $L\in[0.1,10]$~$\mu$m was treated as the fourth unknown, and each co-occurrence in  Tables~\ref{tab8} and  S8  is reported with its own value of $L$.

The quartic characteristic polynomial factorizes into the ordinary factor $\zeta^2-\eps_t+\sin^2\thetainc$ and an extraordinary factor, as it must for any uniaxial dielectric material \cite{Chen}.
An ordinary root $\zeta_o$ coincides with an extraordinary root $\zeta_e$ if and only if $\tilde{\#k}\.\hat{\#c}=\pm\sqrt{\eps_t}$, where $\tilde{\#k}=\ux\sin\thetainc+\uz\zeta_o$; equivalently, $\tilde{\#k}_\perp\.\tilde{\#k}_\perp=0$, where $\tilde{\#k}_\perp=\tilde{\#k}-(\tilde{\#k}\.\hat{\#c})\,\hat{\#c}$. Unless $\tilde{\#k}\parallel \hat{\#c}$ (i.e., $\tilde{\#k}_\perp=\#0$),  the ordinary and extraordinary eigenvectors then coincide and $\mattP$ is defective. Because $\hat{\#c}\in\mathbb{R}^3$, these conditions give
\begin{equation}
\left.\begin{array}{l}
\hat{\#c}\.\uz=\displaystyle{{\Im\lec\sqrt{\eps_t}\ric}/{\Im\lec\zeta_o\ric}}\\[7pt]
\hat{\#c}\.\ux=\displaystyle{\frac{\Re\lec\sqrt{\eps_t}\ric-(\hat{\#c}\.\uz)\,\Re\lec\zeta_o\ric}{\sin\thetainc}}\\[7pt]
\hat{\#c}\.\uy=\pm\sqrt{1-(\hat{\#c}\.\ux)^2-(\hat{\#c}\.\uz)^2}
\end{array}\right\}\,,
\label{hBN-P1}
\end{equation}
with $\Im\lec\zeta_o\ric>0$ for $\calPone$ and $\Im\lec\zeta_o\ric<0$ for $\calPtwo$. Thus, $\calPone$ and $\calPtwo$ hold on curves parametrized by $\thetainc$, provided that $(\hat{\#c}\.\ux)^2+(\hat{\#c}\.\uz)^2<1$. Then $\hat{\#c}\.\uy\ne0$  ensures that the double root is defective, which we confirmed by computing the cofactors of $\mattW$. The plane waves inside the slab are nonuniform 
for oblique incidence, because $\tilde{\#k}$ is complex but not a complex multiple of a real vector. This is consistent with the conclusion that Voigt waves cannot propagate in a uniaxial dielectric material \cite{Gerardin,MLProgOpt}, which concerns uniform plane waves.  

$\calPthree$ would require both $\zeta_o$ and $-\zeta_o$ to satisfy Eqs.~\eqref{hBN-P1}, which demands that $(\hat{\#c}\.\ux)\sin\thetainc=0$ and that $\zeta_o^2/\eps_t$ be real. As a result, both $\calPthree$ and $\calPthreeS$ are impossible for oblique incidence. 

For normal incidence, 
\begin{equation}
[C]=\eps_t[I]+ \frac{\eps_t(\eps_z-\eps_t)}{\eps_t+(\eps_z-\eps_t)\uz\.\hat{\#c}\hat{\#c}\.\uz}\,
\les\begin{matrix}  \ux\.\hat{\#c}\hat{\#c}\.\ux &   \ux\.\hat{\#c}\hat{\#c}\.\uy\\   \uy\.\hat{\#c}\hat{\#c}\.\ux &  \uy\.\hat{\#c}\hat{\#c}\.\uy\end{matrix}\ris\,.
\end{equation} 
That is, $[C]-\eps_t[I]$ is a complex multiple of a real symmetric matrix and cannot be defective. Accordingly,  $\calPthreeS$ is impossible at normal incidence too. Finally, $\calTthree$ is excluded because $\tLL=\tRR$ for this Lorentz-reciprocal material. The searches were accordingly carried out on the curves of Eqs.~\eqref{hBN-P1}, with $\thetainc$ and $L$ as the unknowns.

{\small\setlength{\tabcolsep}{3.5pt}
\begin{longtable}{lcccccl}
\caption{Same as Table~\ref{tab3} but for hBN described by Eq.~\eqref{hBN}, when $\lambdao=12.5$~$\mu$m. The thickness $L$ was an additional unknown in $[0.1,10]$~$\mu$m. See Table~S8 of the Supplementary Material for the complete results of the search.}
\label{tab8}\\
\hline
Combination & $\thetainc$ & $\psi$ & $\betaz$ & $\betay$ & $L$ & Voigt wave \\
 & (deg) & (deg) & (deg) & (deg) & ($\mu$m) & seen in \\
\hline
\endfirsthead
\multicolumn{7}{l}{Table~\ref{tab8} (continued)}\\
\hline
Combination & $\thetainc$ & $\psi$ & $\betaz$ & $\betay$ & $L$ & Voigt wave \\
 & (deg) & (deg) & (deg) & (deg) & ($\mu$m) & seen in \\
\hline
\endhead
$\calPthree\,\&\,\calT$, $\calPthreeS\,\&\,\calT$ & $0$ & -- & -- & -- & -- & impossible \\
$\calPthree\,\&\,\calR$, $\calPthreeS\,\&\,\calR$ & $0$ & -- & -- & -- & -- & impossible \\
$\calPone\,\&\,\calTone$ & $62.57588$ & $180.05552$ & $[0,360)$ & $18.19640$ & $9.2581$ & transmission \\
$\calPone\,\&\,\calTone$ & $64.80864$ & $180.05563$ & $[0,360)$ & $18.56322$ & $4.6393$ & transmission \\
$\calPone\,\&\,\calTone$ & $77.89615$ & $180.05617$ & $[0,360)$ & $20.12016$ & $7.0253$ & transmission \\
$\calPone\,\&\,\calTtwo$ & $62.57588$ & $179.94448$ & $[0,360)$ & $18.19640$ & $9.2581$ & transmission \\
$\calPone\,\&\,\calTtwo$ & $64.80864$ & $179.94437$ & $[0,360)$ & $18.56322$ & $4.6393$ & transmission \\
$\calPone\,\&\,\calTtwo$ & $77.89615$ & $179.94383$ & $[0,360)$ & $20.12016$ & $7.0253$ & transmission \\
$\calPtwo\,\&\,\calTone$ & $69.13305$ & $359.94416$ & $[0,360)$ & $19.19207$ & $1.7134$ & transmission \\
$\calPtwo\,\&\,\calTone$ & $69.14097$ & $0.05584$ & $[0,360)$ & $19.19312$ & $1.7140$ & transmission \\
$\calPtwo\,\&\,\calTtwo$ & $69.13305$ & $0.05584$ & $[0,360)$ & $19.19207$ & $1.7134$ & transmission \\
$\calPtwo\,\&\,\calTtwo$ & $69.14097$ & $359.94416$ & $[0,360)$ & $19.19312$ & $1.7140$ & transmission \\
$\calPone\,\&\,\calRthree$ & $71.10542$ & $179.94407$ & $[0,360)$ & $19.44211$ & $2.3323$ & reflection \\
$\calPone\,\&\,\calRthree$ & $71.65386$ & $180.05595$ & $[0,360)$ & $19.50745$ & $4.6655$ & reflection \\
$\calPtwo\,\&\,\calRthree$ & $71.10542$ & $359.94407$ & $[0,360)$ & $19.44211$ & $2.3323$ & reflection \\
$\calPtwo\,\&\,\calRthree$ & $71.65386$ & $0.05595$ & $[0,360)$ & $19.50745$ & $4.6655$ & reflection \\
\hline
\end{longtable}}

According to Table~\ref{tab8}, a Voigt wave excited in hBN at $\lambdao=12.5$~$\mu$m can be observed in transmission for $\thetainc\in[62,78]$~deg and $L\in[1.7,9.3]$~$\mu$m, and in reflection for $\thetainc\approx71$~deg. Because the dissipation for the ordinary plane wave is weak at this wavelength, Eqs.~\eqref{hBN-P1} yield $(\hat{\#c}\.\ux)^2+(\hat{\#c}\.\uz)^2=1$ to first order in $\Im\lec\eps_t\ric$, and $\vert\hat{\#c}\.\uy\vert\approx3\times10^{-4}$ in every solution. The optic axis then lies within $0.02$~deg of the plane of incidence, with $\psi$ within $0.06$~deg of $180$~deg for $\calPone$ and   within $0.06$~deg of $0$~deg (i.e., $360$~deg) for $\calPtwo$. Consequently, the signature of the Voigt wave is weak:  $\vert\tLL\vert=\vert\tRR\vert\in[0.94,0.99]$
and $\vert\tRL\vert/\vert\tLL\vert\in[2,7]\times10^{-4}$
for  $\calPone\,\&\,\calTone$. 

Every $\calPone\,\&\,\calTone$ solution has a $\calPone\,\&\,\calTtwo$ counterpart with the same $\thetainc$, $\betay$, and $L$ but with $360~{\rm deg}-\psi$ in place of $\psi$. This is because reflection of the optic axis in the plane of incidence interchanges $\calTone$ and $\calTtwo$ but leaves $\calPone$ unchanged. Similarly, each co-occurrence of $\calPone$ and $\calRthree$ has a $\calPtwo\,\&\,\calRthree$ counterpart with $\psi+180$~deg. For the $\calPone\,\&\,\calRthree$ co-occurrence with $\thetainc=71.65386$~deg, $\vert\rLL\vert\approx\vert\rRR\vert=0.027$, $\vert\rRL\vert=0.0008$, and $\vert\rLR\vert=0.0004$.

At $\lambdao=6.6667$~$\mu$m, $\calPone$ and $\calPtwo$ hold on curves on which the optic axis is far from the plane of incidence (e.g., $\psi\approx91.5$~deg or $268.5$~deg and $\betay=21.95$~deg for $\thetainc=60$~deg), but no co-occurrence of $\calP$ with $\calT$ or $\calR$ was found for $\thetainc\in[0.5,89]$~deg and $L\in[0.1,10]$~$\mu$m (Table~S9). The smallest residuals were approached at the edges $\thetainc\to0.5$~deg and $L\to0.1$~$\mu$m of the search domain, where $\Tcirc$ tends to $[I]$. Since $\Re\lec\eps_t\ric<0$ at this wavelength, $\Im\lec\zeta_o\ric\geq2.1$, so that the two plane waves that coalesce into the Voigt wave decay by $e^{-1}$ over a distance less than $0.5$~$\mu$m.

\subsection{Is a defective $\Tcirc$ or $\Rcirc$ essential for observability?}\label{essential}
A Voigt wave excited inside a slab is deemed observable in transmission if $\calP$ and $\calT$ co-occur, and in reflection if $\calP$ and $\calR$ co-occur. The rationale is as follows. Only $\Tcirc$ and $\Rcirc$ are accessible to measurement for a specified direction of incidence $\lec\thetainc,\psi\ric$ and a specified thickness $L$. The defining property of a Voigt wave is the coalescence of two eigenvectors of $\matP$. When $\Tcirc$ ($\Rcirc$) is also defective, the slab regarded as a black box displays the same characteristic: exactly one polarization state is transmitted (reflected) without change of state, and the other polarization states are partially converted into it. This signature is structural, in that it is a qualitative property of a single measured 2$\times$2 matrix and does not require the comparison of slabs of different thicknesses. At normal incidence, its connection with the Voigt wave is also explicit: by Eqs.~\eqref{TR-normal-2}, the nilpotent part $\alpha_\tau(c_a){[\tilde N]}$ of $\Tlin$ descends directly from the Jordan block of $[C]$, and $\alpha_\tau(c_a)\approx i\ko L/2$ for thin slabs, so that the leakage into the eigenpolarization grows linearly with $L$, as Voigt observed. Moreover, $\calT$ then implies $\calPthreeS$ ($\calCtwo$), so that at normal incidence a defective $\Tcirc$ measured for a single value of $L$ suffices to establish that a Voigt wave is excited.

A defective $\Tcirc$ or $\Rcirc$ is nevertheless not essential for the Voigt wave to affect the measured quantities. The elements of $\Tcirc$ and $\Rcirc$ are rational functions of the elements of $\mattM$. When $\mattP$ is defective, $\mattM$ contains terms proportional to $L\exp(igL)$, which generally survive in $\Tcirc$ and $\Rcirc$ whether or not those matrices are defective. In principle, then, the linear dependence on propagation distance could be sought in the variation of $\Tcirc$ or $\Rcirc$ with $L$ for a fixed direction of incidence, without any matrix being defective. 

However, two difficulties beset this approach. First, multiple reflections inside the slab superpose Fabry--P\'erot oscillations on that variation. Second, and more fundamentally, the variation with $L$ cannot discriminate between an exactly defective $\mattP$ and a nearly defective one, because
\begin{equation}
\frac{\exp(ig_1L)-\exp(ig_2L)}{g_1-g_2}\to iL\exp(ig_1L)\quad{\rm as}\quad\vert g_1-g_2\vert L\to0\,.
\end{equation}
Indeed, reflectances and transmittances vary smoothly as the direction of incidence passes through a singular axis \cite{Anatomy}, and the region of near-defectiveness in Fig.~\ref{fig:cono}(c) has a finite angular extent. A linear dependence on $L$ observed in this manner therefore shows only that $\vert g_1-g_2\vert L\ll1$ over the range of thicknesses examined. The same caveat applies to every measurement of finite precision, which is why near misses are recorded in the Supplementary Material.

Conversely, a defective $\Tcirc$ or $\Rcirc$ is not sufficient by itself, as shown by the entries labeled ``no Voigt wave excited'' in Tables~\ref{tab2} and \ref{tab5}. At oblique incidence, a defective $\Tcirc$ or $\Rcirc$ does not discriminate between a Voigt wave and a coincidental exceptional point of the Jones matrix. In summary, a defective $\Tcirc$ or $\Rcirc$ is not necessary for the Voigt wave to leave a trace in the measured quantities, and it is not sufficient by itself for a Voigt wave to be present. Its co-occurrence with $\calP$ is the condition under which the exceptional point of $\matP$ inside the slab is replicated by an exceptional point of an externally measured matrix, and it is in this sense that we use the term observability.

\section{Concluding remarks}\label{cr}

We have formulated a theory of the observability of Voigt waves in a slab of a homogeneous dissipative dielectric material. A Voigt wave is excited inside the slab when the 4$\times$4 propagation matrix $\matP$ is defective, and explicit tests have been provided for the defectiveness types $\calPone$, $\calPtwo$, $\calPthree$, and $\calPthreeS$. Excitation does not by itself ensure observability, because the 2$\times$2 transmission and reflection matrices are extracted from 2$\times$2 blocks of the
 4$\times$4 transfer matrix $\mattT$ and need not inherit its Jordan structure. Conversely, a defective transmission or reflection matrix does not by itself signal the excitation of a Voigt wave. At oblique incidence, the co-occurrence of $\calP$ with $\calT$ or $\calR$ requires four real conditions to be satisfied, and so holds only at isolated points when a fourth parameter is also tuned. Normal incidence is exceptional, since $\calPthreeS$ then implies $\calT$ for almost all thicknesses and $\calT$ implies $\calPthreeS$. That is why Voigt could observe the effect in transmission. Our calculations reproduced his observations and confirmed Pancharatnam's correction of his interpretation. 
 
Every Voigt wave excited in a slab by an obliquely incident
 plane wave is a nonuniform plane wave whose linear dependence is on the distance normal to the faces of the slab rather than along its direction of phase propagation; such waves exist even in materials that have no singular axes. A defective transmission or reflection matrix is neither necessary for a Voigt wave to influence the measured quantities nor sufficient by itself to signal its presence, but its co-occurrence with a defective propagation matrix is our criterion of observability. Dissipation and biaxiality do not, however, guarantee the existence of singular axes. We have shown here that an orthorhombic dielectric material whose principal relative permittivities have collinear reciprocals in the complex plane has none, contrary to statements in the literature \cite{Voigt1908,BornWolf,Sturm}. Its propagation matrix cannot be defective for normal incidence, but it can be for oblique incidence.
 
 For andalusite, the co-occurrence of $\calP$ and $\calR$ was found only for $\thetainc>74$~deg; for $\beta$-Ga$_2$O$_3$, it was not found at all.  For terbium orthoaluminate in a magnetostatic field of $0.2$~T, the Voigt wave is elliptically polarized, so that $\calPthreeS$ co-occurs with $\calTthree$ at normal incidence, and $\calPone$ and $\calPtwo$ co-occur with $\calTthree$ and $\calRthree$ at oblique incidence, whereas $\calTone$, $\calTtwo$, $\calRone$, and $\calRtwo$ were not found and $\calPthree$ does not co-occur with $\calR$. For hexagonal boron nitride, a uniaxial material, a Voigt wave is excited as a nonuniform plane wave for oblique incidence  on explicitly determined curves in the space of $\thetainc$, $\psi$, and $\betay$. It is observable in transmission and reflection at $\lambdao=12.5$~$\mu$m, but no co-occurrence was found at $\lambdao=6.6667$~$\mu$m.

Several extensions of our theory of observability of Voigt waves appear promising. Measurements of reflection at grazing incidence on andalusite slabs could test the predicted co-occurrence of $\calP$ and $\calR$, provided that the direction of incidence can be controlled to within about $10^{-7}$~deg. The effects of finite beamwidth and of the angular spread of the incident light need to be quantified. The theory can be extended to active dielectric materials, to bianisotropic materials, and to the higher-order degeneracies identified by Borzdov \cite{Borzdov1996}, which involve Jordan blocks of size 3 and 4. Engineered composite materials may allow the singular axes to be tuned so that the excitation and the observability of Voigt waves co-occur in configurations of practical use.

\section*{Acknowledgements}
{Both authors were partially supported  by EPSRC (grant number APP68512). A.L. is grateful to the Evan Pugh University Professorships
Endowment at Penn State for ongoing support of his research.}

\section*{Declaration of competing interest}
The authors declare that they have no known competing financial interests or personal relationships that could have
appeared to influence the work reported in this manuscript.

\section*{Declaration of AI use}
A.L.  used ChatGPT and Claude for locating  20th-century papers and translating them
into modern English;   updating a Mathematica\texttrademark~program for a plane-wave reflection/transmission boundary-value problem in order to search efficiently for desired values of  free parameters such as $\thetainc$, $\psi$, $\betaz$, and $\betay$; and
proofreading successive drafts carefully to eliminate typographical errors. T.G.M. validated A.L.'s calculations through spot checks
using an independently created  Mathematica\texttrademark~program. Both authors accept complete responsibility for the contents of this paper.

\clearpage

\begin{center}
Supplementary Material\\
\end{center}
\section*{S1. Derivation of Eqs.~(65) and  the conditions for co-occurrence in Sec. 4.5}
\renewcommand{\theequation}{S\arabic{equation}}
\setcounter{equation}{0}

The transverse electric fields of the incident, reflected, and transmitted plane waves at the faces of the slab are represented by
$[J]\.[\aas,\aap]^T$, $[J]\.[\Sigma]\.[\bbs,\bbp]^T$, and $[J]\.[\ccs,\ccp]^T$, respectively, with $[\Sigma]={\rm Diag}[1,-1]$. The transverse magnetic fields follow from Eq.~(50b). Imposing the continuity of the tangential components of the electric and magnetic fields at $z=0$ and $z=L$ on Eq.~(60) yields the same algebraic equations as for an isotropic slab, but with the scalar $\sqrt{\eps_{\rm iso}}$ replaced by the matrix $[C]^{1/2}$. 

Since all the matrices involved are functions of $[C]$ and therefore commute with one another, these equations are solved exactly as in the isotropic case, with $\tau(\eps_{\rm iso})$, $\rho(\eps_{\rm iso})$, and $\Upsilon(\eps_{\rm iso},L)$ of Eqs.~(61) becoming the 2$\times$2 matrices $[\tau_C]$, $[\rho_C]$, and $[\Upsilon_C]$ of Eqs.~(62). This yields Eqs.~(63).

In the defective case, $[C]=c_a[I]+[N]$ with $c_a\ne0$ and $[N]^2=[0]$. Hence, the Taylor series of any function analytic at $c_a$ terminates after two terms. Thus,
$[C]^{1/2}=\sqrt{c_a}[I]+[N]/(2\sqrt{c_a})$,
$\cos([C]^{1/2}\ko L)=\cos\chi_a[I]-(\ko L\sin\chi_a/2\sqrt{c_a})[N]$, and
$\sin([C]^{1/2}\ko L)=\sin\chi_a[I]+(\ko L\cos\chi_a/2\sqrt{c_a})[N]$, with $\chi_a=\sqrt{c_a}\ko L$; furthermore, $[\Upsilon_C]=u_0[I]+u_1[N]$ and $[\Upsilon_C]^{-1}=[I]/u_0-(u_1/u_0^2)[N]$, where
\begin{equation}
\left.\begin{array}{l}
u_0=\Upsilon(c_a,L)=2\sqrt{c_a}\cos\chi_a-i(1+c_a)\sin\chi_a
\\[5pt]
u_1=\displaystyle{\left.\frac{\partial\Upsilon(\eps_{\rm iso},L)}{\partial\eps_{\rm iso}}\right\vert_{\eps_{\rm iso}=c_a}}=\displaystyle{\frac{\cos\chi_a}{\sqrt{c_a}}-\ko L\sin\chi_a-i\sin\chi_a-i(1+c_a)\frac{\ko L\cos\chi_a}{2\sqrt{c_a}}}
\end{array}\right\}\,,
\end{equation}
provided that $u_0\ne0$. Since $[J]^{-1}\.[I]\.[J]=[I]$ and $[J]^{-1}\.[N]\.[J]=[\tilde N]$, Eqs.~(62) and (63) yield
\begin{equation}
\label{S-TR-explicit}
\left.\begin{array}{l}
\Tlin=\displaystyle{\frac{2\sqrt{c_a}}{u_0}[I]+\left(\frac{1}{\sqrt{c_a}\,u_0}-\frac{2\sqrt{c_a}\,u_1}{u_0^2}\right)[\tilde N]}
\\[12pt]
\Rlin=\displaystyle{[\Sigma]\.\left[\frac{w_0}{u_0}[I]+\left(\frac{w_1}{u_0}-\frac{w_0u_1}{u_0^2}\right)[\tilde N]\right]}
\end{array}\right\}\,,
\end{equation}
where $w_0=i(c_a-1)\sin\chi_a$ and $w_1=i\sin\chi_a+i(c_a-1)\ko L\cos\chi_a/(2\sqrt{c_a})$ are the factor $i(\eps_{\rm iso}-1)\sin(\sqrt{\eps_{\rm iso}}\ko L)$ in the second of Eqs.~(61) and its derivative with respect to $\eps_{\rm iso}$, both evaluated at $\eps_{\rm iso}=c_a$. Since $\tau(c_a)=2\sqrt{c_a}/u_0$ and $\rho(c_a)=w_0/u_0$, the coefficients of $[I]$ in Eqs.~(\ref{S-TR-explicit}) are $\tau(c_a)$ and $\rho(c_a)$, and the coefficients of $[\tilde N]$ are their derivatives $\alpha_\tau(c_a)$ and $\alpha_\rho(c_a)$ by the quotient rule, leading to Eqs.~(65).

The conditions for co-occurrence follow from Eqs.~(65). First,
since ${[\tilde N]}\ne[0]$ but ${[\tilde N]}^2=[0]$, $\Tlin$ is defective if and only if $\alpha_\tau(c_a)\ne0$, which is Eq.~(66). The similarity transformation in Eq.~(20a) carries this conclusion over to $\Tcirc$. Second,
since ${\rm trace}{[\tilde N]}=\det{[\tilde N]}=0$, $\Rlin$ has trace $2\alpha_\rho(c_a) {\tilde N_{11}}=\alpha_\rho(c_a)({\tilde N_{11}}-{\tilde N_{22}})$ and determinant $-\rho^2(c_a)$, so that it has a double eigenvalue if and only if the second of Eqs.~(67) holds. $\Rlin$ is then not a multiple of $[I]$ if and only if $\alpha_\rho(c_a)\ne0$, because $\alpha_\rho(c_a)=0\implies\rho(c_a)=0$ and $\Rlin=[0]$. Equation~(20a) carries this conclusion over to $\Rcirc$.

\section*{S2. Tables S1--S9 for Sec. 5}

Tables~S1--S8 are the complete versions of Tables~1--8 of the main paper, in that numerical order, and Table~S9 has no counterpart in Sec. 5 of  the main paper. Tables~S1--S9 list the outcome of the search for every type of defectiveness and every combination of types, whether or not it was found.  

The search and the acceptance criteria are described in Sec.~5 of the main paper. In Tables~S1, S2, S4, and S5, the entries in the columns headed \textsf{Found}, \textsf{Near miss}, and \textsf{Remark} have the following meanings:

\begin{itemize}

\item \textsf{yes} in the \textsf{Found} column means that the residuals of the type or combination were driven below the thresholds and all the validity tests were passed.
The    values of $\psi$, $\betaz$, and $\betay$ found are stated in that row.

\item \textsf{Near miss} means that (a) $L\vert \gn1-\gn2\vert \leq 0.1$ for $\calPone$, (b) $L\vert\gn3-\gn4\vert\leq0.1$ for $\calPtwo$, (c) $L\vert \gn1-\gn2\vert \leq 0.1$ 
and $L\vert\gn3-\gn4\vert\leq0.1$ for $\calPthree$ and also $L\vert\gn1+\gn2+\gn3+\gn4\vert\leq0.1$ for $\calPthreeS$, (d)  the relative off-diagonal amplitude in $\Tcirc$ is 
$\leq 0.001$ for $\calT$, or (e) the relative off-diagonal amplitude in $\Rcirc$ is 
$\leq 0.001$ for $\calR${, where, for $\calTthree$ and $\calRthree$, $\sqrt{\vert\Delta_{T,R}\vert}/\vert s_{T,R}\vert$ is used in place of the relative off-diagonal amplitude}.

\item \textsf{rejected} means that the residuals were driven below $10^{-8}$ but a validity test failed.
\item \textsf{closest residual} is the smallest Euclidean norm of the normalized residuals reached by the search.
\item \textsf{not searched} means that the combination was not searched because the stated constituent (an individual occurrence or, for a threefold combination, a pair) was not found for the same $\thetainc$ and $L$.
\end{itemize}

In Tables~S3 and S6--S9, \textsf{impossible} means that the combination cannot occur for the reason given in the main paper, and \textsf{not found} means that the search did not find the combination. 

In all tables,
\begin{itemize}
\item a bare $\calP$, $\calT$, or $\calR$ stands for every one of $\calPone$, $\calPtwo$, $\calPthree$, $\calPthreeS$; $\calTone$, $\calTtwo$, $\calTthree$; or $\calRone$, $\calRtwo$, $\calRthree$, respectively. Rows that agree in every column and differ only in one member of a family have been merged.
\end{itemize}

\renewcommand{\thetable}{S\arabic{table}}
\setcounter{table}{0}

\clearpage
\subsection*{Andalusite}

{\small\setlength{\tabcolsep}{4pt}
\begin{longtable}{@{}llcccl>{\raggedright\arraybackslash}p{4.9cm}@{}}
\caption{Occurrences of $\calPone$, $\calPtwo$, $\calPthree$, $\calPthreeS$, $\calTone$, $\calTtwo$, $\calTthree$,
$\calRone$, $\calRtwo$, $\calRthree$, and their combinations for andalusite with stated
values of $\psi$, $\betaz$, and $\betay$ for normal incidence ($\theta\inc=0$~deg) when $\lambdao=450$~nm and $L=5$~mm.}
\label{S1}\\
\hline
Defectiveness & Found & $\psi$  & $\betaz$   & $\betay$  & Near miss & Remark \\
  &   &   (deg) &   (deg) &   (deg) &   &   \\
\hline
\endfirsthead
\multicolumn{7}{l}{Table~\ref{S1} (continued)}\\
\hline
Defectiveness & Found & $\psi$  & $\betaz$   & $\betay$  & Near miss & Remark \\
  &   &   (deg) &   (deg) &   (deg) &   &   \\
\hline
\endhead
$\calPone$ & no & -- & -- & -- & rejected & even $\tilde\Gamma$ at normal incidence \\
$\calPtwo$ & no & -- & -- & -- & rejected & even $\tilde\Gamma$ at normal incidence \\
{$\calPthree$} & {yes} & {$[0,360)$} & {$89.99203$} & {$132.24843$} & {--} & {Voigt wave excited} \\
$\calPthree$ & yes & $[0,360)$ & $90.00797$ & $132.24843$ & -- & Voigt wave excited \\
{$\calPthreeS$} & {yes} & {$[0,360)$} & {$89.99203$} & {$132.24843$} & {--} & {Voigt wave excited} \\
$\calPthreeS$ & yes & $[0,360)$ & $90.00797$ & $132.24843$ & -- & Voigt wave excited \\
$\calTone$ & yes & $[0,360)$ & $89.99203$ & $132.24843$ & -- & {co-occurs with $\calPthreeS$  } \\
$\calTtwo$ & yes & $[0,360)$ & $90.00797$ & $132.24843$ & -- & {co-occurs with $\calPthreeS$ } \\
$\calTthree$ & no & -- & -- & -- & rejected & $t_{\rm LL}=t_{\rm RR}$ (reciprocity) \\
$\calRone$ & no & -- & -- & -- & no & closest residual $2.0\times 10^{-1}$ \\
$\calRtwo$ & no & -- & -- & -- & no & closest residual $2.0\times 10^{-1}$ \\
$\calRthree$ & no & -- & -- & -- & no & closest residual $8.0\times 10^{-2}$ \\
$\calPone\,\&\,\calTone$ & no & -- & -- & -- & -- & even $\tilde\Gamma$ at normal incidence \\
$\calPone\,\&\,\calTtwo$ & no & -- & -- & -- & -- & even $\tilde\Gamma$ at normal incidence \\
$\calP\,\&\,\calTthree$ & no & -- & -- & -- & -- & $t_{\rm LL}=t_{\rm RR}$ (reciprocity) \\
$\calPtwo\,\&\,\calTone$ & no & -- & -- & -- & -- & even $\tilde\Gamma$ at normal incidence \\
$\calPtwo\,\&\,\calTtwo$ & no & -- & -- & -- & -- & even $\tilde\Gamma$ at normal incidence \\
$\calPthree\,\&\,\calTone$ & yes & $[0,360)$ & $89.99203$ & $132.24843$ & -- & Voigt wave seen in transmission \\
$\calPthree\,\&\,\calTtwo$ & yes & $[0,360)$ & $90.00797$ & $132.24843$ & -- & Voigt wave seen in transmission \\
$\calPthreeS\,\&\,\calTone$ & yes & $[0,360)$ & $89.99203$ & $132.24843$ & -- & Voigt wave seen in transmission \\
$\calPthreeS\,\&\,\calTtwo$ & yes & $[0,360)$ & $90.00797$ & $132.24843$ & -- & Voigt wave seen in transmission \\
$\calPone\,\&\,\calR$ & no & -- & -- & -- & -- & even $\tilde\Gamma$ at normal incidence \\
$\calPtwo\,\&\,\calR$ & no & -- & -- & -- & -- & even $\tilde\Gamma$ at normal incidence \\
$\calPthree\,\&\,\calR$ & no & -- & -- & -- & -- & not searched: no $\calR$ found \\
$\calPthreeS\,\&\,\calR$ & no & -- & -- & -- & -- & not searched: no $\calR$ found \\
$\calTone\,\&\,\calR$ & no & -- & -- & -- & -- & not searched: no $\calR$ found \\
$\calTtwo\,\&\,\calR$ & no & -- & -- & -- & -- & not searched: no $\calR$ found \\
$\calTthree\,\&\,\calR$ & no & -- & -- & -- & -- & $t_{\rm LL}=t_{\rm RR}$ (reciprocity) \\
$\calPone\,\&\,\calTone\,\&\,\calR$ & no & -- & -- & -- & -- & even $\tilde\Gamma$ at normal incidence \\
$\calPone\,\&\,\calTtwo\,\&\,\calR$ & no & -- & -- & -- & -- & even $\tilde\Gamma$ at normal incidence \\
$\calP\,\&\,\calTthree\,\&\,\calR$ & no & -- & -- & -- & -- & $t_{\rm LL}=t_{\rm RR}$ (reciprocity) \\
$\calPtwo\,\&\,\calTone\,\&\,\calR$ & no & -- & -- & -- & -- & even $\tilde\Gamma$ at normal incidence \\
$\calPtwo\,\&\,\calTtwo\,\&\,\calR$ & no & -- & -- & -- & -- & even $\tilde\Gamma$ at normal incidence \\
$\calPthree\,\&\,\calTone\,\&\,\calR$ & no & -- & -- & -- & -- & not searched: no $\calR$ found \\
$\calPthree\,\&\,\calTtwo\,\&\,\calR$ & no & -- & -- & -- & -- & not searched: no $\calR$ found \\
$\calPthreeS\,\&\,\calTone\,\&\,\calR$ & no & -- & -- & -- & -- & not searched: no $\calR$ found \\
$\calPthreeS\,\&\,\calTtwo\,\&\,\calR$ & no & -- & -- & -- & -- & not searched: no $\calR$ found \\
\hline
\end{longtable}}

\newpage
{\small\setlength{\tabcolsep}{4pt}
\begin{longtable}{@{}llcccl>{\raggedright\arraybackslash}p{4.9cm}@{}}
\caption{Same as Table~\ref{S1} but for oblique incidence ($\theta\inc=45$~deg).}
\label{S2}\\
\hline
Defectiveness & Found & $\psi$  & $\betaz$   & $\betay$  & Near miss & Remark \\
  &   &   (deg) &   (deg) &   (deg) &   &   \\
\hline
\endfirsthead
\multicolumn{7}{l}{Table~\ref{S2} (continued)}\\
\hline
Defectiveness & Found & $\psi$  & $\betaz$   & $\betay$  & Near miss & Remark \\
  &   &   (deg) &   (deg) &   (deg) &   &   \\
\hline
\endhead
$\calPone$ & yes & $339.12697$ & $101.99317$ & $108.80073$ & -- & Voigt wave excited \\
$\calPtwo$ & yes & $350.11693$ & $84.26687$ & $72.83438$ & -- & Voigt wave excited \\
$\calPthree$ & no & -- & -- & -- & no & closest residual $1.2\times 10^{-1}$ \\
$\calPthreeS$ & no & -- & -- & -- & no & closest residual $1.2\times 10^{-1}$ \\
$\calTone$ & yes & $339.14016$ & $101.97211$ & $108.80773$ & -- & no Voigt wave excited \\
$\calTtwo$ & yes & $339.09364$ & $101.99770$ & $108.80519$ & -- & no Voigt wave excited \\
$\calTthree$ & no & -- & -- & -- & rejected & $t_{\rm LL}=t_{\rm RR}$ (reciprocity) \\
$\calRone$ & no & -- & -- & -- & no & closest residual $1.4\times 10^{-2}$ \\
$\calRtwo$ & no & -- & -- & -- & yes & closest residual $6.1\times 10^{-4}$ \\
$\calRthree$ & yes & $24.68554$ & $144.85993$ & $151.11252$ & -- & no Voigt wave excited \\
$\calPone\,\&\,\calTone$ & no & -- & -- & -- & no & closest residual $4.9\times 10^{-8}$ \\
$\calPone\,\&\,\calTtwo$ & no & -- & -- & -- & yes & closest residual $1.1\times 10^{-8}$ \\
$\calP\,\&\,\calTthree$ & no & -- & -- & -- & -- & $t_{\rm LL}=t_{\rm RR}$ (reciprocity) \\
$\calPtwo\,\&\,\calTone$ & no & -- & -- & -- & no & closest residual $2.6\times 10^{-4}$ \\
$\calPtwo\,\&\,\calTtwo$ & no & -- & -- & -- & no & closest residual $6.3\times 10^{-5}$ \\
$\calPthree\,\&\,\calTone$ & no & -- & -- & -- & -- & not searched: $\calPthree$ not found \\
$\calPthree\,\&\,\calTtwo$ & no & -- & -- & -- & -- & not searched: $\calPthree$ not found \\
$\calPthreeS\,\&\,\calTone$ & no & -- & -- & -- & -- & not searched: $\calPthreeS$ not found \\
$\calPthreeS\,\&\,\calTtwo$ & no & -- & -- & -- & -- & not searched: $\calPthreeS$ not found \\
$\calPone\,\&\,\calRone$ & no & -- & -- & -- & -- & not searched: $\calRone$ not found \\
$\calPone\,\&\,\calRtwo$ & no & -- & -- & -- & -- & not searched: $\calRtwo$ not found \\
$\calPone\,\&\,\calRthree$ & no & -- & -- & -- & no & closest residual $1.7\times 10^{-2}$ \\
$\calPtwo\,\&\,\calRone$ & no & -- & -- & -- & -- & not searched: $\calRone$ not found \\
$\calPtwo\,\&\,\calRtwo$ & no & -- & -- & -- & -- & not searched: $\calRtwo$ not found \\
$\calPtwo\,\&\,\calRthree$ & no & -- & -- & -- & no & closest residual $1.4\times 10^{-2}$ \\
$\calPthree\,\&\,\calRone$ & no & -- & -- & -- & -- & not searched: $\calPthree$ and $\calRone$ not found \\
$\calPthree\,\&\,\calRtwo$ & no & -- & -- & -- & -- & not searched: $\calPthree$ and $\calRtwo$ not found \\
$\calPthree\,\&\,\calRthree$ & no & -- & -- & -- & -- & not searched: $\calPthree$ not found \\
$\calPthreeS\,\&\,\calRone$ & no & -- & -- & -- & -- & not searched: $\calPthreeS$ and $\calRone$ not found \\
$\calPthreeS\,\&\,\calRtwo$ & no & -- & -- & -- & -- & not searched: $\calPthreeS$ and $\calRtwo$ not found \\
$\calPthreeS\,\&\,\calRthree$ & no & -- & -- & -- & -- & not searched: $\calPthreeS$ not found \\
$\calTone\,\&\,\calRone$ & no & -- & -- & -- & -- & not searched: $\calRone$ not found \\
$\calTone\,\&\,\calRtwo$ & no & -- & -- & -- & -- & not searched: $\calRtwo$ not found \\
$\calTone\,\&\,\calRthree$ & no & -- & -- & -- & no & closest residual $1.2\times 10^{-1}$ \\
$\calTtwo\,\&\,\calRone$ & no & -- & -- & -- & -- & not searched: $\calRone$ not found \\
$\calTtwo\,\&\,\calRtwo$ & no & -- & -- & -- & -- & not searched: $\calRtwo$ not found \\
$\calTtwo\,\&\,\calRthree$ & no & -- & -- & -- & no & closest residual $1.1\times 10^{-1}$ \\
$\calTthree\,\&\,\calR$ & no & -- & -- & -- & -- & $t_{\rm LL}=t_{\rm RR}$ (reciprocity) \\
$\calPone\,\&\,\calTone\,\&\,\calRone$ & no & -- & -- & -- & -- & not searched: $\calRone$ not found \\
$\calPone\,\&\,\calTone\,\&\,\calRtwo$ & no & -- & -- & -- & -- & not searched: $\calRtwo$ not found \\
$\calPone\,\&\,\calTone\,\&\,\calRthree$ & no & -- & -- & -- & -- & not searched: $\calPone\,\&\,\calTone$, $\calPone\,\&\,\calRthree$, and $\calTone\,\&\,\calRthree$ not found \\
$\calPtwo\,\&\,\calTone\,\&\,\calRone$ & no & -- & -- & -- & -- & not searched: $\calRone$ not found \\
$\calPtwo\,\&\,\calTone\,\&\,\calRtwo$ & no & -- & -- & -- & -- & not searched: $\calRtwo$ not found \\
$\calPtwo\,\&\,\calTone\,\&\,\calRthree$ & no & -- & -- & -- & -- & not searched: $\calPtwo\,\&\,\calTone$, $\calPtwo\,\&\,\calRthree$, and $\calTone\,\&\,\calRthree$ not found \\
$\calPthree\,\&\,\calTone\,\&\,\calRone$ & no & -- & -- & -- & -- & not searched: $\calPthree$ and $\calRone$ not found \\
$\calPthree\,\&\,\calTone\,\&\,\calRtwo$ & no & -- & -- & -- & -- & not searched: $\calPthree$ and $\calRtwo$ not found \\
$\calPthree\,\&\,\calTone\,\&\,\calRthree$ & no & -- & -- & -- & -- & not searched: $\calPthree$ not found \\
$\calPthreeS\,\&\,\calTone\,\&\,\calRone$ & no & -- & -- & -- & -- & not searched: $\calPthreeS$ and $\calRone$ not found \\
$\calPthreeS\,\&\,\calTone\,\&\,\calRtwo$ & no & -- & -- & -- & -- & not searched: $\calPthreeS$ and $\calRtwo$ not found \\
$\calPthreeS\,\&\,\calTone\,\&\,\calRthree$ & no & -- & -- & -- & -- & not searched: $\calPthreeS$ not found \\
$\calPone\,\&\,\calTtwo\,\&\,\calRone$ & no & -- & -- & -- & -- & not searched: $\calRone$ not found \\
$\calPone\,\&\,\calTtwo\,\&\,\calRtwo$ & no & -- & -- & -- & -- & not searched: $\calRtwo$ not found \\
$\calPone\,\&\,\calTtwo\,\&\,\calRthree$ & no & -- & -- & -- & -- & not searched: $\calPone\,\&\,\calTtwo$, $\calPone\,\&\,\calRthree$, and $\calTtwo\,\&\,\calRthree$ not found \\
$\calPtwo\,\&\,\calTtwo\,\&\,\calRone$ & no & -- & -- & -- & -- & not searched: $\calRone$ not found \\
$\calPtwo\,\&\,\calTtwo\,\&\,\calRtwo$ & no & -- & -- & -- & -- & not searched: $\calRtwo$ not found \\
$\calPtwo\,\&\,\calTtwo\,\&\,\calRthree$ & no & -- & -- & -- & -- & not searched: $\calPtwo\,\&\,\calTtwo$, $\calPtwo\,\&\,\calRthree$, and $\calTtwo\,\&\,\calRthree$ not found \\
$\calPthree\,\&\,\calTtwo\,\&\,\calRone$ & no & -- & -- & -- & -- & not searched: $\calPthree$ and $\calRone$ not found \\
$\calPthree\,\&\,\calTtwo\,\&\,\calRtwo$ & no & -- & -- & -- & -- & not searched: $\calPthree$ and $\calRtwo$ not found \\
$\calPthree\,\&\,\calTtwo\,\&\,\calRthree$ & no & -- & -- & -- & -- & not searched: $\calPthree$ not found \\
$\calPthreeS\,\&\,\calTtwo\,\&\,\calRone$ & no & -- & -- & -- & -- & not searched: $\calPthreeS$ and $\calRone$ not found \\
$\calPthreeS\,\&\,\calTtwo\,\&\,\calRtwo$ & no & -- & -- & -- & -- & not searched: $\calPthreeS$ and $\calRtwo$ not found \\
$\calPthreeS\,\&\,\calTtwo\,\&\,\calRthree$ & no & -- & -- & -- & -- & not searched: $\calPthreeS$ not found \\
$\calP\,\&\,\calTthree\,\&\,\calR$ & no & -- & -- & -- & -- & $t_{\rm LL}=t_{\rm RR}$ (reciprocity) \\
\hline
\end{longtable}}

\newpage
{\small\setlength{\tabcolsep}{4pt}
\begin{longtable}{lccccl}
\caption{Representative values of $\thetainc$, $\psi$, $\betaz$, and $\betay$ for which a Voigt wave can be seen in transmission ($\calP\,\&\,\calT$) or reflection ($\calP\,\&\,\calR$) for andalusite, when $\lambdao=450$~nm and $L=5$~mm.}
\label{S3}\\
\hline
Combination & $\thetainc$ & $\psi$ & $\betaz$ & $\betay$ & Voigt wave \\
 & (deg) & (deg) & (deg) & (deg) & seen in \\
\hline
\endfirsthead
\multicolumn{6}{l}{Table~\ref{S3} (continued)}\\
\hline
Combination & $\thetainc$ & $\psi$ & $\betaz$ & $\betay$ & Voigt wave \\
 & (deg) & (deg) & (deg) & (deg) & seen in \\
\hline
\endhead
{$\calPthree\,\&\,\calTone$} & {$0$} & {$[0,360)$} & {$89.99203$} & {$132.24843$} & {transmission} \\
{$\calPthree\,\&\,\calTtwo$} & {$0$} & {$[0,360)$} & {$90.00797$} & {$132.24843$} & {transmission} \\
$\calPthreeS\,\&\,\calTone$ & $0$ & $[0,360)$ & $89.99203$ & $132.24843$ & transmission \\
$\calPthreeS\,\&\,\calTtwo$ & $0$ & $[0,360)$ & $90.00797$ & $132.24843$ & transmission \\
{$\calPthree\,\&\,\calR$} & {$0$} & {--} & {--} & {--} & {impossible} \\
$\calPthreeS\,\&\,\calR$ & $0$ & -- & -- & -- & impossible \\
$\calPone\,\&\,\calTone$ & $2.69040$ & $113.54460$ & $87.95953$ & $132.92210$ & transmission \\
$\calPone\,\&\,\calTone$ & $5.40308$ & $79.93596$ & $85.60914$ & $131.75601$ & transmission \\
$\calPone\,\&\,\calTone$ & $44.59719$ & $169.16521$ & $83.74697$ & $72.54991$ & transmission \\
$\calPtwo\,\&\,\calTone$ & $55.66771$ & $176.86352$ & $92.12620$ & $102.05168$ & transmission \\
$\calPtwo\,\&\,\calTone$ & $64.93410$ & $22.97800$ & $106.93727$ & $77.88882$ & transmission \\
$\calP\,\&\,\calTtwo$ & -- & -- & -- & -- & not found (near misses only) \\
$\calPone\,\&\,\calRthree$ & $76.61650$ & $336.86124$ & $108.35846$ & $99.59717$ & reflection \\
$\calPone\,\&\,\calRthree$ & $84.64196$ & $104.71313$ & $37.47307$ & $157.18123$ & reflection \\
$\calPtwo\,\&\,\calRthree$ & $74.72141$ & $124.33726$ & $131.04467$ & $117.96669$ & reflection \\
$\calPtwo\,\&\,\calRthree$ & $78.24458$ & $343.59750$ & $76.83644$ & $82.53609$ & reflection \\
\hline
\end{longtable}}

\clearpage
\subsection*{Gallium oxide ($\beta$-Ga$_2$O$_3$)}

{\small\setlength{\tabcolsep}{4pt}
\begin{longtable}{@{}llcccl>{\raggedright\arraybackslash}p{4.9cm}@{}}
\caption{Occurrences of $\calPone$, $\calPtwo$, $\calPthree$, $\calPthreeS$, $\calTone$, $\calTtwo$, $\calTthree$,
$\calRone$, $\calRtwo$, $\calRthree$, and their combinations for gallium oxide ($\beta$-Ga$_2$O$_3$)  with stated
values of $\psi$, $\betaz$, and $\betay$ for normal incidence ($\theta\inc=0$~deg) when $\lambdao=200$~nm and $L=110$~nm.
}
\label{S4}\\
\hline
Defectiveness & Found & $\psi$  & $\betaz$   & $\betay$  & Near miss & Remark \\
  &   &   (deg) &   (deg) &   (deg) &   &   \\
\hline
\endfirsthead
\multicolumn{7}{l}{Table~\ref{S4} (continued)}\\
\hline
Defectiveness & Found & $\psi$  & $\betaz$   & $\betay$  & Near miss & Remark \\
  &   &   (deg) &   (deg) &   (deg) &   &   \\
\hline
\endhead
$\calPone$ & no & -- & -- & -- & rejected & even $\tilde\Gamma$ at normal incidence \\
$\calPtwo$ & no & -- & -- & -- & rejected & even $\tilde\Gamma$ at normal incidence \\
{$\calPthree$} & {yes} & {$[0,360)$} & {$54.80706$} & {$119.83239$} & {--} & {Voigt wave excited} \\
$\calPthree$ & yes & $[0,360)$ & $55.64033$ & $79.45063$ & -- & Voigt wave excited \\
{$\calPthreeS$} & {yes} & {$[0,360)$} & {$54.80706$} & {$119.83239$} & {--} & {Voigt wave excited} \\
$\calPthreeS$ & yes & $[0,360)$ & $55.64033$ & $79.45063$ & -- & Voigt wave excited \\
$\calTone$ & yes & $[0,360)$ & $54.80706$ & $119.83239$ & -- & {co-occurs with $\calPthreeS$  } \\
$\calTtwo$ & yes & $[0,360)$ & $55.64033$ & $79.45063$ & -- & {co-occurs with $\calPthreeS$  } \\
$\calTthree$ & no & -- & -- & -- & rejected & $\tLL=\tRR$ (reciprocity) \\
$\calRone$ & no & -- & -- & -- & no & closest residual $4.7\times 10^{-1}$ \\
$\calRtwo$ & no & -- & -- & -- & no & closest residual $4.7\times 10^{-1}$ \\
$\calRthree$ & no & -- & -- & -- & no & closest residual $8.9\times 10^{-1}$ \\
$\calPone\,\&\,\calTone$ & no & -- & -- & -- & -- & even $\tilde\Gamma$ at normal incidence \\
$\calPone\,\&\,\calTtwo$ & no & -- & -- & -- & -- & even $\tilde\Gamma$ at normal incidence \\
$\calP\,\&\,\calTthree$ & no & -- & -- & -- & -- & $\tLL=\tRR$ (reciprocity) \\
$\calPtwo\,\&\,\calTone$ & no & -- & -- & -- & -- & even $\tilde\Gamma$ at normal incidence \\
$\calPtwo\,\&\,\calTtwo$ & no & -- & -- & -- & -- & even $\tilde\Gamma$ at normal incidence \\
$\calPthree\,\&\,\calTone$ & yes & $[0,360)$ & $54.80706$ & $119.83239$ & -- & Voigt wave seen in transmission \\
$\calPthree\,\&\,\calTtwo$ & yes & $[0,360)$ & $55.64033$ & $79.45063$ & -- & Voigt wave seen in transmission \\
$\calPthreeS\,\&\,\calTone$ & yes & $[0,360)$ & $54.80706$ & $119.83239$ & -- & Voigt wave seen in transmission \\
$\calPthreeS\,\&\,\calTtwo$ & yes & $[0,360)$ & $55.64033$ & $79.45063$ & -- & Voigt wave seen in transmission \\
$\calPone\,\&\,\calR$ & no & -- & -- & -- & -- & even $\tilde\Gamma$ at normal incidence \\
$\calPtwo\,\&\,\calR$ & no & -- & -- & -- & -- & even $\tilde\Gamma$ at normal incidence \\
$\calPthree\,\&\,\calR$ & no & -- & -- & -- & -- & not searched: no $\calR$ found \\
$\calPthreeS\,\&\,\calR$ & no & -- & -- & -- & -- & not searched: no $\calR$ found \\
$\calTone\,\&\,\calR$ & no & -- & -- & -- & -- & not searched: no $\calR$ found \\
$\calTtwo\,\&\,\calR$ & no & -- & -- & -- & -- & not searched: no $\calR$ found \\
$\calTthree\,\&\,\calR$ & no & -- & -- & -- & -- & $\tLL=\tRR$ (reciprocity) \\
$\calPone\,\&\,\calTone\,\&\,\calR$ & no & -- & -- & -- & -- & even $\tilde\Gamma$ at normal incidence \\
$\calPone\,\&\,\calTtwo\,\&\,\calR$ & no & -- & -- & -- & -- & even $\tilde\Gamma$ at normal incidence \\
$\calP\,\&\,\calTthree\,\&\,\calR$ & no & -- & -- & -- & -- & $\tLL=\tRR$ (reciprocity) \\
$\calPtwo\,\&\,\calTone\,\&\,\calR$ & no & -- & -- & -- & -- & even $\tilde\Gamma$ at normal incidence \\
$\calPtwo\,\&\,\calTtwo\,\&\,\calR$ & no & -- & -- & -- & -- & even $\tilde\Gamma$ at normal incidence \\
$\calPthree\,\&\,\calTone\,\&\,\calR$ & no & -- & -- & -- & -- & not searched: no $\calR$ found \\
$\calPthree\,\&\,\calTtwo\,\&\,\calR$ & no & -- & -- & -- & -- & not searched: no $\calR$ found \\
$\calPthreeS\,\&\,\calTone\,\&\,\calR$ & no & -- & -- & -- & -- & not searched: no $\calR$ found \\
$\calPthreeS\,\&\,\calTtwo\,\&\,\calR$ & no & -- & -- & -- & -- & not searched: no $\calR$ found \\
\hline
\end{longtable}}

\newpage
{\small\setlength{\tabcolsep}{4pt}
\begin{longtable}{@{}llcccl>{\raggedright\arraybackslash}p{4.9cm}@{}}
\caption{Same as Table~\ref{S4} but for oblique incidence ($\theta\inc=45$~deg).}
\label{S5}\\
\hline
Defectiveness & Found & $\psi$  & $\betaz$   & $\betay$  & Near miss & Remark \\
  &   &   (deg) &   (deg) &   (deg) &   &   \\
\hline
\endfirsthead
\multicolumn{7}{l}{Table~\ref{S5} (continued)}\\
\hline
Defectiveness & Found & $\psi$  & $\betaz$   & $\betay$  & Near miss & Remark \\
  &   &   (deg) &   (deg) &   (deg) &   &   \\
\hline
\endhead
$\calPone$ & yes & $134.25433$ & $112.94510$ & $74.79045$ & -- & Voigt wave excited \\
$\calPtwo$ & yes & $314.25609$ & $112.94564$ & $74.79082$ & -- & Voigt wave excited \\
$\calPthree$ & no & -- & -- & -- & no & closest residual $3.9\times 10^{-2}$ \\
$\calPthreeS$ & no & -- & -- & -- & no & closest residual $1.9\times 10^{-1}$ \\
$\calTone$ & yes & $45.08679$ & $90.93329$ & $36.80185$ & -- & no Voigt wave excited \\
$\calTtwo$ & yes & $136.83906$ & $65.11573$ & $92.15653$ & -- & no Voigt wave excited \\
$\calTthree$ & no & -- & -- & -- & rejected & $\tLL=\tRR$ (reciprocity) \\
$\calRone$ & no & -- & -- & -- & no & closest residual $3.6\times 10^{-1}$ \\
$\calRtwo$ & no & -- & -- & -- & no & closest residual $3.6\times 10^{-1}$ \\
$\calRthree$ & no & -- & -- & -- & no & closest residual $5.2\times 10^{-1}$ \\
$\calPone\,\&\,\calTone$ & no & -- & -- & -- & no & closest residual $4.7\times 10^{-2}$ \\
$\calPone\,\&\,\calTtwo$ & no & -- & -- & -- & no & closest residual $3.4\times 10^{-2}$ \\
$\calP\,\&\,\calTthree$ & no & -- & -- & -- & -- & $\tLL=\tRR$ (reciprocity) \\
$\calPtwo\,\&\,\calTone$ & no & -- & -- & -- & no & closest residual $7.7\times 10^{-3}$ \\
$\calPtwo\,\&\,\calTtwo$ & no & -- & -- & -- & no & closest residual $7.7\times 10^{-3}$ \\
$\calPthree\,\&\,\calTone$ & no & -- & -- & -- & -- & not searched: $\calPthree$ not found \\
$\calPthree\,\&\,\calTtwo$ & no & -- & -- & -- & -- & not searched: $\calPthree$ not found \\
$\calPthreeS\,\&\,\calTone$ & no & -- & -- & -- & -- & not searched: $\calPthreeS$ not found \\
$\calPthreeS\,\&\,\calTtwo$ & no & -- & -- & -- & -- & not searched: $\calPthreeS$ not found \\
$\calPone\,\&\,\calR$ & no & -- & -- & -- & -- & not searched: no $\calR$ found \\
$\calPtwo\,\&\,\calR$ & no & -- & -- & -- & -- & not searched: no $\calR$ found \\
$\calPthree\,\&\,\calR$ & no & -- & -- & -- & -- & not searched: $\calPthree$ and $\calR$ not found \\
$\calPthreeS\,\&\,\calR$ & no & -- & -- & -- & -- & not searched: $\calPthreeS$ and $\calR$ not found \\
$\calTone\,\&\,\calR$ & no & -- & -- & -- & -- & not searched: no $\calR$ found \\
$\calTtwo\,\&\,\calR$ & no & -- & -- & -- & -- & not searched: no $\calR$ found \\
$\calTthree\,\&\,\calR$ & no & -- & -- & -- & -- & $\tLL=\tRR$ (reciprocity) \\
$\calPone\,\&\,\calTone\,\&\,\calR$ & no & -- & -- & -- & -- & not searched: no $\calR$ found \\
$\calPtwo\,\&\,\calTone\,\&\,\calR$ & no & -- & -- & -- & -- & not searched: no $\calR$ found \\
$\calPthree\,\&\,\calTone\,\&\,\calR$ & no & -- & -- & -- & -- & not searched: $\calPthree$ and $\calR$ not found \\
$\calPthreeS\,\&\,\calTone\,\&\,\calR$ & no & -- & -- & -- & -- & not searched: $\calPthreeS$ and $\calR$ not found \\
$\calPone\,\&\,\calTtwo\,\&\,\calR$ & no & -- & -- & -- & -- & not searched: no $\calR$ found \\
$\calPtwo\,\&\,\calTtwo\,\&\,\calR$ & no & -- & -- & -- & -- & not searched: no $\calR$ found \\
$\calPthree\,\&\,\calTtwo\,\&\,\calR$ & no & -- & -- & -- & -- & not searched: $\calPthree$ and $\calR$ not found \\
$\calPthreeS\,\&\,\calTtwo\,\&\,\calR$ & no & -- & -- & -- & -- & not searched: $\calPthreeS$ and $\calR$ not found \\
$\calP\,\&\,\calTthree\,\&\,\calR$ & no & -- & -- & -- & -- & $\tLL=\tRR$ (reciprocity) \\
\hline
\end{longtable}}

\newpage
{\small\setlength{\tabcolsep}{4pt}
\begin{longtable}{lccccl}
\caption{Same as Table~\ref{S3} but for $\beta$-Ga$_2$O$_3$, when $\lambdao=200$~nm and $L=110$~nm.}
\label{S6}\\
\hline
Combination & $\thetainc$ & $\psi$ & $\betaz$ & $\betay$ & Voigt wave \\
 & (deg) & (deg) & (deg) & (deg) & seen in \\
\hline
\endfirsthead
\multicolumn{6}{l}{Table~\ref{S6} (continued)}\\
\hline
Combination & $\thetainc$ & $\psi$ & $\betaz$ & $\betay$ & Voigt wave \\
 & (deg) & (deg) & (deg) & (deg) & seen in \\
\hline
\endhead
{$\calPthree\,\&\,\calTone$} & {$0$} & {$[0,360)$} & {$54.80706$} & {$119.83239$} & {transmission} \\
{$\calPthree\,\&\,\calTtwo$} & {$0$} & {$[0,360)$} & {$55.64033$} & {$79.45063$} & {transmission} \\
{$\calPthreeS\,\&\,\calTone$} & {$0$} & {$[0,360)$} & {$54.80706$} & {$119.83239$} & {transmission} \\
{$\calPthreeS\,\&\,\calTtwo$} & {$0$} & {$[0,360)$} & {$55.64033$} & {$79.45063$} & {transmission} \\
{$\calPthree\,\&\,\calR$} & {$0$} & {--} & {--} & {--} & {impossible} \\
$\calPthreeS\,\&\,\calR$ & $0$ & -- & -- & -- & impossible \\
$\calPone\,\&\,\calTone$ & $3.88077$ & $87.21476$ & $52.81390$ & $120.06205$ & transmission \\
$\calPone\,\&\,\calTone$ & $55.02777$ & $68.85288$ & $284.03730$ & $109.93957$ & transmission \\
$\calPone\,\&\,\calTtwo$ & $55.02777$ & $111.14712$ & $104.03730$ & $70.06043$ & transmission \\
$\calPone\,\&\,\calTtwo$ & $58.25981$ & $93.64501$ & $282.10842$ & $84.29836$ & transmission \\
$\calPtwo\,\&\,\calTone$ & $47.52530$ & $231.09885$ & $290.77002$ & $105.80787$ & transmission \\
$\calPtwo\,\&\,\calTone$ & $65.87349$ & $109.49068$ & $148.80419$ & $97.63551$ & transmission \\
$\calPtwo\,\&\,\calTtwo$ & $63.75773$ & $342.33450$ & $113.16642$ & $122.13639$ & transmission \\
$\calPtwo\,\&\,\calTtwo$ & $69.31938$ & $152.87416$ & $252.13568$ & $38.90741$ & transmission \\
$\calP\,\&\,\calR$ & -- & -- & -- & -- & not found \\
\hline
\end{longtable}}

\newpage
\subsection*{Terbium orthoaluminate (TbAlO$_3$) in a magnetostatic field}

Table~S7 is the complete version of Table~7 of the main paper. Every entry listed as found passed the tests (a)--(d) of Sec.~5 of the main paper. The magnetostatic field $\Bdc$ is directed along $\=R_y\.\=R_z\.\uz$, so that $\betay$ is reported in $[0,180]$~deg by using the equivalence of the orientations $\lec\psi,\betaz,\betay\ric$ and $\lec\psi+180~{\rm deg},\betaz,-\betay\ric$. Near misses for $\calPtwo\,\&\,\calTone$ and $\calPtwo\,\&\,\calTtwo$ have relative off-diagonal amplitudes below $10^{-7}$, but $\vert\tLL-\tRR\vert/\vert\tLL+\tRR\vert\approx2\times10^{-5}$ to $7\times10^{-5}$ there because of the non-reciprocity, so that $\Tcirc$ is not defective. Likewise, near misses for $\calPone\,\&\,\calRone$ and $\calPone\,\&\,\calRtwo$ have relative off-diagonal amplitudes of about $2\times10^{-5}$, but $\vert\rLL-\rRR\vert/\vert\rLL+\rRR\vert\approx3\times10^{-4}$. For $\calPthree\,\&\,\calTthree$ and $\calPthree\,\&\,\calRthree$ at oblique incidence, $L\in[0.1,3]$~mm and $\Bdc\in[-1,1]$~T were also treated as unknowns, and every search converged to $\thetainc\to0$.

{\small\setlength{\tabcolsep}{4pt}
\begin{longtable}{lccccl}
\caption{Same as Table~\ref{S3} but for TbAlO$_3$  with $\Bdc=0.2$~T, when $\lambdao=485$~nm and $L=1$~mm.}
\label{S7}\\
\hline
Combination & $\thetainc$ & $\psi$ & $\betaz$ & $\betay$ & Voigt wave \\
 & (deg) & (deg) & (deg) & (deg) & seen in \\
\hline
\endfirsthead
\multicolumn{6}{l}{Table~\ref{S7} (continued)}\\
\hline
Combination & $\thetainc$ & $\psi$ & $\betaz$ & $\betay$ & Voigt wave \\
 & (deg) & (deg) & (deg) & (deg) & seen in \\
\hline
\endhead
$\calPthree\,\&\,\calTthree$ & $0$ & $[0,360)$ & $0.02630$ & $45.21871$ & transmission \\
$\calPthree\,\&\,\calTthree$ & $0$ & $[0,360)$ & $0.02630$ & $134.78129$ & transmission \\
$\calPthree\,\&\,\calTthree$ & $0$ & $[0,360)$ & $179.97370$ & $45.21871$ & transmission \\
$\calPthree\,\&\,\calTthree$ & $0$ & $[0,360)$ & $179.97370$ & $134.78129$ & transmission \\
$\calPthreeS\,\&\,\calTthree$ & $0$ & $[0,360)$ & $0.02630$ & $45.21871$ & transmission \\
$\calPthreeS\,\&\,\calTthree$ & $0$ & $[0,360)$ & $0.02630$ & $134.78129$ & transmission \\
$\calPthreeS\,\&\,\calTthree$ & $0$ & $[0,360)$ & $179.97370$ & $45.21871$ & transmission \\
$\calPthreeS\,\&\,\calTthree$ & $0$ & $[0,360)$ & $179.97370$ & $134.78129$ & transmission \\
$\calPthree\,\&\,\calTone$, $\calPthree\,\&\,\calTtwo$ & $0$ & -- & -- & -- & impossible (Voigt wave not circular) \\
$\calPthreeS\,\&\,\calTone$, $\calPthreeS\,\&\,\calTtwo$ & $0$ & -- & -- & -- & impossible (Voigt wave not circular) \\
$\calPthree\,\&\,\calR$ & $0$ & -- & -- & -- & impossible \\
$\calPthreeS\,\&\,\calR$ & $0$ & -- & -- & -- & impossible \\
$\calPone\,\&\,\calTone$ & -- & -- & -- & -- & not found \\
$\calPone\,\&\,\calTtwo$ & -- & -- & -- & -- & not found \\
$\calPone\,\&\,\calTthree$ & $20.83493$ & $153.35062$ & $172.72458$ & $55.38430$ & transmission \\
$\calPone\,\&\,\calTthree$ & $32.96580$ & $204.11720$ & $10.22891$ & $151.68260$ & transmission \\
$\calPone\,\&\,\calTthree$ & $41.78026$ & $106.08769$ & $149.36107$ & $47.34809$ & transmission \\
$\calPone\,\&\,\calTthree$ & $41.99498$ & $173.29872$ & $176.46031$ & $67.25724$ & transmission \\
$\calPone\,\&\,\calTthree$ & $81.76929$ & $248.33166$ & $47.09199$ & $159.56268$ & transmission \\
$\calPone\,\&\,\calTthree$ & $82.83557$ & $9.62627$ & $172.37777$ & $101.28038$ & transmission \\
$\calPone\,\&\,\calTthree$ & $87.81297$ & $93.68122$ & $127.36973$ & $151.05253$ & transmission \\
$\calPone\,\&\,\calTthree$ & $88.97145$ & $277.83047$ & $52.07758$ & $26.41978$ & transmission \\
$\calPtwo\,\&\,\calTone$ & -- & -- & -- & -- & not found (near misses only) \\
$\calPtwo\,\&\,\calTtwo$ & -- & -- & -- & -- & not found (near misses only) \\
$\calPtwo\,\&\,\calTthree$ & $31.53697$ & $140.00930$ & $15.55956$ & $30.80769$ & transmission \\
$\calPtwo\,\&\,\calTthree$ & $31.80424$ & $277.43434$ & $155.37022$ & $44.80847$ & transmission \\
$\calPtwo\,\&\,\calTthree$ & $37.74422$ & $201.20062$ & $169.86855$ & $116.25100$ & transmission \\
$\calPtwo\,\&\,\calTthree$ & $40.44342$ & $131.17357$ & $22.91897$ & $28.31785$ & transmission \\
$\calPtwo\,\&\,\calTthree$ & $41.01942$ & $325.97385$ & $163.01367$ & $62.26452$ & transmission \\
$\calPtwo\,\&\,\calTthree$ & $44.27969$ & $59.05027$ & $28.46441$ & $53.99074$ & transmission \\
$\calPtwo\,\&\,\calTthree$ & $45.73088$ & $110.55718$ & $32.26003$ & $130.73427$ & transmission \\
$\calPtwo\,\&\,\calTthree$ & $45.98793$ & $106.65313$ & $33.26779$ & $32.83625$ & transmission \\
$\calPtwo\,\&\,\calTthree$ & $52.47694$ & $297.10039$ & $145.79853$ & $153.07667$ & transmission \\
$\calPtwo\,\&\,\calTthree$ & $53.07592$ & $284.44616$ & $141.94678$ & $45.62705$ & transmission \\
$\calPtwo\,\&\,\calTthree$ & $53.83055$ & $329.14615$ & $160.71480$ & $67.34152$ & transmission \\
$\calPtwo\,\&\,\calTthree$ & $74.26686$ & $245.96067$ & $135.57353$ & $129.46717$ & transmission \\
$\calPtwo\,\&\,\calTthree$ & $76.21707$ & $98.67111$ & $49.92397$ & $27.32874$ & transmission \\
$\calPtwo\,\&\,\calTthree$ & $77.39116$ & $229.21793$ & $143.90888$ & $117.46019$ & transmission \\
$\calPtwo\,\&\,\calTthree$ & $80.83409$ & $29.56998$ & $22.81170$ & $73.21982$ & transmission \\
$\calPtwo\,\&\,\calTthree$ & $81.64448$ & $326.66320$ & $154.35676$ & $167.17434$ & transmission \\
$\calPtwo\,\&\,\calTthree$ & $81.84180$ & $279.29298$ & $128.92395$ & $153.90186$ & transmission \\
$\calPone\,\&\,\calRone$ & -- & -- & -- & -- & not found (near misses only) \\
$\calPone\,\&\,\calRtwo$ & -- & -- & -- & -- & not found (near misses only) \\
$\calPone\,\&\,\calRthree$ & $63.19634$ & $330.75283$ & $20.35563$ & $16.34958$ & reflection \\
$\calPone\,\&\,\calRthree$ & $81.77287$ & $249.16033$ & $47.44176$ & $159.24732$ & reflection \\
$\calPone\,\&\,\calRthree$ & $82.85379$ & $9.68658$ & $172.33005$ & $101.28677$ & reflection \\
$\calPone\,\&\,\calRthree$ & $87.79575$ & $93.04251$ & $127.32147$ & $150.64464$ & reflection \\
$\calPtwo\,\&\,\calRone$ & -- & -- & -- & -- & not found \\
$\calPtwo\,\&\,\calRtwo$ & -- & -- & -- & -- & not found \\
$\calPtwo\,\&\,\calRthree$ & $64.33726$ & $176.45250$ & $2.52060$ & $14.59444$ & reflection \\
$\calPtwo\,\&\,\calRthree$ & $67.41680$ & $90.67209$ & $47.37934$ & $33.90861$ & reflection \\
$\calPtwo\,\&\,\calRthree$ & $67.68767$ & $194.06658$ & $169.70668$ & $13.98036$ & reflection \\
$\calPtwo\,\&\,\calRthree$ & $72.48937$ & $357.30907$ & $177.92929$ & $77.80643$ & reflection \\
$\calPtwo\,\&\,\calRthree$ & $74.33956$ & $241.74487$ & $137.51756$ & $126.31125$ & reflection \\
$\calPtwo\,\&\,\calRthree$ & $76.21066$ & $99.16789$ & $49.82968$ & $27.06083$ & reflection \\
$\calPtwo\,\&\,\calRthree$ & $80.89328$ & $30.01118$ & $23.14293$ & $73.04573$ & reflection \\
$\calPtwo\,\&\,\calRthree$ & $81.58475$ & $339.04862$ & $163.65798$ & $168.18734$ & reflection \\
$\calPtwo\,\&\,\calRthree$ & $81.85179$ & $279.19248$ & $128.90189$ & $153.84821$ & reflection \\
$\calPtwo\,\&\,\calRthree$ & $84.68558$ & $85.02128$ & $52.23637$ & $35.07752$ & reflection \\
$\calPthree\,\&\,\calTthree$, $\calPthree\,\&\,\calRthree$ & $>0$ & -- & -- & -- & not found, even with $L$ and $B$ free \\
$\calPthreeS\,\&\,\calT$, $\calPthreeS\,\&\,\calR$ & $>0$ & -- & -- & -- & not searched ($\calPthree$ not found) \\
\hline
\end{longtable}}

\newpage
\subsection*{Hexagonal boron nitride (hBN)}

Table~S8 is the complete version of Table~8   of the main paper. Since hBN is uniaxial, $\betaz$ is arbitrary, and the thickness $L\in[0.1,10]$~$\mu$m was taken as an additional unknown. The searches were carried out on the curves of Eqs.~(77) of the main paper, on which $\calPone$ or $\calPtwo$ holds, with $\thetainc$ and $L$ as the unknowns; each solution was accepted only if the normalized residuals of $\calT$ or $\calR$ were below $10^{-10}$, the double root  was double to within $10^{-15}$, some cofactor of $\matW$ exceeded $10^{-12}$, and the other two roots were distinct. The optic axis is reported with $\hat{\#c}\.\uz\geq0$, so that $\betay\in[0,90]$~deg. At $\lambdao=6.6667$~$\mu$m, the closest residuals for $\calP\,\&\,\calT$ were approached at the edges $\thetainc\to0.5$~deg and $L\to0.1$~$\mu$m of the search domain, where $\Tcirc$ tends to $[I]$. Note that the closest residuals are not near misses.

{\small\setlength{\tabcolsep}{3.5pt}
\begin{longtable}{@{}lccccc>{\raggedright\arraybackslash}p{3.6cm}@{}}
\caption{Same as Table~\ref{S3} but for hBN, when $\lambdao=12.5$~$\mu$m. The thickness $L$ was an additional unknown in $[0.1,10]$~$\mu$m, and $\betaz$ is arbitrary.}
\label{S8}\\
\hline
Combination & $\thetainc$ & $\psi$ & $\betaz$ & $\betay$ & $L$ & Voigt wave \\
 & (deg) & (deg) & (deg) & (deg) & ($\mu$m) & seen in \\
\hline
\endfirsthead
\multicolumn{7}{l}{Table~\ref{S8} (continued)}\\
\hline
Combination & $\thetainc$ & $\psi$ & $\betaz$ & $\betay$ & $L$ & Voigt wave \\
 & (deg) & (deg) & (deg) & (deg) & ($\mu$m) & seen in \\
\hline
\endhead
$\calPthree\,\&\,\calT$, $\calPthreeS\,\&\,\calT$ & $0$ & -- & -- & -- & -- & impossible \\
$\calPthree\,\&\,\calR$, $\calPthreeS\,\&\,\calR$ & $0$ & -- & -- & -- & -- & impossible \\
$\calPone\,\&\,\calTone$ & $62.57588$ & $180.05552$ & $[0,360)$ & $18.19640$ & $9.2581$ & transmission \\
$\calPone\,\&\,\calTone$ & $62.57904$ & $179.94448$ & $[0,360)$ & $18.19694$ & $9.2584$ & transmission \\
$\calPone\,\&\,\calTone$ & $63.59323$ & $180.05557$ & $[0,360)$ & $18.36703$ & $6.9506$ & transmission \\
$\calPone\,\&\,\calTone$ & $63.59434$ & $179.94443$ & $[0,360)$ & $18.36721$ & $6.9508$ & transmission \\
$\calPone\,\&\,\calTone$ & $64.80376$ & $179.94437$ & $[0,360)$ & $18.56245$ & $4.6394$ & transmission \\
$\calPone\,\&\,\calTone$ & $64.80864$ & $180.05563$ & $[0,360)$ & $18.56322$ & $4.6393$ & transmission \\
$\calPone\,\&\,\calTone$ & $66.13770$ & $179.94430$ & $[0,360)$ & $18.76809$ & $2.3230$ & transmission \\
$\calPone\,\&\,\calTone$ & $66.16133$ & $180.05570$ & $[0,360)$ & $18.77165$ & $2.3229$ & transmission \\
$\calPone\,\&\,\calTone$ & $74.49900$ & $179.94394$ & $[0,360)$ & $19.81683$ & $2.3373$ & transmission \\
$\calPone\,\&\,\calTone$ & $76.60201$ & $180.05613$ & $[0,360)$ & $20.01318$ & $4.6804$ & transmission \\
$\calPone\,\&\,\calTone$ & $77.89615$ & $180.05617$ & $[0,360)$ & $20.12016$ & $7.0253$ & transmission \\
$\calPone\,\&\,\calTtwo$ & $62.57588$ & $179.94448$ & $[0,360)$ & $18.19640$ & $9.2581$ & transmission \\
$\calPone\,\&\,\calTtwo$ & $62.57904$ & $180.05552$ & $[0,360)$ & $18.19694$ & $9.2584$ & transmission \\
$\calPone\,\&\,\calTtwo$ & $63.59323$ & $179.94443$ & $[0,360)$ & $18.36703$ & $6.9506$ & transmission \\
$\calPone\,\&\,\calTtwo$ & $63.59434$ & $180.05557$ & $[0,360)$ & $18.36721$ & $6.9508$ & transmission \\
$\calPone\,\&\,\calTtwo$ & $64.80376$ & $180.05563$ & $[0,360)$ & $18.56245$ & $4.6394$ & transmission \\
$\calPone\,\&\,\calTtwo$ & $64.80864$ & $179.94437$ & $[0,360)$ & $18.56322$ & $4.6393$ & transmission \\
$\calPone\,\&\,\calTtwo$ & $66.13770$ & $180.05570$ & $[0,360)$ & $18.76809$ & $2.3230$ & transmission \\
$\calPone\,\&\,\calTtwo$ & $66.16133$ & $179.94430$ & $[0,360)$ & $18.77165$ & $2.3229$ & transmission \\
$\calPone\,\&\,\calTtwo$ & $74.50161$ & $179.94394$ & $[0,360)$ & $19.81710$ & $2.3375$ & transmission \\
$\calPone\,\&\,\calTtwo$ & $76.60443$ & $180.05613$ & $[0,360)$ & $20.01339$ & $4.6802$ & transmission \\
$\calPone\,\&\,\calTtwo$ & $77.89896$ & $180.05617$ & $[0,360)$ & $20.12038$ & $7.0251$ & transmission \\
$\calPone\,\&\,\calTthree$ & -- & -- & -- & -- & -- & impossible ($\tLL=\tRR$) \\
$\calPtwo\,\&\,\calTone$ & $69.13305$ & $359.94416$ & $[0,360)$ & $19.19207$ & $1.7134$ & transmission \\
$\calPtwo\,\&\,\calTone$ & $69.14097$ & $0.05584$ & $[0,360)$ & $19.19312$ & $1.7140$ & transmission \\
$\calPtwo\,\&\,\calTtwo$ & $69.13305$ & $0.05584$ & $[0,360)$ & $19.19207$ & $1.7134$ & transmission \\
$\calPtwo\,\&\,\calTtwo$ & $69.14097$ & $359.94416$ & $[0,360)$ & $19.19312$ & $1.7140$ & transmission \\
$\calPtwo\,\&\,\calTthree$ & -- & -- & -- & -- & -- & impossible ($\tLL=\tRR$) \\
$\calPone\,\&\,\calRone$ & -- & -- & -- & -- & -- & not found; closest residual $6\times10^{-3}$ \\
$\calPone\,\&\,\calRtwo$ & -- & -- & -- & -- & -- & not found; closest residual $6\times10^{-3}$ \\
$\calPone\,\&\,\calRthree$ & $71.10542$ & $179.94407$ & $[0,360)$ & $19.44211$ & $2.3323$ & reflection \\
$\calPone\,\&\,\calRthree$ & $71.65386$ & $180.05595$ & $[0,360)$ & $19.50745$ & $4.6655$ & reflection \\
$\calPtwo\,\&\,\calRone$ & -- & -- & -- & -- & -- & not found; closest residual $6\times10^{-3}$ \\
$\calPtwo\,\&\,\calRtwo$ & -- & -- & -- & -- & -- & not found; closest residual $6\times10^{-3}$ \\
$\calPtwo\,\&\,\calRthree$ & $71.10542$ & $359.94407$ & $[0,360)$ & $19.44211$ & $2.3323$ & reflection \\
$\calPtwo\,\&\,\calRthree$ & $71.65386$ & $0.05595$ & $[0,360)$ & $19.50745$ & $4.6655$ & reflection \\
$\calPthree\,\&\,\calT$, $\calPthreeS\,\&\,\calT$ & $>0$ & -- & -- & -- & -- & impossible \\
$\calPthree\,\&\,\calR$, $\calPthreeS\,\&\,\calR$ & $>0$ & -- & -- & -- & -- & impossible \\
\hline
\end{longtable}}

\newpage
{\small\setlength{\tabcolsep}{3.5pt}
\begin{longtable}{@{}lccccc>{\raggedright\arraybackslash}p{3.6cm}@{}}
\caption{Record of attempts to find co-occurrences of $\calP$ with either $\calT$ or $\calR$ for hBN, when $\lambdao=6.6667$~$\mu$m. The thickness $L$ was an additional unknown in $[0.1,10]$~$\mu$m, and $\betaz$ is arbitrary.}
\label{S9}\\
\hline
Combination & $\thetainc$ & $\psi$ & $\betaz$ & $\betay$ & $L$ & Voigt wave \\
 & (deg) & (deg) & (deg) & (deg) & ($\mu$m) & seen in \\
\hline
\endfirsthead
\multicolumn{7}{l}{Table~\ref{S9} (continued)}\\
\hline
Combination & $\thetainc$ & $\psi$ & $\betaz$ & $\betay$ & $L$ & Voigt wave \\
 & (deg) & (deg) & (deg) & (deg) & ($\mu$m) & seen in \\
\hline
\endhead
$\calPthree\,\&\,\calT$, $\calPthreeS\,\&\,\calT$ & $0$ & -- & -- & -- & -- & impossible \\
$\calPthree\,\&\,\calR$, $\calPthreeS\,\&\,\calR$ & $0$ & -- & -- & -- & -- & impossible \\
$\calPone\,\&\,\calTone$ & -- & -- & -- & -- & -- & not found; closest residual $7\times10^{-6}$ \\
$\calPone\,\&\,\calTtwo$ & -- & -- & -- & -- & -- & not found; closest residual $7\times10^{-6}$ \\
$\calPone\,\&\,\calTthree$ & -- & -- & -- & -- & -- & impossible ($\tLL=\tRR$) \\
$\calPtwo\,\&\,\calTone$ & -- & -- & -- & -- & -- & not found; closest residual $8\times10^{-6}$ \\
$\calPtwo\,\&\,\calTtwo$ & -- & -- & -- & -- & -- & not found; closest residual $8\times10^{-6}$ \\
$\calPtwo\,\&\,\calTthree$ & -- & -- & -- & -- & -- & impossible ($\tLL=\tRR$) \\
$\calPone\,\&\,\calRone$ & -- & -- & -- & -- & -- & not found; closest residual $2\times10^{-2}$ \\
$\calPone\,\&\,\calRtwo$ & -- & -- & -- & -- & -- & not found; closest residual $2\times10^{-2}$ \\
$\calPone\,\&\,\calRthree$ & -- & -- & -- & -- & -- & not found; closest residual $2\times10^{-3}$ \\
$\calPtwo\,\&\,\calRone$ & -- & -- & -- & -- & -- & not found; closest residual $2\times10^{-2}$ \\
$\calPtwo\,\&\,\calRtwo$ & -- & -- & -- & -- & -- & not found; closest residual $2\times10^{-2}$ \\
$\calPtwo\,\&\,\calRthree$ & -- & -- & -- & -- & -- & not found; closest residual $2\times10^{-3}$ \\
$\calPthree\,\&\,\calT$, $\calPthreeS\,\&\,\calT$ & $>0$ & -- & -- & -- & -- & impossible \\
$\calPthree\,\&\,\calR$, $\calPthreeS\,\&\,\calR$ & $>0$ & -- & -- & -- & -- & impossible \\
\hline
\end{longtable}}


\begin{thebibliography}{99}
\bibitem{real-condition}
A real condition is one equation between real-valued quantities. A complex condition is equivalent to two real conditions.
\end{thebibliography}
\end{document}